\documentclass[review]{elsarticle}

\usepackage[hidelinks,hyperfootnotes=false, bookmarks]{hyperref}
\usepackage{graphicx} 
\usepackage{mathtools}
\usepackage{amsmath}
\usepackage{amssymb}
\usepackage{amsthm}
\usepackage{bm}
\usepackage{placeins}
\usepackage{cases}
\usepackage{algorithm}
\usepackage{empheq}
\usepackage{algpseudocode}
\usepackage[utf8]{inputenc}
\usepackage{multirow}
\usepackage{booktabs}
\usepackage{colortbl}
\usepackage{color,soul}
\usepackage{array}
\biboptions{sort&compress}
\usepackage{nicefrac}
\usepackage{import}
\usepackage{epstopdf}
\usepackage{mathrsfs}
\usepackage[figurename=Fig.]{caption}
\usepackage[numbers]{natbib}
\usepackage{array}
\usepackage{amsfonts}
\usepackage{scalerel}
\usepackage[labelformat=simple]{subcaption}
\usepackage[scientific-notation=true]{siunitx}
\newtheorem{remark}{Remark}
\usepackage{color,soul}
\usepackage{blindtext, xcolor}
\usepackage{comment}

\newcommand{\grayrule}{\arrayrulecolor{gray!20} \midrule \arrayrulecolor{black}}

\usepackage{tikz}
\usepackage{amstext}
\usetikzlibrary{calc,arrows,shapes,positioning,backgrounds,fit}

\usepackage{svg}
\svgpath{{ps/}}

\DeclareMathOperator{\tr}{tr}

\newcolumntype{M}[1]{>{\centering\arraybackslash}m{#1}}

\newcolumntype{L}{>{\centering\arraybackslash}m{3cm}}
\usepackage[margin=1in]{geometry}
\usepackage{lineno,hyperref}

\algblock{Input}{EndInput}
\algnotext{EndInput}
\algblock{Output}{EndOutput}

\newcommand{\Tau}{\mathrm{T}}
\newcommand{\Bu}{\bm{u}}
\newcommand{\Be}{\bm{\varepsilon}}
\newcommand{\Bec}{\bm{\varepsilon}^\mathrm{p}}
\newcommand{\Bep}{\bm{\varepsilon}^\mathrm{p}}
\newcommand{\Bee}{\bm{\varepsilon}^\mathrm{e}}
\newcommand{\Bsig}{\bm{\sigma}}
\newcommand{\Bn}{\bm{n}}
\newcommand{\Bsp}{\bm{s}^\mathrm{p}}
\newcommand{\Bsc}{\bm{s}^\mathrm{p}}

\newcommand{\squad}{\hspace{0.5em}}

\newcommand{\overbar}[1]{\mkern 1.5mu\overline{\mkern-1.5mu#1\mkern-1.5mu}\mkern 1.5mu}

\usepackage[color=gray!20, backgroundcolor=yellow, textwidth=2cm, textsize=footnotesize]{todonotes}

\definecolor{myred}{RGB}{163,0,0}
\definecolor{myblue}{RGB}{0,67,104}
\definecolor{mygreen}{RGB}{0,104,0}
\newcommand{\tC}[1]{\textcolor{black}{#1}}
\usepackage{setspace}
\renewcommand*{\today}{ }

\begin{document}

\begin{frontmatter}

\title{A micromechanics-based variational phase-field model for fracture in geomaterials with brittle-tensile and compressive-ductile~behavior\tnoteref{t1,t2}}

\author[add1]{Jacinto Ulloa}
\cortext[JU]{Corresponding author: Jacinto Ulloa}
\ead{jacintoisrael.ulloa@kuleuven.be}
\author[add1]{Jef Wambacq}
\ead{jef.wambacq@kuleuven.be}
\author[add2]{Roberto Alessi}
\ead{roberto.alessi@unipi.it}
\author[add3]{Esteban Samaniego}
\ead{esteban.samaniego@ucuenca.edu.ec}
\author[add1]{Geert Degrande}
\ead{geert.degrande@kuleuven.be}
\author[add1]{Stijn François}
\ead{stijn.francois@kuleuven.be}

\address[add1]{KU Leuven, Department of Civil Engineering, Kasteelpark Arenberg 40, B-3001 Leuven, Belgium}
\address[add2]{Università di Pisa, Department of Civil and Industrial Engineering, Largo Lucio Lazzarino 2, 56122 Pisa, Italy}
\address[add3]{Universidad de Cuenca, Facultad de Ingeniería y Departamento de Recursos Hídricos y Ciencias Ambientales, Av. 12 de Abril y Av. Loja, 010151 Cuenca, Ecuador}

% . 104684, In press
\tnotetext[t1]{{\itshape Postprint version.}}
\tnotetext[t2]{{\itshape Published version:}
       J.~Ulloa,  J.~Wambacq, R.~Alessi, E.~Samaniego, G.~Degrande, and S.~François.
       \newblock A micromechanics-based variational phase-field model for fracture in geomaterials with brittle-tensile and compressive-ductile~behavior.
       \newblock {\em Journal of the Mechanics and Physics of Solids},  2021.\\
       {\itshape DOI: \tt\url{https://doi.org/10.1016/j.jmps.2021.104684}} \vspace{3mm}}

\begin{abstract}
This paper presents a framework for modeling failure in quasi-brittle geomaterials under different loading conditions. A micromechanics-based model is proposed in which the field variables are linked to physical mechanisms at the microcrack level: damage is related to the growth of microcracks, while plasticity is related to the frictional sliding of closed microcracks. Consequently, the hardening/softening functions and parameters entering the free energy follow from the definition of a single degradation function and the elastic material properties. The evolution of opening microcracks in tension leads to brittle behavior and mode I fracture, while the evolution of closed microcracks under frictional sliding in compression/shear leads to ductile behavior and mode II fracture. Frictional sliding is endowed with a non-associative law, a crucial aspect of the model that considers the effect of dilation and allows for realistic material responses with non-vanishing frictional energy dissipation. Despite the non-associative law, a variationally consistent formulation is presented using notions of energy balance and stability, following the energetic formulation for rate-independent systems. The material response of the model is first described, followed by the numerical implementation procedure and several benchmark finite element simulations. The results highlight the ability of the model to describe tensile, shear, and mixed-mode fracture, as well as responses with brittle-to-ductile transition. A key result is that, by virtue of the micromechanical arguments, realistic failure modes can be captured, without resorting to the usual heuristic modifications considered in the phase-field literature. The numerical results are thoroughly discussed with reference to previous numerical studies, experimental evidence, and analytical fracture criteria. 

\end{abstract}


\end{frontmatter}

\clearpage 

{\small\tableofcontents}

\clearpage 

\section{Introduction}

Failure in quasi-brittle geomaterials such as rocks and concrete is mostly driven by the growth and coalescence of microcracks. Depending on the confining pressure (among other factors such as temperature and loading rate), distinctive failure modes can be observed, with notably different behavior in tension and compression~\citep{andrieux1986,borja2004,choo2018}. Under tensile loading at low confining pressure, opening microcracks lead to brittle fracture with mode I kinematics. Conversely, compressive loading (or tensile loading at higher confinement) favors the closing of microcracks, which are subject to frictional sliding under deviatoric stress. The frictional sliding of closed microcracks manifests macroscopically as dilative plastic strains, while their growth and coalescence lead to mode II failure in the form of localized {shear bands} and shear fractures. This behavior entails a macroscopic hardening-softening response that results from the competition between friction-induced hardening and damage-induced softening~\citep{andrieux1986,zhu2011} and exhibits  pressure-dependent residual strength attributed to the frictional sliding of fracture surfaces~\citep{hajiabdolmajid2002,renani2018,peng2019}. With excessive sliding, surface asperities may deteriorate, resulting in a reduction of residual strength~\citep{zhao2018}. Finally, the macroscopic response becomes increasingly ductile and diffuse under higher confining pressure, where plastic strains may shift from dilation to compaction.
 
Several studies have been devoted to the micromechanical modeling of failure mechanisms in rock-like materials~\citep{andrieux1986, pensee2002, zhu2008, zhu2011, zhu2016, xie2011, zhao2018, jia2020}, where coupling between microcrack growth and frictional sliding is usually considered to derive constitutive equations based on homogenization schemes. Thereby, opening microcracks result in a macroscopically brittle response, while the frictional sliding of closed microcracks is assumed to obey a Coulomb-type friction law, leading to macroscopic pressure-dependent plasticity coupled to damage. This micromechanical framework presents analogies with cohesive-frictional elastoplasticity models and continuum damage mechanics: Mohr-Coulomb--based plasticity models with hardening can be considered to account for the frictional sliding of microcracks, while the macroscopic loss of stiffness and cohesive strength predicted by continuum damage mechanics accounts for microcrack growth in the bonding material. As such, phenomenological damage models coupled to plasticity at the continuum scale appear as a viable option to describe failure in rock-like materials~\citep{vermeer1984, lubliner1989, chazallon1998, chiarelli2003, salari2004, wu2006, parisio2015}. While models of this type have shown great predictive ability, in most cases, phenomenological assumptions are made with no link to the underlying micromechanical processes,  resulting in a large number of parameters with no clear physical interpretation. Some works~\citep{lanoye2013,marigo2019} which are still developed at the continuum scale overcome this limitation, presenting plastic-damage models that are directly related to micromechanics and are thus particularly interesting from a physical standpoint.

The plastic-damage models discussed above adopt a constitutive framework with local internal variables. It is well established that this modeling approach is unable to handle softening responses with strain localization, where the loss of ellipticity of the governing equations leads to pathological mesh-dependence and vanishing energy dissipation in finite element simulations. Moreover, it has been reported~\cite{rudnicki1975,needleman1979,sabet2019} that the use of non-associative models, the de facto approach for geomaterials, is also an underlying cause of such issues, even in the absence of material softening. The modeling of geomaterials under failure conditions thus requires a delicate treatment that allows for the description of localized responses, for which different families of methods have been proposed. An example is the explicit introduction of deformation modes in the finite element technology~\cite{jirasek2000comparative}, either by considering jumps in the strain field (weak discontinuities)~\cite{ortiz1987finite,belytschko1988finite} or jumps in the displacement field (strong discontinuities)~\cite{oliver1999, OliHueSamCha2004, regueiro2001, foster2007, ChenAndSam2011, wu2016,zhao2018b}. Another well-established approach is the use of enhanced continuum theories with regularized kinematics, including rate-dependent~\cite{oka1995,cervera1996,de2020}, Cosserat continuum~\cite{muhlhaus1987thickness,de1991localisation}, non-local~\cite{bazant1988,bazant2002,jirasek2004,grassl2006,yoshioka2019}, and gradient-enhanced~\citep{muhlhaus1991,de1996,peerlings1998,comi1999,pamin2003,zreid2018} models. While these works have mainly focused on material softening, surprisingly, little attention has been given to the regularization of non-associative models. Recent works~\cite{sabet2019,de2020,hageman2021} on this topic employ rate-dependent and/or Cosserat continuum models to obtain mesh-objectivity in non-associative plasticity. Another instance can be found in~\citet{ulloa2021a}, where gradient-enhanced plasticity is considered for the same purpose.

The present study fits within the framework of gradient-enhanced theories in the context of the phase-field approach to fracture, which has received significant attention in the past decade due to its ability to naturally describe crack nucleation and complex crack topologies. In particular, we aim at developing a gradient-damage/phase-field model capable of predicting complex failure mechanisms in rock-like materials, including non-associative laws, while preserving a tight link to the underlying micromechanical processes.

As initially conceived by~\citet{BourFrancMar2000}, the phase-field approach to brittle fracture represents a smooth approximation in the sense of $\Gamma$ convergence to the variational formulation of Griffith's fracture, such that the regularized variational problem of~\citet{BourFrancMar2000} recovers the free discontinuity problem of~\citet{FrancMar1998} as the regularization parameter goes to zero. The regularized functional depends on a continuous phase-field variable that characterizes broken and unbroken material states. For intermediate states, the phase-field variable acts as a degradation mechanism on the elastic strain energy density. Thus, although not initially viewed as such~\cite{BourFrancMar2000,BourFrancMar2008}, the phase-field approach to fracture was later interpreted and widely accepted as a gradient-enhanced damage model in the mechanics community~\cite{amor2009,MieHofWel2010,pham2011,marigo2016,kristensen2021}. The regularization parameter is then viewed as a material internal length scale related to the strength, allowing for a quantitative prediction of crack nucleation~\cite{tanne2018,delorenzis2021}.

During the past decade, several modifications of the phase-field model were proposed to account for different failure mechanisms. For instance, noting that the original phase-field model predicts crack interpenetration (at the same critical stress as fracture in tension), various decompositions of the strain energy density were proposed to preclude fracture in compression~\cite{amor2009,MieWelHof2010,van2020}. Energy decomposition was also considered to model shear fracture~\cite{lancioni2009}, where tensile fracture is also precluded. A more general formulation was proposed by~\citet{freddi2010regularized}. This formulation was recently extended by~\citet{delorenzis2021}, including a Drucker-Prager--type strength surface in order to capture crack nucleation in a multiaxial setting. Further, energy decomposition has been applied to model plastic slip bands~\cite{freddi2016}. In all these references, the variational structure of the underlying theory is preserved. However, other works abandon the variational structure in favor of flexibility; examples in the context of  tension/compression asymmetry with energy~decomposition~\cite{ambati2015review,wu2018length} and without energy~decomposition~\cite{kumar2020}  can be found in the~literature.

Distinguishing between tensile and compressive/shear states in phase-field models is not sufficient to capture the failure mechanisms observed in quasi-brittle geomaterials. For instance, phase-field models typically consider a single toughness parameter. However, in rock-like materials, the critical energy release rate (or fracture toughness) for mode I (tensile) fracture is significantly lower than the critical energy release rate  for mode II (shear) fracture~\cite{shen1994}. Motivated by these shortcomings, modifications of the phase-field evolution equations to incorporate distinctive mode I and mode II toughness parameters have been recently proposed, aiming to account for tensile, shear, and mixed-mode fracture. This idea was first proposed in~\citet{zhang2017}, where the ratio between the crack driving force and the fracture toughness is additively decomposed into mode I and mode II contributions based on a spectral decomposition of the strain tensor (see also~\citet{spetz2020}, where fracture in compression was further considered). The resulting fracture criterion is analogous to the $F$-criterion of~\citet{shen1994} in fracture mechanics. This approach was further developed in~\citet{bryant2018}, including anisotropic fracture and a decomposition of the ratio between the crack driving force and the fracture toughness based on the crack kinematics. Therein, the crack orientation is obtained from an optimization problem based on the $F$-criterion. A limitation of these works is that pressure-dependent frictional behavior, a well-known phenomenon in geomaterials, is~neglected. Therefore, \citet{fei2021} proposed a double phase-field model based on their recent work on cracks with frictional contact~\cite{fei2020b}, where the $F$-criterion is employed to determine the dominant failure mode. Therein, the orientation of tensile cracks is given by the major principal direction in the slip plane, while the orientation of shear cracks is given by Mohr-Coulomb's failure angle with respect to the major~principal~direction.

The frictional behavior of geomaterials has also been considered by incorporating plasticity in the formulation. Although less studied than the brittle case, by now, several ductile phase-field models have been proposed, with initial works including  variational~\cite{alessi2015,kuhn2016,MieTeiAld2016,rodriguez2018} and non-variational~\cite{AmbGerDeL2015,miehe2015b} approaches (see references~\cite{alessi2018comparison,noii2021} for overviews). These works focused on von-Mises plasticity and are thus not applicable to geomaterials. However, recent works have proposed phase-field models coupled to frictional plasticity, usually of the Drucker-Prager type. For instance, a double phase-field model for tensile, shear, and mixed-mode fracture with plasticity was proposed in~\citet{you2020}, while other studies~\cite{choo2018,you2021} focus on modeling the brittle-to-ductile transition phenomenon. Phase-field models coupled to frictional plasticity have been further considered in the finite strain setting~\cite{choo2018,kienle2019}, as well as in the context of multiphase materials~\cite{spath2021} and fluid-driven fracture~\cite{aldakheel2020,kienle2021}. Moreover, building upon the phase-field modeling of frictional interfaces~\cite{fei2020b},~\citet{bryant2021} proposed a model that embeds rate-, size-, and temperature-dependent friction.

It can thus be concluded that fruitful progress on the phase-field modeling of fracture has been made in the past few years, paving the way for developments in computational geomechanics. Nevertheless, several issues remain open. For instance, modifications of the original brittle phase-field model, e.g., based on energy splits, generally involve assumptions that lose track of the underlying micromechanical processes. Moreover, ductile phase-field models are based on phenomenological plasticity, inheriting the lack of physical significance of internal variables and material parameters at lower scales. In addition, to our knowledge, a model that accounts for mixed-mode fracture, as well as non-associative pressure-dependent plasticity has not been proposed, although it is well established that the dilation angle plays a crucial role in realistically capturing the stiffness, load-carrying capacity, volumetric plastic strains, and shear band orientation in geomaterials~\cite{arthur1977,vardoulakis1980,vermeer1984}. Finally, a common feature of models that embed complex behaviors is that the variational structure inherent to the original phase-field theory is lost in favor of greater flexibility. This has been the case in most (if not all) extensions to mixed-mode fracture and frictional plasticity. 

In the present contribution, we propose a coupled plasticity--phase-field model for tensile, shear, and mixed-mode fracture in quasi-brittle geomaterials, including a physically meaningful brittle-to-ductile transition.  As a key feature, the model is derived from micromechanical arguments, revealing a clear link between the field variables at the macroscale and the dissipative mechanisms at the microcrack level, namely, microcrack growth and frictional sliding. The advantages of this approach include the incorporation of tension-compression asymmetry without heuristic energy decompositions; the distinction between mode I and mode II fracture regimes, each embedding its own evolution laws and parameters; and the definition of parameters and degradation functions that naturally follow from micromechanical arguments. As such, a single degradation function must be chosen to derive a constitutive model able to describe a variety of failure mechanisms in quasi-brittle materials. As another main feature, the proposed model admits a variational formulation in the sense of the energetic formulation for rate-independent systems~\cite{Mielke2006,Mielke2015}, where all governing equations are derived from principles of energy balance and stability. Further, dilatancy effects are considered, for which the plastic strains are assumed to obey a non-associative friction law. As will be discussed, this assumption is not only realistic but also necessary for a physically meaningful non-vanishing plastic dissipation due to frictional sliding. The non-associative evolution law is carefully introduced in the variational formulation using a recently developed generalization of the principle of maximum dissipation~\citep{ulloa2021a}.

\section{Micromechanics-based variational phase-field model}

This section presents the proposed micromechanics-based phase-field model for fracture in quasi-brittle geomaterials. In order to introduce the notations and the modeling framework, a brief account on  thermomechanical modeling with gradient-enhanced internal variables is provided in section~\ref{framework}, where the constitutive model is defined in terms of a free energy and a dissipation potential. Then, section~\ref{local_micro} describes the local micromechanics-based model from which the present work is inspired, focusing on the micromechanical origin of damage and plasticity in microcracked solids. The proposed phase-field model is presented in section~\ref{proposed}, where the free energy and the dissipation potential are derived in agreement with the micromechanical arguments of section~\ref{local_micro}. Therein, the governing equations are derived in variational form using notions of energy balance and stability. Finally, section~\ref{homog} presents an illustrative description of the homogeneous response. 

\subsection{Problem outline and thermomechanical framework}\label{framework}

Consider the evolution of an arbitrary solid of mass density $\rho$ occupying a domain $\Omega\subset\mathbb{R}^3$ (figures~\ref{scheme0} and~\ref{scheme1}) during a pseudo-time (loading) interval $\Tau\coloneqq[0,t_\mathrm{max}]$. The boundary, denoted as $\Gamma$, consists of a Dirichlet part ${\Gamma_\mathrm{D}}$ with imposed displacements $\overbar{\bm{u}}(\bm{x},t)\in\mathbb{R}^3$ and a Neumann part $\Gamma_\mathrm{N}$ with imposed tractions $\bar{\bm{t}}(\bm{x},t)\in\mathbb{R}^3$, such that ${\Gamma_\mathrm{D} \cup \Gamma_\mathrm{N} = \Gamma}$ and ${\Gamma_\mathrm{D} \cap \Gamma_\mathrm{N} =\varnothing}$. The solid may be subjected to body forces per unit mass, denoted as $\bm{b}(\bm{x},t)\in\mathbb{R}^3$.

In the framework of thermomechanics with internal variables, the response of the solid is characterized by the displacement field $\Bu\colon\Omega\times\Tau\to\mathbb{R}^3$ and a generic set of internal variables with $m$ degrees of freedom, which we shall denote for now as  $\mathbf{a}\colon\Omega\times\mathrm{T}\to \mathbb{R}^{m}$. Moreover, in order to accommodate gradient-enhanced theories, e.g., phase-field models, the first-order spatial gradients contained in $\nabla\mathbf{a}\colon\Omega\times\mathrm{T}\to \mathbb{R}^{3m}$ are included in the formulation. Throughout this work, we assume the small-strain hypothesis, such that the compatible strain tensor $\boldsymbol{\varepsilon}\colon\Omega\times\mathrm{T}\to\mathbb{R}^{3\times 3}_\mathrm{sym}\coloneqq\{\bm{e}\in \mathbb{R}^{3\times 3} \, | \,  \bm{e}=\bm{e}^\mathrm{T}\}$ obeys the linear relation $\boldsymbol{\varepsilon}=\nabla^{\mathrm{s}}\boldsymbol{u}$.

To introduce dissipative behavior, the proposed variational model adopts the theory of generalized standard materials~\cite{halphen1975}, where the evolution of the system is characterized by two basic energy quantities: an internal energy density and a dissipation potential. Concerning the former, we let $\psi\coloneqq\psi(\bm{\varepsilon},\mathbf{a},\nabla\mathbf{a})$ denote a Helmholtz-type free energy density. The second law of thermodynamics is taken as an a priori restriction, given for isothermal processes by the Clausius-Planck inequality 
\begin{equation}
\delta\coloneqq\bm{\sigma}:\dot{\bm{\varepsilon}} -\dot{\psi}(\bm{\varepsilon},\mathbf{a},\nabla\mathbf{a})\geq0.
\label{CD}
\end{equation}
Applying the Coleman-Noll procedure, the constitutive stress-strain relation
\begin{equation}
\bm{\sigma}=\frac{\partial\psi}{\partial{\bm{\varepsilon}}}(\bm{\varepsilon},\mathbf{a},\nabla\mathbf{a})
\label{constitutive}
\end{equation}
follows from equation~\eqref{CD}, where $\bm{\sigma}\colon\Omega\times\mathrm{T}\to\mathbb{R}^{3\times 3}_\mathrm{sym}$ is the Cauchy stress tensor. Static admissibility  is then enforced for all $t\in \mathrm{T}$ through the equilibrium equations
\begin{equation}
\mathrm{div}\,\bm{\sigma}+\rho\bm{b} = \boldsymbol{0} \quad \text{in} \squad \Omega \qquad \text{and} \qquad  \bm{\sigma}\cdot\boldsymbol{n}=\bar{\bm{t}} \quad \text{on} \squad \Gamma_{\mathrm{N}}, \qquad	\text{with} \qquad \bm{u}=\overbar{\bm{u}} \quad \text{on} \squad \Gamma_{\mathrm{D}}.
\label{st_adm}
\end{equation}
On the other hand, the generalized stresses conjugate to $\mathbf{a}$ read
\begin{equation}
\mathbf{s}=-\delta_{\mathbf{a}}\psi(\bm{\varepsilon},\mathbf{a},\nabla\mathbf{a}),
\label{dualforces} 
\end{equation}
where $\delta_{\Box}\coloneqq\partial_{\Box}-\mathrm{div}[\partial_{\nabla{\Box}}]$ denotes the spatial Euler-Lagrange operator. Using equations~\eqref{constitutive} and~\eqref{dualforces}, the Clausius-Planck inequality~\eqref{CD} yields the conditions
\begin{equation}
\mathbf{s}\cdot\dot{\mathbf{a}}\geq0\quad\text{in}\squad\Omega\qquad\text{and}\qquad-\big(\Bn\cdot\partial_{\nabla\mathbf{a}}\psi(\bm{\varepsilon},\mathbf{a},\nabla\mathbf{a})\big)\cdot\dot{\mathbf{a}}\geq0\quad\text{on}\squad\Gamma.
\label{red_diss}
\end{equation}
For rate-independent systems admitting a variational formulation, the dissipation rate may be cast as a thermodynamically admissible primal dissipation potential of the form $\delta\coloneqq
\phi(\dot{\mathbf{a}},\nabla{\dot{\mathbf{a}}};\mathbf{c},\mathbf{s})\geq0$. Following our recent work~\cite{ulloa2021a}, the dissipation potential is allowed to depend on the generalized stresses $\mathbf{s}$ to account for non-associative flow, a key aspect of the model proposed in the present study. Moreover, in the rate-independent setting, $\phi$ is a convex homogeneous function of first degree in $\{\dot{\mathbf{a}},\nabla{\dot{\mathbf{a}}}\}$. As a consequence, $\phi$ is not differentiable at null rates (where $\phi$ also vanishes). From equation~\eqref{CD} and the identification $\delta= \phi(\dot{\mathbf{a}},\nabla{\dot{\mathbf{a}}};\mathbf{c},\mathbf{s})$, it follows that $$\mathbf{s}\in\delta_{\dot{\mathbf{a}}}\phi(\dot{\mathbf{a}},\nabla{\dot{\mathbf{a}}};\mathbf{c},\mathbf{s})=\partial_{\dot{\mathbf{a}}}\phi(\dot{\mathbf{a}},\nabla{\dot{\mathbf{a}}};\mathbf{c},\mathbf{s})-\mathrm{div}\big[\partial_{\nabla\dot{\mathbf{a}}}\phi(\dot{\mathbf{a}},\nabla{\dot{\mathbf{a}}};\mathbf{c},\mathbf{s})\big],$$ where, in the context of convex analysis, the operator $\partial_\bullet\Box(\bullet)$ denotes the multivalued subdifferential of $\Box$ at $\bullet$ (hereinafter, the argument in the subscript will only be included when required for clarity).
In view of equation~\eqref{dualforces}, the evolution equation for the internal variables $\mathbf{a}$ takes the form of a non-local Biot-type inclusion~(cf.~\citep[chapter~6, equation~2.18]{biot1965} for the local rate-dependent case):
%%
%\begin{equation}
%\delta_{{\mathbf{a}}}\psi(\bm{\varepsilon},\mathbf{a},\nabla\mathbf{a}) + \delta_{\dot{\mathbf{a}}}\phi(\dot{\mathbf{a}},\nabla\dot{\mathbf{a}};\mathbf{s})\ni \bm{0}.
%\label{biot}
%\end{equation}
%%
\begin{equation}
\delta_{{\mathbf{a}}}\psi(\bm{\varepsilon},\mathbf{a},\nabla\mathbf{a}) + \delta_{\dot{\mathbf{a}}}\phi(\dot{\mathbf{a}},\nabla\dot{\mathbf{a}};\mathbf{c},\mathbf{s})\ni \bm{0}\squad\text{in}\squad\Omega \quad \text{and} \quad \big(\partial_{\nabla\mathbf{a}}\psi(\bm{\varepsilon},\mathbf{a},\nabla\mathbf{a})+\partial_{\nabla\dot{\mathbf{a}}}\phi(\dot{\mathbf{a}},\nabla\dot{\mathbf{a}};\mathbf{c},\mathbf{s})\big)\cdot\bm{n}\ni\bm{0}\squad\text{{on}}\squad\Gamma.
\label{biot}
\end{equation}
Equations \eqref{st_adm} and \eqref{biot} represent the strong form of the evolution problem for a general dissipative model with gradient-enhanced internal variables. The model described in the sequel fits within this general formulation.

\subsection{Micromechanics background}\label{local_micro}

In this section, we introduce the micromechanical framework from which the present work is inspired. The (local) formulation presented here is not new but will serve as the conceptual backbone of the new phase-field fracture model proposed in section~\ref{proposed}. Since the objective is to relate the field variables of the proposed model to lower-scale mechanisms, only a brief summary of micromechanical arguments is provided below. For further details on this topic, we refer the reader to~\citet{zhu2008,zhu2011,zhu2016} and related works.

\begin{figure}[htb!]
  \centering
    \footnotesize
    \includeinkscape[scale=1]{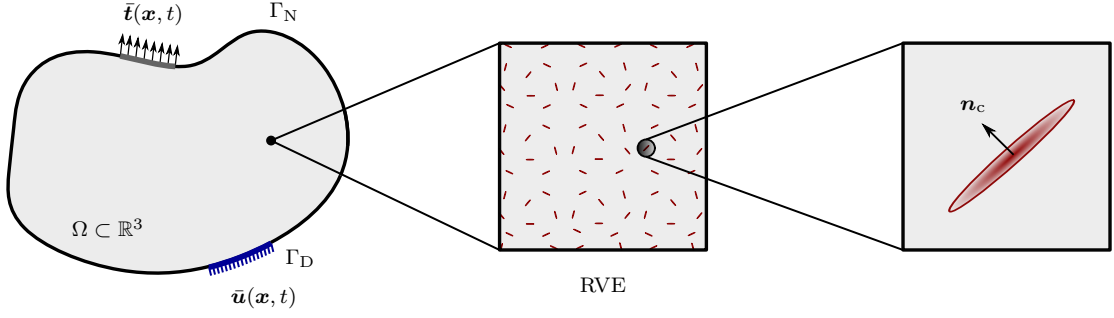}
\caption{Schematic representation of the continuum solid and  the boundary value problem at the macroscale  (left), the RVE consisting of a matrix-inclusion system with penny-shaped microcracks (middle), and the orientation of a microcrack (right).}
\label{scheme0}
\end{figure}

Closely following these references, we consider a solid matrix material with penny-shaped microcracks that are assumed much smaller than the size of the representative volume element (RVE). Figure~\ref{scheme0} schematically depicts a material of this type, where an RVE is described as a matrix-inclusion system. Furthermore, we assume that the microcracks are uniformly distributed in all directions, such that isotropic behavior of the RVE can be considered, and that the matrix material is linear elastic, characterized by a bulk modulus $K$ and a shear modulus $\mu$. Thus, the corresponding fourth-order elasticity tensor reads
\begin{equation}
\bm{\mathsf{C}}=K\bm{1}\otimes\bm{1}+2\mu\bigg(\bm{\mathsf{I}}-\frac{1}{3}\bm{1}\otimes\bm{1}\bigg),
\label{Celas}
\end{equation}
where the second- and fourth-order identity tensors $\bm{1}$ and $\bm{\mathsf{I}}$ have been used. For now, it is assumed that the solid domain $\Omega$ does not present strain localization at the macroscale. The macroscopic response of the microcracked solid at the RVE level can then be described as follows.

In view of the small strain hypothesis, the macroscopic strain tensor $\Be$ can be additively decomposed as
\begin{equation}
\Be=\Bee+\Bec.
\label{strain_decomp}
\end{equation}
The elastic part $\Bee$ is attributed to elastic deformation of the matrix  material, while the inelastic part $\Bec$ is related to the microcrack-induced displacement discontinuities~\cite{zhu2008}. In particular, given a family of microcracks with normal $\bm{n}_\mathrm{c}$, one has 
\begin{equation}
\Bec=\frac{1}{4\pi}\int_{\mathrm{S}^2}\big( \beta \bm{n}_\mathrm{c}\otimes\bm{n}_\mathrm{c} + \bm{\gamma}\otimes^\mathrm{s}\bm{n}_\mathrm{c} \big)\,\mathrm{d}S,
\label{inel_strain_decomp0}
\end{equation}
where $\mathrm{S}^2$ is the surface of a unit sphere, $\beta$ corresponds to the normal microcrack opening, and $\bm{\gamma}$ corresponds to the tangential displacement discontinuities. Further, given the assumption of isotropically distributed microcracks, $\Bec$ admits a volumetric-deviatoric decomposition~\cite{zhu2011}:
\begin{equation}
\Bec=\Bec_\mathrm{dev}+\frac{1}{3}\tr\Bec\bm{1}, \quad \text{with} \quad \Bec_\mathrm{dev}\equiv \frac{1}{4\pi}\int_\mathrm{S^2}\bm{\gamma}\otimes^\mathrm{s}\bm{n}_\mathrm{c}\,\mathrm{d}S, \quad \tr\Bec\equiv \beta.
\label{inel_strain_decomp}
\end{equation}
At this point, an additive decomposition of the mechanical problem in the RVE is typically invoked~\cite{zhu2011}, which, from standard arguments of homogenization theory, reveals that the macroscopic stress equals the homogeneous stress field in the matrix material:
\begin{equation}
\Bsig=\bm{\mathsf{C}}:(\Be-\Bec).
\label{stress_decomp}
\end{equation}

By virtue of the adopted homogenization scheme and the assumption of isotropic behavior at the RVE level, we may view the state of the RVE as corresponding to one of two possible scenarios: (i) a state of open microcracks and (ii) a state of closed microcracks. Let us address these cases below.

\paragraph*{Open microcracks} 
In this case, energy dissipation is solely produced by microcrack growth, and $\Bec$ follows from a linear homogenization procedure as a \emph{state function} of the total strain tensor $\Be$ and the microcrack density. Specifically, adopting the Mori-Tanaka scheme~\citep{mori1973}, one~has
\begin{equation}
\Bec=\bigg[\frac{1}{3}\frac{b_K\omega}{1+b_K\omega}\bm{1}\otimes\bm{1} + \frac{b_\mu\omega}{1+b_\mu\omega}\bigg(\bm{\mathsf{I}}-\frac{1}{3}\bm{1}\otimes\bm{1}\bigg)\bigg]:\Be \quad \text{with} \quad b_K=\frac{16}{9}\frac{1-\nu^2}{1-2\nu}\,, \quad b_\mu=\frac{32}{45}\frac{(1-\nu)(5-\nu)}{2-\nu}\, ,
\label{ec_open}
\end{equation}
where $\nu$ is the Poisson's ratio of the matrix material and $\omega$ is an internal variable directly related to the microcrack density in the RVE~\cite{zhu2011}. This relation allows us to express the stress tensor~\eqref{stress_decomp} as $\Bsig=\bm{\mathsf{C}}^\mathrm{hom}(\omega):\Be$, where $\bm{\mathsf{C}}^\mathrm{hom}(\omega)$ represents an {effective} elasticity tensor given for isotropic materials by
\begin{equation}
\bm{\mathsf{C}}^\mathrm{hom}(\omega)=K^\mathrm{hom}(\omega)\bm{1}\otimes\bm{1}+2\mu^\mathrm{hom}(\omega)\bigg(\bm{\mathsf{I}}-\frac{1}{3}\bm{1}\otimes\bm{1}\bigg),
\label{Chom}
\end{equation}
with the effective bulk and shear moduli $K^\mathrm{hom}(\omega)$ and $\mu^\mathrm{hom}(\omega)$  given by
\begin{equation}
K^\mathrm{hom}(\omega)=\frac{K}{1+b_K\omega} \qquad  \text{and} \qquad \mu^\mathrm{hom}(\omega)=\frac{\mu}{1+b_\mu\omega}\,.
\label{lame_hom}
\end{equation}
Accordingly, the strain energy density of the RVE can be written~as
\begin{equation}
\psi^\mathrm{open}(\Be,\omega)=\frac{1}{2}\,\Be:\bm{\mathsf{C}}^\mathrm{hom}(\omega):\Be.
\label{psi_open}
\end{equation}
The inelastic response of the microfractured material with opening microcracks is thus characterized by the single dissipative internal variable $\omega$ and the stiffness function $\omega\mapsto\bm{\mathsf{C}}^\mathrm{hom}(\omega)$. Note that equation~\eqref{psi_open} has the typical form of a quadratic elastic strain energy density in continuum damage mechanics, revealing the growth of microcracks as the microscopic origin of  damage at the macroscale. Consequently, the microcrack density parameter $\omega$ can be interpreted at the macroscale as a \emph{damage} variable.

\paragraph*{Closed microcracks} 

In this case, the evolution of $\Bec$ must account for the frictional sliding of closed microcracks, where the normal component is due to surface asperities. Thus, $\Bec$ can no longer be defined as a state function of $\Be$ and $\omega$, as in equation~\eqref{ec_open}; instead, it must become a \emph{state variable} obeying a suitable dissipative friction law. The strain energy density of the RVE then takes the  form~\cite{andrieux1986,zhu2011,marigo2019}
\begin{equation}
\psi^\mathrm{close}(\Be,\Bec,\omega)=\frac{1}{2}(\Be-\Bec):\bm{\mathsf{C}}:(\Be-\Bec) + \frac{1}{2}\,\Bec:\bm{\mathsf{H}}^\mathrm{block}(\omega):\Bec.
\label{psi_closed}
\end{equation}
The first term in~\eqref{psi_closed} corresponds to the elastic strain energy stored in the solid matrix material. The second term represents the energy \emph{blocked} by the frictional contact of closed microcracks, where the fourth-order tensor $\bm{\mathsf{H}}^\mathrm{block}(\omega)$  accounts for the coupling between microcrack growth and frictional sliding. Note that~\eqref{psi_closed} has the typical form of a quadratic elastoplastic stored energy density with a kinematic hardening term~\cite{marigo2019,ulloa2021a}. Indeed, the inelastic strains $\Bec$ may now be interpreted at the macroscale as \emph{plastic strains}, revealing the frictional sliding of microcracks as the microscopic origin of plasticity in quasi-brittle materials. However, as opposed to purely phenomenological hardening/softening plasticity models, the function ${\omega\mapsto \bm{\mathsf{H}}^\mathrm{block}(\omega)}$ does not require the definition of phenomenological hardening/softening laws and parameters. Instead, it follows directly from continuity requirements as a function of $\bm{\mathsf{C}}^\mathrm{hom}(\omega)$ and $\bm{\mathsf{C}}$, as shown~below.

\bigskip

Employing the Coleman-Noll procedure, the stress-strain relations at the macroscale follow as
\begin{equation}
\Bsig_\mathrm{open}=\frac{\partial\psi^\mathrm{open}}{\partial\Be}=\bm{\mathsf{C}}^\mathrm{hom}(\omega):\Be \qquad \text{and} \qquad \Bsig_\mathrm{close}=\frac{\partial\psi^\mathrm{close}}{\partial\Be}=\bm{\mathsf{C}}:(\Be-\Bec)
\label{sigma0}
\end{equation}
for the cases of open and closed microcracks, respectively. Moreover,  the generalized stresses conjugate to the inelastic strain tensor read
\begin{equation}
\Bsc_\mathrm{open}=-\frac{\partial\psi^\mathrm{open}}{\partial\Bec}=\bm{0} \qquad \text{and} \qquad  \Bsc_\mathrm{close}=-\frac{\partial\psi^\mathrm{close}}{\partial\Bec}=\bm{\mathsf{C}}:(\Be-\Bec) - \bm{\mathsf{H}}^\mathrm{block}(\omega):\Bec,
\label{sp0}
\end{equation}
which correspond to the local stress field acting on the microcrack surfaces in the RVE~\cite{zhu2011}. The first expression above is in agreement with the physical condition of vanishing contact tractions in opening microcracks. Likewise, the generalized stresses conjugate to the microcrack density read
\begin{equation}
s^\mathrm{d}_\mathrm{open}=-\frac{\partial\psi^\mathrm{open}}{\partial\omega}=-\frac{1}{2}\,\Be:{\bm{\mathsf{C}}^\mathrm{hom}}'(\omega):\Be \qquad \text{and} \qquad s^\mathrm{d}_\mathrm{close}=-\frac{\partial\psi^\mathrm{close}}{\partial\omega}=-\frac{1}{2}\,\Bec:{\bm{\mathsf{H}}^\mathrm{block}}'(\omega):\Bec.
\label{sd0}
\end{equation}

At this point, we shall invoke the requirement that the stresses~\eqref{sigma0}, \eqref{sp0}, and~\eqref{sd0} be continuous during microcrack opening/closure transitions, thus avoiding unphysical jumps in the considered material behavior. Note that fulfilling this requirement implies that the free energy is continuously differentiable. As a consequence, the following conditions hold at the microcrack \emph{opening/closure transition}:
\begin{equation}
\Bsig_\mathrm{close}=\Bsig_\mathrm{open},  \qquad \Bsc_\mathrm{close}=\Bsc_\mathrm{open}, \qquad \text{and}\qquad s^\mathrm{d}_\mathrm{close}=s^\mathrm{d}_\mathrm{open}.
\label{cont_cond}
\end{equation}
This equivalence principle implies two crucial results. The first is the cancellation of the local stress field acting on the microcrack surfaces, that is:
\begin{equation}
\Bsc_\mathrm{close}=\bm{0}
\label{optrans00}
\end{equation}
at the opening/closure transition. The second is that the functional form of the coupling tensor ${\omega\mapsto\bm{\mathsf{H}}^\mathrm{block}(\omega)}$ cannot be arbitrary and must be given by
\begin{equation}
\bm{\mathsf{H}}^\mathrm{block}(\omega)=\Big[ {\bm{\mathsf{C}}^\mathrm{hom}}^{-1}(\omega) - \bm{\mathsf{C}}^{-1} \Big]^{-1}=\frac{K}{b_K\omega}\bm{1}\otimes\bm{1}+\frac{2\mu}{b_\mu\omega}
\bigg(\bm{\mathsf{I}}-\frac{1}{3}\bm{1}\otimes\bm{1}\bigg).
\label{Hblock}
\end{equation}
Finally, we note that, in order to ensure the continuity requirements~\eqref{cont_cond} during both closed-to-open and open-to-closed transitions, condition~\eqref{optrans00} must hold not only at the transition but also during the entire opening regime. This requirement represents a key constraint for the construction of the model. 

\begin{remark}\label{rem1}
Condition~\eqref{optrans00} can be viewed as a constraint for the evolution of $\Bep$ at the opening/closure transition as well as during the entire opening regime, providing
\begin{equation}
\Bec=\big[ \bm{\mathsf{H}}^\mathrm{block}(\omega) + \bm{\mathsf{C}} \big]^{-1}\bm{\mathsf{C}}:\Be\equiv \bigg[\frac{1}{3}\frac{b_K\omega}{1+b_K\omega}\bm{1}\otimes\bm{1} + \frac{b_\mu\omega}{1+b_\mu\omega}\bigg(\bm{\mathsf{I}}-\frac{1}{3}\bm{1}\otimes\bm{1}\bigg)\bigg]:\Be,
\label{et_at_trans}
\end{equation}
in agreement with equation~\eqref{ec_open}. In this case, $\Bec$ is physically linked to  displacement discontinuities of opening microcracks and is thus not a dissipative mechanism. Indeed, as a consequence of continuity, replacing~\eqref{et_at_trans} in the free energy density~\eqref{psi_closed} and the generalized stresses \eqref{sigma0}$_2$, \eqref{sp0}$_2$, and~\eqref{sd0}$_2$ for closed microcracks yields the counterpart expressions for open microcracks~\eqref{psi_open}, \eqref{sigma0}$_1$, \eqref{sp0}$_1$, and~\eqref{sd0}$_1$. Conversely, for closed microcracks in the frictional sliding regime, $\Bec$ is dissipative. Therefore, in that case, its evolution law is separately postulated, but such that~\eqref{et_at_trans} is recovered at the transition in order to ensure continuity.
\end{remark}

To complete the evolution problem, dissipative evolution equations for the internal variables $\Bec$ and $\omega$ are usually defined in a standard thermomechanical framework. Given the physical meaning of the generalized stress~\eqref{sp0} as the local stress field acting on the surfaces of closed microcracks~\cite{zhu2008}, a Coulomb frictional sliding criterion is postulated in terms of $\Bsc_\mathrm{close}$. Invoking, once more, the assumption of isotropic behavior, the normal component of $\Bsc_\mathrm{close}$ is fully determined by $\tr\Bsc_\mathrm{close}$, and the sliding criterion takes the form of a non-cohesive Drucker-Prager yield function in generalized stress space (but cohesive in true stress space). As such, the opening/closure states are characterized by
\begin{equation*}
\Bep:\begin{dcases}\text{non-dissipative, obtained from equation~\eqref{et_at_trans}}\quad &\text{if }  \tr\Bsc_\mathrm{close}=0 \quad \text{(open microcracks)}, \\
\text{dissipative, obtained from the friction law} \quad &\text{if } \tr\Bsc_\mathrm{close}<0 \quad \text{(closed microcracks)} \hfill .\end{dcases}
\label{optrans1}
\end{equation*} 
From the friction criterion, the evolution of $\Bec$ is typically defined in previous works~\citep{zhu2008,zhu2011,zhu2016,marigo2019}  by invoking the normality rule. Similarly, a local damage criterion is postulated in terms of the generalized stresses~\eqref{sd0} to characterize the evolution of the microcrack density~$\omega$.

It is worth mentioning that the evolution equations discussed above in terms of generalized stress criteria can also be cast in the form of equation~\eqref{biot}, that is, in terms of a primal dissipation potential $\phi$. The latter approach will be considered in the model proposed in the sequel. Moreover, it will be noted that the use of the normality law for the evolution of $\Bep$ leads to vanishing energy dissipation due to frictional sliding. Therefore, a non-associative law, which can also be considered in the micromechanical framework~\cite{xie2011}, will be advocated in the proposed model. On the other hand, the local damage variable $\omega$ will be enriched with non-local effects and replaced with a phase-field variable for the description of fracture at the macroscale. 

\subsection{Proposed micromechanics-based phase-field model for macroscopic cracks}\label{proposed}

Being inherently local, the model presented in section~\ref{local_micro} is not suitable for damaging solids with localized responses at the macroscale. To approach this problem, \citet{zhao2018b} proposed a treatment based on the strong discontinuity approach for localized macrocracks. In the present study, we propose an alternative based on regularized kinematics in the context of the gradient-damage/phase-field approach to~fracture. 

The model proposed below follows a rigorous variational framework in agreement with evolution equations of the form of~\eqref{biot}. As such, the free energy density is first derived based on the micromechanical framework presented in section~\ref{local_micro}. Then, to characterize the dissipative evolution of the internal variables, a thermodynamically admissible dissipation potential is defined, including contributions from microcrack growth and frictional sliding. Both the free energy and the dissipation potential will be taken as inputs in the variational formulation of the evolution problem presented in section~\ref{formulation}.

\begin{figure}[H]
  \centering
    \footnotesize
    \includeinkscape[scale=1]{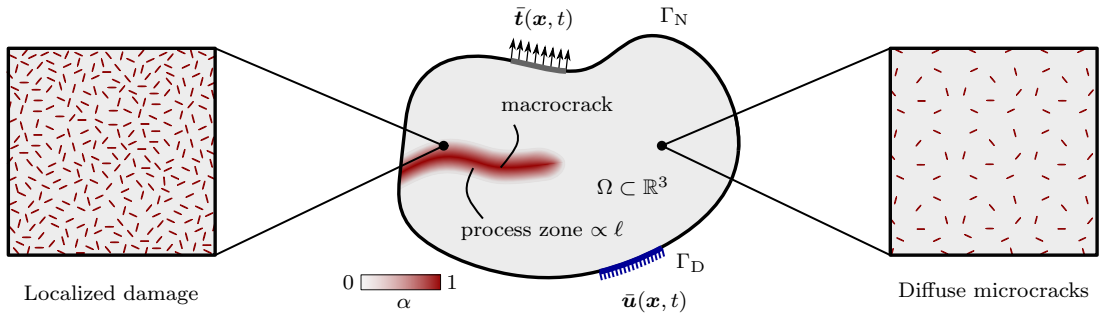}
\caption{Micromechanics-based phase-field description of localized macrocracks. The crack phase-field $\alpha$ characterizes the localization process at the macroscale, occurring in a process zone of width proportional to the internal length scale~$\ell$.}
\label{scheme1}
\end{figure}

\subsubsection{Free energy density}

Consider now the non-homogeneous solid depicted in figure~\ref{scheme1}, where a fracture process zone including a localized macrocrack evolving in $\Omega\times\Tau$ can be distinguished. In standard phase-field models, the coupled evolution of the displacement field $\Bu\colon\Omega\times\Tau\to\mathbb{R}^3$ and the crack phase-field $\alpha\colon\Omega\times\Tau\to[0,1]$ determines the state of the solid, where intact material points and completely fractured material points are given by $\,\alpha(\bm{x},t)=0\,$ and $\,\alpha(\bm{x},t)=1\,$, respectively. Further, in order to incorporate a frictional sliding mechanism, the plastic strain tensor $\Bep\colon\Omega\times\Tau\to\mathbb{R}^{3\times 3}_\mathrm{sym}$ is included in the formulation.

In view of the homogenized energy densities~\eqref{psi_open} and~\eqref{psi_closed} for solids with \emph{open} and \emph{closed} microcracks, and  labeling these cases as {\sf{open}} and {\sf{closed}} hereinafter, we take as a point of departure the free energy  
\begin{equation}
\psi(\bm{\varepsilon},{\color{black}\bm{\varepsilon}^\mathrm{p}},{\color{black}\alpha})\coloneqq\begin{dcases}
\frac{1}{2}\,\Be:\bm{\mathsf{C}}^\mathrm{dam}(\alpha):\Be& \quad \text{if \sf{open}},  \\[5pt] 
\frac{1}{2}(\bm{\varepsilon}-\bm{\varepsilon}^\mathrm{p}):\bm{\mathsf{C}}:(\bm{\varepsilon}-\bm{\varepsilon}^\mathrm{p}) + \frac{1}{2}\,\bm{\varepsilon}^\mathrm{p}:\bm{\mathsf{H}}^\mathrm{kin}(\alpha):\bm{\varepsilon}^\mathrm{p} & \quad \text{if \sf{closed}}.  
\end{dcases}
\label{free}
\end{equation}
The conceptual backbone of the proposed model is then the following ansatz:
\begin{equation}
\bm{\mathsf{C}}^\mathrm{hom}(\omega)\equiv\bm{\mathsf{C}}^\mathrm{dam}(\alpha),
\label{ansatz}
\end{equation}
which relates the damage-dependent elasticity tensor $\bm{\mathsf{C}}^\mathrm{dam}(\alpha)$ to the micromechanics-based characterization of opening microcracks. Recalling the assumption of isotropic behavior, the function $\alpha\mapsto\bm{\mathsf{C}}^\mathrm{dam}(\alpha)$ admits the volumetric-deviatoric decomposition
\begin{equation}
\bm{\mathsf{C}}^\mathrm{dam}(\alpha)\coloneqq g_K(\alpha)K\bm{1}\otimes\bm{1}+2\,g_\mu(\alpha)\mu\bigg(\bm{\mathsf{I}}-\frac{1}{3}\bm{1}\otimes\bm{1}\bigg).
\label{Cdam}
\end{equation}
In view of equations~\eqref{Chom}, \eqref{lame_hom}, and~\eqref{Cdam}, it follows that
% 
%\begin{equation}
%g_K(\alpha)\equiv\frac{1}{1+b_K\omega} \quad \text{and} \quad g_\mu(\alpha)\equiv\frac{1}{1+b_\mu\omega}\, ,
%\label{lame_phf}
%\end{equation}
\begin{equation}
b_K \,\omega(\alpha)=\frac{1-g_K(\alpha)}{g_K(\alpha)} \qquad \text{and} \qquad b_\mu \,\omega(\alpha)=\frac{1-g_\mu(\alpha)}{g_\mu(\alpha)}\, ,
\label{lame_phf}
\end{equation}
providing a clear link between the crack phase-field $\alpha$ and the microcrack density variable $\omega$. From equation~\eqref{lame_phf}, we note that the degradation functions $g_K(\alpha)$ and $g_\mu(\alpha)$ may not be defined independently in order to preserve a one-to-one relationship between $\alpha$ and $\omega$. In the present study, we employ the function~\cite{alessi2015}
\begin{equation}
g_K(\alpha)\coloneqq\frac{(1-\alpha)^2}{1+(b-1)[1-(1-\alpha)^2]}\, ,
\label{gk}
\end{equation}
which provides a single tuning parameter $b$, \tC{allowing us to recover the more common quadratic version $(1-\alpha)^2$ for $b=1$ and to consider a smoother softening stage, i.e., a less brittle response, for $b>1$}. Equation~\eqref{lame_phf} then yields the shear degradation function
\begin{equation}
g_\mu(\alpha)=\frac{g_K(\alpha)}{g_K(\alpha)+\frac{b_\mu}{b_K}[1-g_K(\alpha)]}\, .
\label{gmu}
\end{equation}
Finally, from equations~\eqref{psi_closed}, \eqref{free}, and \eqref{ansatz},	we make the association 
\begin{equation}
\bm{\mathsf{H}}^\mathrm{kin}(\alpha) \equiv \bm{\mathsf{H}}^\mathrm{block}(\omega).
\end{equation}
Then, in view of equation~\eqref{Hblock}, the kinematic hardening function $\alpha\mapsto\bm{\mathsf{H}}^\mathrm{kin}(\alpha)$ takes the form
\begin{equation}
\bm{\mathsf{H}}^\mathrm{kin}(\alpha)=\Big[ {\bm{\mathsf{C}}^\mathrm{dam}}^{-1}(\alpha) - \bm{\mathsf{C}}^{-1} \Big]^{-1},
\label{Hkin}
\end{equation}
which admits the volumetric-deviatoric decomposition 
\begin{equation}
\bm{\mathsf{H}}^\mathrm{kin}(\alpha)= H^\mathrm{kin}_K(\alpha)\bm{1}\otimes\bm{1}+H^\mathrm{kin}_\mu(\alpha)\bigg(\bm{\mathsf{I}}-\frac{1}{3}\bm{1}\otimes\bm{1}\bigg),
\label{Hkin2}
\end{equation}
where the damage-dependent kinematic hardening moduli read
\begin{equation}
H^\mathrm{kin}_K(\alpha)=\frac{g_K(\alpha)K}{1-g_K(\alpha)} \qquad \text{and} \qquad H^\mathrm{kin}_\mu(\alpha)=\frac{2\,g_\mu(\alpha)\mu}{1-g_\mu(\alpha)}\,.
\label{Hkin_K_mu}
\end{equation}

% for vanishing closed microcracks 
At this point, it is worth highlighting that in contrast with conventional phase-field models, the present model provides a direct link between the microcrack density parameter $\omega$ and the crack phase-field $\alpha$ (equation~\eqref{lame_phf}), such that a complete damage process $\alpha\to 1$ and vanishing damage $\alpha\to0$ correspond to $\omega\to\infty$ and $\omega\to0$, respectively. Moreover, in equation~\eqref{Hkin_K_mu}, the hardening moduli tend to infinity as $\alpha\to 0$, rendering an elastic response in the compressive/shear regime  when $\alpha=0$. The physical significance of this result is that no frictional sliding can occur in the absence of existing microcracks. Additionally, in previous phase-field models coupled to plasticity, damaging hardening moduli are defined from a purely phenomenological basis. Indeed, for the sake of simplicity, the same degradation function is often chosen for both elastic and plastic parameters (although more general choices are also possible~\cite{AleMarMauVid2017,samaniego2021}). Conversely, in the present model, it is neither necessary to define hardening moduli nor plastic degradation functions, since the damage-dependent kinematic hardening moduli follow in~\eqref{Hkin_K_mu} as functions of the elastic properties. It is thus only required to define a single suitable degradation function (equation~\eqref{gk} in the present case), from which all constitutive functions in the free energy~\eqref{free} are consequently defined. Finally, in the original phase-field formulation, a degradation function such as~\eqref{gk} (most often the quadratic version; see references~\citep{kuhn2015,wu2018phase} for overviews) multiplies the elasticity tensor $\bm{\mathsf{C}}$, rendering symmetric behavior in tension and compression. The conventional approach to overcome this shortcoming is the introduction of energy splits, where the degradation function acts on specific parts of the elasticity tensor defined as functions of either strain or stress. In the present work, an energy split of this type is not required since asymmetric behavior is naturally included in the micromechanics-based free energy~\eqref{free}.

With the free energy density~\eqref{free} fully defined, the Coleman-Noll  procedure yields 
\begin{equation}
\bm{\sigma}(\bm{\varepsilon},\bm{\varepsilon}^\mathrm{p},\alpha)=\dfrac{\partial\psi}{\partial\bm{\varepsilon}}=\begin{dcases}
\bm{\mathsf{C}}^\mathrm{dam}(\alpha):\bm{\varepsilon}& \quad \text{if \sf{open}},  \\
\bm{\mathsf{C}}:(\bm{\varepsilon}-\bm{\varepsilon}^\mathrm{p}) & \quad \text{if \sf{closed}}, 
\end{dcases}
\label{sig}
\end{equation}
providing the stress-strain relation, along with the generalized stresses conjugate to the plastic strains
\begin{equation}
\Bsp(\bm{\varepsilon},\bm{\varepsilon}^\mathrm{p},\alpha)=-\frac{\partial\psi}{\partial\bm{\varepsilon}^\mathrm{p}}=\begin{dcases}
\bm{0}& \quad \text{if \sf{open}},  \\
\bm{\mathsf{C}}:(\bm{\varepsilon}-\bm{\varepsilon}^\mathrm{p}) - \bm{\mathsf{H}}^\mathrm{kin}(\alpha):\bm{\varepsilon}^\mathrm{p} & \quad \text{if \sf{closed}}.
\end{dcases} 
\label{sp}
\end{equation}
In agreement with the micromechanical arguments in section~\ref{local_micro}, the microcrack opening/closure condition is naturally embedded in equation~\eqref{sp}. Thus, in view of isotropic behavior, we employ the~conditions
\begin{equation}
\begin{dcases}\tr\Bsp(\bm{\varepsilon},\bm{\varepsilon}^\mathrm{p},\alpha)=0 \quad \text{if \sf{open}}, \\ \tr\Bsp(\bm{\varepsilon},\bm{\varepsilon}^\mathrm{p},\alpha) < 0 \quad \text{if \sf{closed}}\end{dcases}
\label{optrans}
\end{equation}
to signal open and closed microcrack states, hereinafter referred to as the \emph{tensile regime} and the \emph{compressive/shear regime}, respectively. Note from equations~\eqref{sig} and~\eqref{sp} that this terminology applies to the generalized stress tensor $\Bsp$ and not to the true stress tensor $\bm{\sigma}$. Finally, the generalized stress conjugate to the crack phase-field reads
\begin{equation}
{s}^\mathrm{d}(\bm{\varepsilon},\bm{\varepsilon}^\mathrm{p},\alpha)=-\dfrac{\partial\psi}{\partial\alpha}=\begin{dcases}
-\frac{1}{2}\,\bm{\varepsilon}:{\bm{\mathsf{C}}^\mathrm{dam}}'(\alpha):\bm{\varepsilon}& \quad \text{if \sf{open}},  \\[5pt]
-\frac{1}{2}{\,\bm{\varepsilon}^\mathrm{p}:\bm{\mathsf{H}}^{\mathrm{kin}}}'(\alpha):\bm{\varepsilon}^\mathrm{p}  & \quad \text{if \sf{closed}}.  
\end{dcases} 
\label{sd}
\end{equation}

In conventional phase-field models for ductile fracture, a distinction between tensile and compressive/shear regimes is not made, in the sense that plasticity  is always active, and degradation is usually applied to both elastic and plastic energy terms. In contrast, in the present model, according to equation~\eqref{free} and the resulting generalized stresses~\eqref{sig}, \eqref{sp}, and~\eqref{sd}, the tensile regime associated to opening microcracks corresponds to brittle damage. Therefore,  the plastic driving force in~equation~\eqref{sp} vanishes, while the damage driving force in equation~\eqref{sd} is due to elastic strain energy. On the other hand, the compressive/shear regime associated to sliding microcracks entails ductile damage, where only the blocked energy term in~\eqref{free} is degraded. Accordingly, plastic strains are driven by the kinematic-hardening--type force with a damage-dependent back-stress in~equation~\eqref{sp}, while the damage driving force in~equation~\eqref{sd} is due to the blocked plastic energy that results from frictional sliding.

\begin{remark}\label{rem2}
In agreement with Remark~\ref{rem1}, equation~\eqref{Hkin} implies the continuity of the free energy~\eqref{free} and the generalized stress fields~\eqref{sig}, \eqref{sp}, and~\eqref{sd}, which can, in turn, be equivalently written in compact form in terms of their corresponding closed-microcrack expressions. In particular, at the opening/closure transition (and further in the tensile regime), $\Bep$ may be computed from the cancellation of $\Bsp$ as $$\Bep=\big[ \bm{\mathsf{H}}^\mathrm{kin}(\alpha) + \bm{\mathsf{C}} \big]^{-1}\bm{\mathsf{C}}:\Be\quad \iff \quad \Bsp=\bm{0}.$$ Using equations~\eqref{Hkin} and~\eqref{Hkin_K_mu}, we obtain the relations
\begin{equation}
\Bep_\mathrm{dev}= \big(1-g_\mu(\alpha)\big)\Be_\mathrm{dev} \quad \text{and} \quad \tr\Bep= \big(1-g_K(\alpha)\big)\tr\Be \quad \iff \quad \Bsp=\bm{0} .
\label{ep_reqs}
\end{equation}
In this case, $\Bep$ is physically linked to the displacement discontinuities of opening microcracks and is thus not a dissipative mechanism. Note that replacing the last expressions in equations~\eqref{free}, \eqref{sig}, \eqref{sp}, and~\eqref{sd} confirms the continuity conditions at the opening/closure transition. In the present work, we choose to preserve the piecewise representations of the free energy and generalized stresses, such that $\Bep$ is not involved in the tensile regime. Consequently, $\Bep$ is exclusively viewed as a dissipative mechanism evolving in the compressive/shear regime, for which a suitable evolution law is defined in the sequel. Nevertheless, to ensure continuity, the evolution of $\Bep$ must be such that conditions~\eqref{ep_reqs} are satisfied.
\end{remark}

\subsubsection{Dissipation potential}

Recalling the thermomechanical arguments discussed in section~\ref{framework}, a thermodynamically admissible dissipation potential is now proposed to characterize the evolution of the internal variables  $\Bep$ and $\alpha$. The dissipation potential is additively decomposed into a plastic contribution $\phi^\mathrm{p}$ and a fracture contribution $\phi^\mathrm{d}$:
\begin{equation}
{\phi(\dot{\bm{\varepsilon}}^\mathrm{p},\dot{\alpha},\nabla\dot{\alpha};\alpha,\nabla\alpha,{\bm{s}^\mathrm{p}})=\phi^\mathrm{p}(\dot{\bm{\varepsilon}}^\mathrm{p};{\bm{s}^\mathrm{p}})+\phi^\mathrm{d}(\dot{\alpha},\nabla\dot{\alpha};\alpha,\nabla\alpha,{\bm{s}^\mathrm{p}}).}
\label{phi}
\end{equation}

To determine the plastic contribution, we first define the evolution equations in generalized stress space, i.e., in \emph{dual form}, as typically done in plasticity theory. The first item is the generalized stress~constraint  
\begin{equation}
\Bsp\in\mathbb{K} \coloneqq \big\{\Bsp\in\mathbb{R}^{3\times 3}_{\mathrm{sym}} \ \ \vert \ \  f^\mathrm{p}(\Bsp)\leq 0\big\},
\label{sc_setK}
\end{equation}
where $\mathbb{K}$ is a non-empty, closed, and convex set of admissible generalized stresses, given in terms of the Drucker-Prager--type yield function linked to frictional sliding and governing the compressive/shear regime:
\begin{equation}
f^\mathrm{p}(\bm{s}^\mathrm{p})\coloneqq\Vert\Bsp_{{\mathrm{dev}}}\Vert + \sqrt\frac{2}{3}A_\varphi \mathrm{tr} \Bsp.
\label{fp}
\end{equation} 
The condition $f^\mathrm{p}(\Bsp)\leq0$ can be viewed as a friction criterion on the (isotropic) local stress field $\Bsp$, where $A_\varphi$ is the friction coefficient, such that $\mathbb{K}$ can be viewed as a cone with the apex at the origin in generalized stress space. Thus, in agreement with~\citet{zhu2011}, $f^\mathrm{p}(\Bsp)$ does not include a cohesive term  in generalized stress space, since such a term would not allow $\Bsp_\mathrm{dev}$ to vanish at the opening/closure transition (equation~\eqref{optrans}). However, by virtue of the back-tress term in equation~\eqref{sp}, a damage-dependent cohesion is indeed attained in terms of $\Bsig$, i.e., in true stress space. 

At this point, a possibility to fully define the evolution of the plastic strain tensor $\Bep$ is to invoke the normality condition through the principle of maximum dissipation~\citep{hill1948,moreau1974}. From standard convex analysis~\citep{rockafellar1970,han1999}, one then obtains an associative flow rule of the form $\dot{\bm{\varepsilon}}^\mathrm{p}\in \partial I_\mathbb{K}(\Bsp)$, that is, $\Bep$ lies in the subdifferential of the indicator function of $\mathbb{K}$ at $\Bsp$. The volumetric plastic strains are then modulated by the friction coefficient $A_\varphi$. Experimental evidence suggests that this modeling choice overestimates the amount of dilation observed in geomaterials, further resulting in unrealistic material responses and unrealistic shear band orientations~\cite{vardoulakis1980,vermeer1984}. Moreover, as we shall see below, the use of an associative flow rule in the present model implies a vanishing energy dissipation due to frictional sliding. Consequently, we consider a non-associative flow rule, which may be written as 
\begin{equation}
\dot{\bm{\varepsilon}}^\mathrm{p}\in\mathbb{Q}(\Bsp)\coloneqq\big\{\lambda\,\hat{\bm{n}}\in\mathbb{R}^{3\times 3}_{\mathrm{sym}}  \ \ \vert \ \  \hat{\bm{n}}\in\partial g^\mathrm{p}(\Bsp);\, \lambda\geq 0, \,\, \lambda= 0 \,\,\, \text{if} \,\,\, f^\mathrm{p}(\Bsp)<0\big\},
\label{nonass}
\end{equation}
in terms of the \emph{plastic potential} 
\begin{equation}
g^\mathrm{p}(\Bsp)\coloneqq\Vert\Bsp_{{\mathrm{dev}}}\Vert + \sqrt\frac{2}{3}A_\theta \mathrm{tr}\Bsp,
\label{gp}
\end{equation}
where $A_\theta$ is the dilation constant ($0\leq A_\theta <A_\varphi$). Clearly, if $A_\theta=A_\varphi$, we recover the associative model. 

It is worth mentioning that both sets $\mathbb{K}$ and $\mathbb{Q}$ are fixed in $\Bsp$ space, but vary in $\Bsig$ space as damage evolves through the back-stress $\bm{\mathsf{H}}^\mathrm{kin}(\alpha):\bm{\varepsilon}^\mathrm{p}$. A straightforward extension is to consider damage-dependent coefficients $A_\varphi$ and $A_\theta$, as discussed in appendix~\ref{dam_fric_dil}, such that $\mathbb{K}$ and $\mathbb{Q}$ are no longer fixed in $\Bsp$ space. 

The main implication of adopting a non-associative flow rule is that the variational structure inherent to associative models is apparently lost. However, as presented in~\citet{ulloa2021a}, a variational structure can be recovered for non-associative models by employing a state-dependent set of generalized stresses (see also~\citet{francfort2018} and references therein). For the present model, we define the convex set 
\begin{equation}
\mathbb{L}(\Bsp)\coloneqq\bigg\{\tilde{\bm{s}}^\mathrm{p}\in\mathbb{R}^{3\times 3}_{\mathrm{sym}} \ \ \vert \ \  \Vert\tilde{\bm{s}}^\mathrm{p}_{{\mathrm{dev}}}\Vert + \sqrt\frac{2}{3}A_\theta\tr\tilde{\bm{s}}^\mathrm{p}\leq \sqrt{\frac{2}{3}}(A_\theta-A_\varphi)\tr\bm{s}^\mathrm{p} \bigg\}.
\end{equation}
Employing a result of~\citet[Proposition 4]{laborde1987}, we recover the following conditions:
\begin{equation}
\begin{dcases}
\Bsp\in\mathbb{K} 
  &\iff  \quad   
  \Bsp \in \mathbb{L}(\Bsp),\\
\dot{\bm{\varepsilon}}^\mathrm{p}\in\mathbb{Q}(\Bsp)  &\iff \quad  \dot{\bm{\varepsilon}}^\mathrm{p}\in \partial I_{\mathbb{L}(\Bsp)}(\Bsp).
\end{dcases}
\label{stdepcond}
\end{equation}
These conditions imply that the generalized stress constraint~\eqref{sc_setK} and the non-associative flow rule~\eqref{nonass} can be equivalently written in terms of the state-dependent set $\mathbb{L}(\Bsp)$. Moreover, the role of $\mathbb{L}(\Bsp)$ in the non-associative model is analogous to the role of $\mathbb{K}$ in the associative model. Thus, the non-associative evolution equations~\eqref{sc_setK} and~\eqref{nonass} correspond to the necessary conditions of the variational principle
\begin{equation}
\phi^\mathrm{p}(\dot{\bm{\varepsilon}}^\mathrm{p};\Bsp)=\mathrm{sup}\big\{\tilde{\bm{s}}^\mathrm{p}:\dot{\bm{\varepsilon}}^\mathrm{p} - I_{\mathbb{L}(\bm{s}^\mathrm{p})}(\tilde{\bm{s}}^\mathrm{p})\big\}.
\label{disspotnona}
\end{equation}
This expression represents a \emph{generalized principle of maximum dissipation}~\cite{ulloa2021a} in the sense that dissipation is maximum with respect to generalized stresses within $\mathbb{L}(\Bsp)$, but not necessarily within $\mathbb{K}$. As a distinctive feature of non-associative models, the dissipation potential $\phi^\mathrm{p}(\dot{\bm{\varepsilon}}^\mathrm{p};\Bsp)$ inherits the dependence on the generalized stress and follows as the support function of $\mathbb{L}(\Bsp)$. For $f=g$, the associative case is recovered and the dissipation potential becomes state-independent. Evaluating the supremum~\eqref{disspotnona} for all $\dot{\bm{\varepsilon}}^\mathrm{p}\in\mathbb{R}^{3\times 3}_{\mathrm{sym}}$~yields
\begin{equation}
\begin{aligned}
\phi^\mathrm{p}(\dot{\bm{\varepsilon}}^\mathrm{p};\bm{s}^\mathrm{p})
&=\sup\bigg\{{\tilde{\bm{s}}^\mathrm{p}}:\dot{\bm{\varepsilon}}^\mathrm{p} \ \ \vert \  \ \Vert\tilde{\bm{s}}^\mathrm{p}_{{\mathrm{dev}}}\Vert + \sqrt{\frac{2}{3}}A_\theta\tr\tilde{\bm{s}}^\mathrm{p}\leq \sqrt{\frac{2}{3}}(A_\theta-A_\varphi)\tr\bm{s}^\mathrm{p}\bigg\}\\
&=\sup\bigg\{\Vert{\tilde{\bm{s}}_{\mathrm{dev}}^\mathrm{p}}\Vert \, \Vert\dot{\bm{\varepsilon}}_{\mathrm{dev}}^\mathrm{p}\Vert+\frac{1}{3}\tr{\tilde{\bm{s}}^\mathrm{p}}\tr \dot{\bm{\varepsilon}}^\mathrm{p}\ \  \vert \ \  A_\theta\tr\tilde{\bm{s}}^\mathrm{p}  \leq (A_\theta-A_\varphi)\tr\bm{s}^\mathrm{p}-\sqrt{\frac{3}{2}}\Vert\tilde{\bm{s}}^\mathrm{p}_{{\mathrm{dev}}}\Vert\bigg\} \\
&=\sup\bigg\{\dfrac{\tr\dot{\bm{\varepsilon}}^\mathrm{p}}{3A_\theta}(A_\theta-A_\varphi)\tr\bm{s}^\mathrm{p} + \Vert{\tilde{\bm{s}}_{\mathrm{dev}}^\mathrm{p}}\Vert\bigg(\Vert\dot{\bm{\varepsilon}}_{\mathrm{dev}}^\mathrm{p}\Vert - \frac{1}{\sqrt{6}A_\theta}\tr\dot{\bm{\varepsilon}}^\mathrm{p}\bigg)\bigg\}.
\end{aligned}
\label{disspotnona_dp}
\end{equation}
Noting that the expression inside the supremum is unbounded when the term multiplying $\Vert{\tilde{\bm{s}}_{\mathrm{dev}}^\mathrm{p}}\Vert$ is positive, the plastic dissipation potential is written as\footnote{This derivation holds for $A_\theta>0$. However, this assumption can be straightforwardly relaxed by expressing the plastic dissipation potential in terms of $\Vert\dot{\bm{\varepsilon}}^\mathrm{p}_\mathrm{dev}\Vert$ instead of $\tr\dot{\bm{\varepsilon}}^\mathrm{p}$.}
\begin{equation}
\phi^\mathrm{p}(\dot{\bm{\varepsilon}}^\mathrm{p};{\Bsp})=\begin{dcases}
\frac{\tr\dot{\bm{\varepsilon}}^\mathrm{p} }{3A_\theta}{(A_\theta-A_\varphi)\tr\Bsp} \quad \text{if } \tr\dot{\bm{\varepsilon}}^\mathrm{p}\geq\sqrt{6}A_\theta\Vert \dot{\bm{\varepsilon}}^\mathrm{p}_\mathrm{dev}\Vert,\\
+\infty \quad \text{otherwise}.\end{dcases}
\label{phi_p}
\end{equation}
This function corresponds to the dissipation power of the frictional sliding mechanism and contributes to the total dissipation potential~\eqref{phi}. Remarkably, for $A_\theta=A_\varphi$, the plastic dissipation potential vanishes. Note that this observation can also be made from the fact that ${\bm{s}}^\mathrm{p}\in\partial\mathbb{K}$ and $\dot{\bm{\varepsilon}}^\mathrm{p}$ are always orthogonal by virtue of the yield function~\eqref{fp}. As such, the associative model is not consistent with the interpretation of $\Bep$ as a dissipative frictional mechanism, highlighting the crucial role of non-associativity in the present study. 

Finally, the fracture contribution to the dissipation potential~\eqref{phi} is defined as follows. In view of the damage driving force~\eqref{sd}, the fracture dissipation potential is endowed with independent parameters for the tensile and compressive/shear regimes. We thus define
\begin{equation}
\phi^\mathrm{d}(\dot{\alpha},\nabla\dot{\alpha};\alpha,\nabla\alpha,{\bm{s}^\mathrm{p}})\coloneqq\begin{dcases}\frac{{G}_\mathrm{c}(\bm{s}^\mathrm{p})}{{\ell}(\bm{s}^\mathrm{p})}\big(\alpha\,\dot{\alpha}+{\ell}^{2}(\bm{s}^\mathrm{p})\nabla\alpha\cdot\nabla\dot{\alpha}\big) \quad \text{if } \dot{\alpha}\geq 0 ,\\
+\infty \quad \text{otherwise}.\end{dcases} 
\label{phi_d}
\end{equation}
The damage irreversibility condition $\dot{\alpha}\geq0$ is automatically enforced in this definition. Moreover, the fracture toughness and the internal length scale read
\begin{equation}
{G}_\mathrm{c}(\bm{s}^\mathrm{p})\coloneqq\begin{dcases}
G_\mathrm{cI}  \quad &\text{if }  \tr\bm{s}^\mathrm{p}=0, \\
G_\mathrm{cII} \quad &\text{if }  \tr\bm{s}^\mathrm{p}<0,
\end{dcases}
\qquad \text{and} \qquad
{\ell}(\bm{s}^\mathrm{p})\coloneqq\begin{dcases}
\ell_\mathrm{I}  \quad &\text{if }  \tr\bm{s}^\mathrm{p}=0, \\
\ell_\mathrm{II} \quad &\text{if }  \tr\bm{s}^\mathrm{p}<0.
\end{dcases}
\label{Gcsp}
\end{equation}
Accordingly, fracture in the tensile regime and fracture in the compressive/shear regime are governed by the mode I fracture toughness $G_\mathrm{cI}$ and the mode II fracture toughness $G_\mathrm{cII}$, respectively.  This feature of the model plays a crucial role in capturing different failure modes including mixed-mode fracture. For the sake of generality, a distinction has also been made between mode I and mode II length scales $\ell_\mathrm{I}$ and $\ell_\mathrm{II}$. \tC{Note that, owing to the micromechanics-based free energy~\eqref{free}, the distinction between fracture modes is intrinsic to the present model and therefore  does not require additional phase-field variables (e.g.,~\citet{bleyer2018phase}), or modifications of the phase-field evolution equations, as considered in phase-field models for mixed-mode fracture~\cite{zhang2017}. Moreover, mode II fracture is naturally coupled to friction-induced plasticity.}

In the present study, the influence of distinctive length scales is not addressed. Thus, without losing generality, we consider hereafter $\ell_\mathrm{I}=\ell_\mathrm{II}=\ell$.

\begin{remark}
As done for plasticity, the fracture dissipation potential~\eqref{phi_d} could have also been derived by first postulating a generalized stress constraint of the form
\begin{equation}
s^\mathrm{d}\in\mathbb{K}^\mathrm{d} \coloneqq \big\{s^\mathrm{d}\in\mathbb{R}_{+} \ \ \vert \ \  f^\mathrm{d}(s^\mathrm{d};\Bsp)\leq 0\big\},
\label{sc_setKd}
\end{equation}
in terms of the damage yield function
\begin{equation}
f^\mathrm{d}(s^\mathrm{d};\Bsp)\coloneqq s^\mathrm{d}-\frac{{G}_\mathrm{c}(\Bsp)}{{\ell}}\alpha+{\ell}\,\mathrm{div}[{G}_\mathrm{c}(\Bsp)\nabla\alpha].
\label{fd}
\end{equation} 
Invoking the principle of maximum dissipation for damage evolution, condition~\eqref{sc_setKd} and the associative flow rule 
\begin{equation}
\dot{\alpha}\in \partial I_{\mathbb{K}^\mathrm{d}}(s^\mathrm{d})
\label{flow_d}
\end{equation} 
follow as necessary conditions of the non-local principle%the principle of maximum dissipation for damage evolution
\begin{equation}
\int_\Omega\phi^\mathrm{d}(\dot{\alpha},\nabla\dot{\alpha};\alpha,\nabla\alpha,{\Bsp})\,\mathrm{d}\bm{x}=\mathrm{sup}\bigg\{\int_\Omega\Big(\tilde{s}^\mathrm{d}\,\dot{\alpha} - I_{\mathbb{K}^\mathrm{d}}(\tilde{s}^\mathrm{d})\Big)\,\mathrm{d}\bm{x}\bigg\},
\end{equation}
from which the fracture dissipation contribution~\eqref{phi_d} is recovered by applying integration by parts and the boundary condition $\dot{\alpha}(\nabla{\alpha}\cdot\Bn)=0$ on $\Gamma$. 
\end{remark}

\begin{remark}
Along with the equilibrium equations~\eqref{st_adm}, the evolution problem in terms of the generalized stress constraint~\eqref{sc_setK}/\eqref{sc_setKd} and the flow rule in generalized stress space~\eqref{nonass}/\eqref{flow_d} corresponds to the so-called \emph{dual formulation}. On the other hand, in the \emph{primal formulation}, the generalized stress constraint and the flow rule follow as consequences of equation~\eqref{biot} for a given dissipation potential. Given the free energy density~\eqref{free} and the dissipation potential~\eqref{phi} for the present model, the primal formulation in strong form consists of (i)~the equilibrium equations~\eqref{st_adm} with $\Bsig$ given in~\eqref{sig}, and (ii) the differential inclusion~\eqref{biot} with ${\mathbf{a}\coloneqq\{\Bep,\alpha\}}$ and $\,{\mathbf{s}\coloneqq\{\Bsp,s^\mathrm{d}\}}$, which specializes to the system
\begin{numcases}{ }
{\partial_{\Bep}\psi(\bm{\varepsilon},{\color{black}\bm{\varepsilon}^\mathrm{p}},{\color{black}\alpha})}+\partial_{\dot{\bm{\varepsilon}}^\mathrm{p}}\phi^\mathrm{p}(\dot{\bm{\varepsilon}}^\mathrm{p};\Bsp)\ni \bm{0}, \\[1pt]
\label{biot_model_p}
{\partial_\alpha\psi(\bm{\varepsilon},{\color{black}\bm{\varepsilon}^\mathrm{p}},{\color{black}\alpha})}
 +\partial_{\dot{\alpha}}\phi^\mathrm{d}(\dot{\alpha},\nabla\dot{\alpha};\alpha,\nabla\alpha,{\bm{s}^\mathrm{p}})-\mathrm{div}\big[\partial_{\nabla\dot{\alpha}}\phi^\mathrm{d}(\dot{\alpha},\nabla\dot{\alpha};\alpha,\nabla\alpha,{\bm{s}^\mathrm{p}})\big] \ni 0.
\label{biot_model_d}
\end{numcases}
The variational evolution problem presented in the sequel corresponds to this (primal) formulation.
\end{remark}

\subsubsection{Variational formulation and governing equations}\label{formulation}

In this section, we adopt the energetic formulation for rate-independent systems, where the evolution problem is recovered in variational form using notions of energy balance and stability~\cite{Mielke2006,Mielke2015}. Moreover, to ensure thermodynamic consistency, a dissipation inequality is included in the formulation. While the energy balance and the dissipation inequality correspond to statements of the first and second laws of thermodynamics, the stability condition represents an additional restriction for solutions to attain a minimal energy state at a given time. In its most general form, the energetic formulation employs a notion of global stability, requiring no regularity assumptions for solutions. However, a more physical notion of local stability is often preferred, in particular for non-convex energies, at the cost of assuming sufficient regularity~\cite{Alessi2016}. 

Herein, we consider the formulation based on local stability and do not dwell on mathematical concepts involving the regularity of admissible states. For a thorough mathematical survey on the energetic formulation,  see~\citet{Mielke2015}. Some applications of the theory in solid mechanics can be found in the literature~\citep{BourFrancMar2008, alessi2015, alessi2015shape, rokovs2016, pham2011, rodriguez2018, Alessi2017fatigue, luege2018, lancioni2020}. In this context, the formulation was recently outlined in the general framework of gradient-enhanced internal variables~\cite{ulloa2021b} and generalized to non-associative models~\cite{ulloa2021a}.

We begin by defining the global internal stored energy functional
\begin{equation}
\mathcal{E}(\Bu,\Bep,\alpha)\coloneqq\int_\Omega\psi(\bm{\varepsilon},\Bep,\alpha)\,\mathrm{d}\bm{x},
\end{equation}
while the work of external actions is defined as the time integral of the external power:
\begin{equation}
\mathcal{L}\big(\bm{u};[0,t]\big)\coloneqq \int_{0}^{t}\bigg[\int_\Omega\rho\bm{b}(\tau)\cdot\dot{\bm{u}}(\tau)\,\mathrm{d}\bm{x} + \int_{\Gamma_\mathrm{N}}\bar{\bm{t}}(\tau)\cdot\dot{\bm{u}}(\tau)\,\mathrm{d}S + \int_{\Gamma_{\mathrm{D}}}\bm{t}_\mathrm{r}(\tau)\cdot\dot{\bar{\bm{u}}}(\tau)\, \mathrm{d}S\,\bigg]\mathrm{d}\tau,
\label{stored_ext1}
\end{equation}
where $\bm{t}_\mathrm{r}$ is the traction vector on $\Gamma_{\mathrm{D}}$. On the other hand, the global dissipative power functional reads
\begin{equation}
\mathcal{R}(\dot{\bm{\varepsilon}}^\mathrm{p},\dot{\alpha};\alpha,{\bm{s}^\mathrm{p}})\coloneqq\int_\Omega\phi(\dot{\bm{\varepsilon}}^\mathrm{p},\dot{\alpha},\nabla\dot{\alpha};\alpha,\nabla\alpha,{\bm{s}^\mathrm{p}}\big)\,\mathrm{d}\bm{x}.
\end{equation}
We note that in standard phase-field models, the dissipated energy follows from the time integral of the dissipative power as a state function. However, in the present case, the dissipated energy is a path-dependent quantity due to the dependence of the dissipation potential on the generalized stress $\Bsp$. With the above definitions, we are now in position to derive the governing equations of the proposed model in variational~form. 

A process $\{\Bu,\Bep,\alpha\}\colon\Tau\to\mathscr{Q}$ satisfies energy balance if the following condition holds for all $t\in\Tau$:
\begin{equation}
\mathcal{E}\big(\Bu(t),\Bep(t),\alpha(t)\big) +  \int_0^t\mathcal{R}\big(\dot{\Be}^\mathrm{p}(s),\dot{\alpha}(s);\alpha(s),\Bsp(s)\big)\,\mathrm{d}s = \mathcal{E}\big(\Bu(0),\Bep(0),\alpha(0)\big)  +  \mathcal{L}\big(\bm{u};[0,t]\big).
\label{EB}
\end{equation}
Provided that the energy functionals are sufficiently regular in $\Tau$, the time derivative of~\eqref{EB} yields the first-order energy balance, given by the power balance equation
\begin{equation}
\frac{\mathrm{d}}{\mathrm{d}t}{\mathcal{E}}\big(\Bu(t),\Bep(t),\alpha(t)\big) + \mathcal{R}\big(\dot{\Be}^\mathrm{p}(t),\dot{\alpha}(t);\alpha(t),\Bsp(t)\big) - \frac{\mathrm{d}}{\mathrm{d}t}\mathcal{L}\big(\bm{u};[0,t]\big)=0.
\label{EB-1}
\end{equation}
Focusing on local stability, we consider solutions that fulfill a \emph{local directional stability condition}. As such, at any time, the energy state is enforced to be minimum with respect to energy states reached by taking admissible variations in the neighborhood of the current state. This condition can be expanded into Taylor terms of increasing order, which may be enforced as necessary conditions for local directional stability. In this context, it can be shown~\cite{ulloa2021b} that a process $\{\Bu,\Bep,\alpha\}\colon\Tau\to\mathscr{Q}$ satisfies \emph{first-order stability} if the following condition holds for all $t\in\Tau$:
\begin{equation}
\begin{aligned}
\delta\mathcal{E}\big(\Bu(t),\Bep(t),\alpha(t)\big)\big(\tilde{\Bu},\tilde{\bm{\varepsilon}}^\mathrm{p},\tilde{\alpha}\big)+\mathcal{R}\big(\tilde{\bm{\varepsilon}}^\mathrm{p},\tilde{\alpha};\alpha(t),{\bm{s}^\mathrm{p}}(t)\big) &-\int_\Omega\rho\bm{b}(t)\cdot\tilde{\bm{u}}\,\mathrm{d}\bm{x} \\ &- \int_{\Gamma_\mathrm{N}}\bar{\bm{t}}(t)\cdot\tilde{\bm{u}}\,\mathrm{d}S \geq0 \quad \forall\,\{\tilde{\Bu},\tilde{\Be}^\mathrm{p},\tilde{\alpha}\}\in\tilde{\mathscr{Q}},
\label{DS-1}
\end{aligned}
\end{equation}
where $\delta\mathcal{E}(\Box)(\tilde{\Box})$ is the Gâteaux derivative of $\mathcal{E}$ in the direction $\tilde{\Box}$. 

Above, $\mathscr{Q}\coloneqq\mathscr{U}\times\mathscr{B}\times\mathscr{D}$ and $\tilde{\mathscr{Q}}\coloneqq\tilde{\mathscr{U}}\times\tilde{\mathscr{B}}\times\tilde{\mathscr{D}}$ denote suitable function spaces for the primary fields and the corresponding test functions. In particular, we consider
\begingroup
\allowdisplaybreaks
\begin{align}
&\mathscr{U}\coloneqq\{\bm{w} \in \mathrm{BD}(\Omega;\mathbb{R}^3) \  \ \vert \ \ \bm{w}=\bar{\bm{u}} \,\,\, \text{on} \,\,\, \Gamma_\mathrm{D}\}, && \tilde{\mathscr{U}}\coloneqq\{\bm{w} \in \mathrm{BD}(\Omega;\mathbb{R}^3) \ \ \vert \ \ \tilde{\bm{w}}=\bm{0} \,\,\,  \text{on} \,\,\, \Gamma_\mathrm{D}\},  \label{funscspace_u}\\
&\mathscr{B}\coloneqq \mathrm{M_b}(\Omega\cup\Gamma_\mathrm{D};\mathbb{R}^{3\times 3}_{\mathrm{sym}}), &&\tilde{\mathscr{B}}\coloneqq \{\bm{e}\in\mathscr{B}  \ \ \vert \ \  \tr{\bm{e}} \geq \sqrt{6}A_\theta\Vert\bm{e}_\mathrm{dev}\Vert\},  \label{funscspace_p}\\
&\mathscr{D}\coloneqq \mathrm{H}^1(\Omega;[0,1]), &&\tilde{\mathscr{D}}\coloneqq \mathrm{H}^1(\Omega;\mathbb{R}_+).
 \label{funscspace_d}
\end{align}
\endgroup
Here, $\mathrm{H}^1$ is the Sobolev space of functions with square-integrable first derivatives. On the other hand, $\mathrm{BD}$ is the space of bounded deformations, while $\mathrm{M_b}$ is a space of Radon measures. The reason for employing these function spaces is that the hardening-softening response in the compressive/shear regime approaches perfect plasticity as $\alpha\to1$, as can be noted from equations~\eqref{Hkin_K_mu}, \eqref{sp}, and~\eqref{fp}, resulting in plastic strain localization.\footnote{In view of localized responses in perfect plasticity, a more rigorous treatment of the formulation may consider the energy functionals split over the regular and singular parts of the domain. An example of this treatment can be found in~\citet{alessi2015}. We do not include these details in the present formulation for the sake of simplicity.} Moreover, we have embedded in~\eqref{funscspace_p} and~\eqref{funscspace_d} the constraints present in the dissipation potentials~\eqref{phi_p} and~\eqref{phi_d}, such that the global dissipative power entering the stability condition~\eqref{DS-1} remains finite. As such, we address only the non-trivial conditions for the fulfillment of~\eqref{DS-1}. It is worth noting that if~\eqref{DS-1} holds as an equality, the study of higher-order conditions is required to ensure local stability; see, for instance, \citet{pham2011} or \citet{alessi2015shape}, where second-order stability conditions play a crucial role in describing size effects. In the present study, only the first-order stability conditions are considered, and its consequences in conjunction with energy balance are discussed~below.

With the above definitions, the generalized stresses~\eqref{sig}, \eqref{sp}, and~\eqref{sd}, and the dissipation potentials~\eqref{phi_p} and~\eqref{phi_d}, the power balance~\eqref{EB-1} yields
\begin{equation}
\begin{aligned}
\int_\Omega \bigg( \Bsig:\nabla^\mathrm{s}\dot{\Bu} - \Bsp:\dot{\Be}^\mathrm{p} -s^\mathrm{d}\dot{\alpha}  +  \frac{\tr\dot{\bm{\varepsilon}}^\mathrm{p} }{3A_\theta}{(A_\theta-A_\varphi)\tr\Bsp}+I_\mathbb{R_+}\big(\tr\dot{\bm{\varepsilon}}^\mathrm{p}-\sqrt{6}A_\theta\Vert \dot{\bm{\varepsilon}}^\mathrm{p}_\mathrm{dev}\Vert\big)\\
 +\frac{{G}_\mathrm{c}(\bm{s}^\mathrm{p})}{{\ell}}\big(\alpha\,\dot{\alpha}+{\ell}^{2}\nabla\alpha\cdot\nabla\dot{\alpha}\big)  + I_{\mathbb{R}_+}(\dot{\alpha})
\bigg)\,\mathrm{d}\bm{x} \\ -\int_\Omega\rho\bm{b}\cdot\dot{\bm{u}}\,\mathrm{d}\bm{x} - \int_{\Gamma_\mathrm{N}}\bar{\bm{t}}\cdot\dot{\bm{u}}\,\mathrm{d}S -\int_{\Gamma_{\mathrm{D}}}\bm{t}_\mathrm{r}\cdot\dot{\bar{\bm{u}}}\, \mathrm{d}S =0,
\label{EB-1_2}
\end{aligned}
\end{equation}
where the indicator function $I_{\mathbb{R}_+}$ is employed to enforce the constraints of the dissipation potentials~\eqref{phi_p} and~\eqref{phi_d}. Similarly, the first-order stability condition~\eqref{DS-1} yields
\begin{equation}
\begin{aligned}
\int_\Omega \bigg( \Bsig:\nabla^\mathrm{s}\tilde{\Bu} - \Bsp:\tilde{\Be}^\mathrm{p} -s^\mathrm{d}\tilde{\alpha}  +  \frac{\tr\tilde{\bm{\varepsilon}}^\mathrm{p} }{3A_\theta}{(A_\theta-A_\varphi)\tr\Bsp} +\frac{{G}_\mathrm{c}(\bm{s}^\mathrm{p})}{{\ell}}\big(\alpha\,\tilde{\alpha}+{\ell}^{2}\nabla\alpha\cdot\nabla\tilde{\alpha}\big)\bigg)\,\mathrm{d}\bm{x} \\ -\int_\Omega\rho\bm{b}\cdot\tilde{\bm{u}}\,\mathrm{d}\bm{x} - \int_{\Gamma_\mathrm{N}}\bar{\bm{t}}\cdot\tilde{\bm{u}}\,\mathrm{d}S \geq0 \quad \forall\,\{\tilde{\Bu},\tilde{\Be}^\mathrm{p},\tilde{\alpha}\}\in\tilde{\mathscr{Q}}.
\label{DS-1_2}
\end{aligned}
\end{equation}
The following conditions directly follow.

\paragraph*{Mechanical balance and yield criteria}

Letting $\tilde{\Be}^\mathrm{p}=\bm{0}$ and $\tilde{\alpha}=0$ in the first-order stability condition~\eqref{DS-1_2}, and noting that the displacement field $\Bu$  is only  constrained by the Dirichlet boundary conditions, we obtain the mechanical balance equation corresponding to the weak form of the equilibrium equations~\eqref{st_adm}:
\begin{equation}
\int_\Omega \big( \Bsig:\nabla^\mathrm{s}\tilde{\Bu}-\rho\bm{b}\cdot\tilde{\bm{u}}\big)\,\mathrm{d}\bm{x} - \int_{\Gamma_\mathrm{N}}\bar{\bm{t}}\cdot\tilde{\bm{u}}\,\mathrm{d}S =0 \quad \forall\,\tilde{\Bu}\in\tilde{\mathscr{U}}.
\label{weak_u}
\end{equation}
On the other hand, setting $\tilde{\Bu}=\bm{0}$ and $\tilde{\alpha}=0$, and using the relation $\Bsp:\tilde{\Be}^\mathrm{p}=\bm{s}^\mathrm{p}_\mathrm{dev}:\tilde{\bm{\varepsilon}}^\mathrm{p}_\mathrm{dev}+(1/3)\tr\bm{s}^\mathrm{p}\tr\tilde{\bm{\varepsilon}}^\mathrm{p}$, equation~\eqref{DS-1_2} yields
\begin{equation}
\int_\Omega \bigg( \bm{s}^\mathrm{p}_\mathrm{dev}:\tilde{\bm{\varepsilon}}^\mathrm{p}_\mathrm{dev} +  \frac{A_\varphi}{3A_\theta}\tr\Bsp\tr\tilde{\bm{\varepsilon}}^\mathrm{p}\bigg)\,\mathrm{d}\bm{x} \leq0 \quad \forall\,\tilde{\Be}^\mathrm{p}\in\tilde{\mathscr{B}}.
\label{weak_p0}
\end{equation}
In agreement with the function spaces~\eqref{funscspace_p}, we seek to maximize the left-hand side by taking $\tilde{\bm{\varepsilon}}^\mathrm{p}_\mathrm{dev}$ collinear to $\bm{s}^\mathrm{p}_\mathrm{dev}$ and, in view of $\tr\Bsp\leq0$, setting the smallest possible value $\tr\tilde{\bm{\varepsilon}}^\mathrm{p}=\sqrt{6}A_\theta\Vert\tilde{\bm{\varepsilon}}^\mathrm{p}_\mathrm{dev}\Vert$. Equation~\eqref{weak_p0} then gives
\begin{equation}
\int_\Omega \bigg( \Vert\bm{s}^\mathrm{p}_{{\mathrm{dev}}}\Vert + \sqrt\frac{2}{3}A_\varphi \tr \bm{s}^\mathrm{p}\bigg)\Vert\tilde{\bm{\varepsilon}}^\mathrm{p}_\mathrm{dev}\Vert\,\mathrm{d}\bm{x} \leq0 \quad \forall\,\tilde{\Be}^\mathrm{p}\in\tilde{\mathscr{B}},
\label{weak_p}
\end{equation}
from which the generalized stress constraint~\eqref{sc_setK} with the plastic yield function~\eqref{fp} is recovered as
\begin{equation}
f^\mathrm{p} \leq 0  \quad \text{in} \squad \Omega.
\label{strong_p_0}
\end{equation}
Finally,  for $\tilde{\Bu}=\bm{0}$ and $\tilde{\Be}^\mathrm{p}=\bm{0}$, equation~\eqref{DS-1_2} yields
\begin{equation}
\int_\Omega \bigg( s^\mathrm{d}\tilde{\alpha}  - \frac{{G}_\mathrm{c}(\bm{s}^\mathrm{p})}{{\ell}}\big(\alpha\,\tilde{\alpha}+{\ell}^{2}\nabla\alpha\cdot\nabla\tilde{\alpha}\big) \bigg) \,\mathrm{d}\bm{x}   \leq0 \quad \forall\,\tilde{\alpha}\in\tilde{\mathscr{D}},
\label{weak_d}
\end{equation}
representing the weak form of the damage stress constraint~\eqref{sc_setKd} with the damage yield function~\eqref{fd}. After integrating by parts, we recover the criteria
\begin{equation}
f^\mathrm{d} \leq 0  \quad \text{in} \squad \Omega \qquad \text{and} \qquad  \nabla\alpha\cdot\bm{n}\geq 0 \quad \text{on} \squad \Gamma.
\label{strong_d_0}
\end{equation}

\paragraph*{Consistency conditions and flow rule} We now take the power balance~\eqref{EB-1_2} as a point of departure, which demands that $\tr\dot{\bm{\varepsilon}}^\mathrm{p}\geq\sqrt{6}A_\theta\Vert \dot{\bm{\varepsilon}}^\mathrm{p}_\mathrm{dev}\Vert$ and $\dot{\alpha}\geq0$. After integrating the gradient terms by parts, we obtain
\begin{equation}
\begin{aligned}
\int_\Omega \big( \mathrm{div}\Bsig + \rho\bm{b}\big)\cdot\dot{\bm{u}} \,\mathrm{d}\bm{x}   - \int_{\Gamma_\mathrm{N}}\big(\Bsig\cdot\bm{n}&-\bar{\bm{t}}\,\big)\cdot\dot{\bm{u}}\,\mathrm{d}S  -\int_{\Gamma_{\mathrm{D}}}\big((\Bsig\cdot\bm{n})\cdot\dot{\bm{u}} -\bm{t}_\mathrm{r}\cdot\dot{\bar{\bm{u}}}\big)\, \mathrm{d}S \\  
& \hspace*{1.8cm} + \int_\Omega\bigg(\Bsp:\dot{\Be}^\mathrm{p} - \frac{\tr\dot{\bm{\varepsilon}}^\mathrm{p} }{3A_\theta}{(A_\theta-A_\varphi)\tr\Bsp}\bigg)\,\mathrm{d}\bm{x} \\
& \hspace*{1.8cm}+ \int_\Omega f^\mathrm{d}\dot{\alpha}\,\mathrm{d}\bm{x} - \int_{\Gamma}{\ell}\,{G}_\mathrm{c}(\Bsp)\nabla\alpha\cdot\bm{n}\,\dot{\alpha}\,\mathrm{d}S =0.
\label{EB-1_3}
\end{aligned}
\end{equation}
Recall that, from first-order stability, we have recovered the equilibrium equations~\eqref{st_adm} along with the plastic-damage criteria~\eqref{strong_p_0} and~\eqref{strong_d_0}. Thus, in view of $\Bu\in\mathscr{U}$, the first three integrals vanish, while the remaining integrals are non-positive and, therefore, must vanish individually for~\eqref{EB-1_3} to hold.  

The fourth integral in~\eqref{EB-1_3} yields
\begin{equation}
\Bsp:\dot{\Be}^\mathrm{p} - \frac{\tr\dot{\bm{\varepsilon}}^\mathrm{p} }{3A_\theta}{(A_\theta-A_\varphi)\tr\Bsp} = 0.
\label{comp_p00}
\end{equation} 
We now let, without loss of generality, $\dot{\bm{\varepsilon}}^\mathrm{p}\coloneqq \lambda\hat{\bm{n}}$, with $\lambda\geq0$ and $\hat{\bm{n}}\in\mathbb{R}^{3\times 3}_\mathrm{sym}$ such that $\tr\hat{\bm{n}}=\sqrt{6}A_\theta$. Then, the constraint $\tr\dot{\bm{\varepsilon}}^\mathrm{p}\geq\sqrt{6}A_\theta\Vert \dot{\bm{\varepsilon}}^\mathrm{p}_\mathrm{dev}\Vert$ implies that $\Vert\hat{\bm{n}}_\mathrm{dev}\Vert\leq 1$.  As a consequence, 
\begin{equation}
\begin{aligned}
0=\Bsp:\dot{\Be}^\mathrm{p} - \frac{\tr\dot{\bm{\varepsilon}}^\mathrm{p} }{3A_\theta}{(A_\theta-A_\varphi)\tr\Bsp} &= \, \lambda\,\bigg(\bm{s}^\mathrm{p}_\mathrm{dev}:\hat{\bm{n}}_\mathrm{dev} + \sqrt{\frac{2}{3}}A_\varphi\tr\bm{s}^\mathrm{p}\bigg)\\
&\leq \, \lambda\,\bigg(\Vert\bm{s}^\mathrm{p}_{\mathrm{dev}}\Vert \,\Vert \hat{\bm{n}}_\mathrm{dev}\Vert + \sqrt{\frac{2}{3}}A_\varphi\tr\bm{s}^\mathrm{p}\bigg)\leq \lambda\,f^\mathrm{p}.
\end{aligned}
\label{comp_p0}
\end{equation} 
Thus, in view of~\eqref{strong_p_0}, we recover the consistency conditions for plasticity 
\begin{equation}
\lambda\,f^\mathrm{p}=0  \quad \text{in} \squad \Omega.
\label{comp_p}
\end{equation}
Note that, at this point, $\hat{\bm{n}}_\mathrm{dev}$ has not been determined. For $\lambda>0$, we may solve $f^\mathrm{p}=0$ for  $\tr\Bsp$, which is replaced in~\eqref{comp_p00} along with $\dot{\bm{\varepsilon}}^\mathrm{p}=\lambda\hat{\bm{n}}$ to obtain the relation
\begin{equation}
\Bsp_\mathrm{dev}:\hat{\bm{n}}_\mathrm{dev}=\Vert\bm{s}^\mathrm{p}_\mathrm{dev}\Vert \quad \implies \quad \hat{\bm{n}}_\mathrm{dev}\in\partial\,\Vert\bm{s}^\mathrm{p}_\mathrm{dev}\Vert.
\label{fr_00}
\end{equation}
Recalling the definition of the plastic potential~\eqref{gp}, it follows that 
\begin{equation}
\dot{\bm{\varepsilon}}^\mathrm{p}=\lambda\,\hat{\bm{n}} \quad \text{with} \quad \hat{\bm{n}}\in\partial\,\Vert\bm{s}^\mathrm{p}_\mathrm{dev}\Vert+\sqrt{\frac{2}{3}}A_\theta\bm{1}\equiv \partial g^\mathrm{p}(\Bsp).
\label{fr_p}
\end{equation}
Thus, by virtue of equations~\eqref{comp_p} and~\eqref{fr_p}, we have recovered the non-associative flow rule~\eqref{nonass} including the plasticity consistency conditions as a consequence of energy balance.

We finally consider the cancellation of the last two integrals in equation~\eqref{EB-1_3}, which yield the damage consistency conditions 
\begin{equation}
\dot{\alpha}\, f^\mathrm{d} = 0  \quad \text{in} \squad \Omega \qquad \text{and} \qquad  \dot{\alpha} (\nabla\alpha\cdot\bm{n}) = 0 \quad \text{on} \squad \Gamma.
\label{comp_d}
\end{equation}

\paragraph*{Dissipation inequality}

In order to ensure thermodynamic consistency, the dissipation inequality is included in the formulation. This condition is guaranteed by the non-negativity of the dissipation potential:
\begin{equation}
\phi^\mathrm{p}(\dot{\bm{\varepsilon}}^\mathrm{p};{\bm{s}^\mathrm{p}})+\phi^\mathrm{d}(\dot{\alpha},\nabla\dot{\alpha};\alpha,\nabla\alpha,{\bm{s}^\mathrm{p}})\geq0 \quad \text{in} \squad  \Omega\times\Tau.
\label{DI}
\end{equation}
In the present model, it is easy to see from equation~\eqref{phi_d} that the local term in $\phi^\mathrm{d}$ is a priori non-negative, but the non-local term in $\phi^\mathrm{d}$ is not. Nevertheless, $\phi^\mathrm{d}$ is non-negative as a consequence of first-order stability. Specifically, from equation~\eqref{weak_d}, it follows that
\begin{equation}
\phi^\mathrm{d}(\dot{\alpha},\nabla\dot{\alpha};\alpha,\nabla\alpha,{\bm{s}^\mathrm{p}}) \geq  s^\mathrm{d}\dot{\alpha} \geq0 \quad \text{in} \squad \Omega\times\Tau \quad \forall\,\dot{\alpha}\in\tilde{\mathscr{D}}.
\end{equation}
Similarly, from equation~\eqref{phi_p}  alone, the non-negativity of the plastic contribution $\phi^\mathrm{p}$ is not observed a priori. However, when considering the stability condition, it follows from equation~\eqref{weak_p} that $\tr\Bsp\leq 0$ such that, in view of $A_\theta<A_\varphi$,  $\phi^\mathrm{p}$ is non-negative (and finite for non-vanishing $\Bsp$ and $\dot{\bm{\varepsilon}}^\mathrm{p}$). 

\paragraph*{Overview}

At this point, we have obtained from the principles of the energetic formulation the governing equations of the coupled multi-field system, consisting of mechanical balance, the plasticity evolution problem, and the damage evolution problem. Thereby, depending on the opening/closure transition~\eqref{optrans}, the model can be viewed as either a modified brittle phase-field model (open or tensile regime), or as a ductile damage model where fracture is solely driven by plastic strains (closed or compressive/shear regime). At the microscale, the former is associated with opening microcracks, while the latter is associated with a coupling between microcrack growth and frictional sliding. Tables~\ref{overview1} and~\ref{overview2} present an overview of the proposed~model.

\begin{table}
\renewcommand{\arraystretch}{1.1}
\small
\centering
\caption{Energy quantities and state equations.}
\begin{tabular}{ll}
\toprule
\multicolumn{2}{l}{\sffamily\textbf{Free energy and state equations}}
\\
\midrule
Stored energy            & $\psi(\bm{\varepsilon},\bm{\varepsilon}^\mathrm{p},\alpha)=\begin{dcases}
\frac{1}{2}\bm{\varepsilon}:\bm{\mathsf{C}}^\mathrm{dam}(\alpha):\bm{\varepsilon}& \quad \text{if \sf{open}},  \\[5pt] 
\frac{1}{2}(\bm{\varepsilon}-\bm{\varepsilon}^\mathrm{p}):\bm{\mathsf{C}}:(\bm{\varepsilon}-\bm{\varepsilon}^\mathrm{p}) + \frac{1}{2}\bm{\varepsilon}^\mathrm{p}:\bm{\mathsf{H}}^\mathrm{kin}(\alpha):\bm{\varepsilon}^\mathrm{p} & \quad \text{if \sf{closed}}\end{dcases}$
\\
\grayrule
Generalized stresses  & $\bm{\sigma}(\bm{\varepsilon},\bm{\varepsilon}^\mathrm{p},\alpha)=\dfrac{\partial\psi}{\partial\bm{\varepsilon}}, \quad\Bsp(\bm{\varepsilon},\bm{\varepsilon}^\mathrm{p},\alpha)=-\dfrac{\partial\psi}{\partial\bm{\varepsilon}^\mathrm{p}}, \quad {s}^\mathrm{d}(\bm{\varepsilon},\bm{\varepsilon}^\mathrm{p},\alpha)=-\dfrac{\partial\psi}{\partial\alpha}$
\\
\grayrule
Opening/closure transition & $\begin{dcases}\tr\Bsp(\bm{\varepsilon},\bm{\varepsilon}^\mathrm{p},\alpha)=0 \quad \text{if \sf{open}}, \\ \tr\Bsp(\bm{\varepsilon},\bm{\varepsilon}^\mathrm{p},\alpha) < 0 \quad \text{if \sf{closed}}\end{dcases}$
\\
\midrule
\multicolumn{2}{l}{\sffamily\textbf{Dissipation potential:} $\phi=\phi^\mathrm{p}+\phi^\mathrm{d}\geq 0$}
\\
\midrule
Plastic dissipation potential        & $\phi^\mathrm{p}(\dot{\bm{\varepsilon}}^\mathrm{p};{\Bsp})=\begin{dcases}
\frac{\tr\dot{\bm{\varepsilon}}^\mathrm{p} }{3A_\theta}{(A_\theta-A_\varphi)\tr\Bsp} \quad \text{if } \tr\dot{\bm{\varepsilon}}^\mathrm{p}\geq\sqrt{6}A_\theta\Vert \dot{\bm{\varepsilon}}^\mathrm{p}_\mathrm{dev}\Vert,\\
+\infty \quad \text{otherwise}\end{dcases}$ 
\\ \grayrule
Damage dissipation potential        & $\phi^\mathrm{d}(\dot{\alpha},\nabla\dot{\alpha};\alpha,\nabla\alpha,{\bm{s}^\mathrm{p}})=\begin{dcases}\frac{{G}_\mathrm{c}(\bm{s}^\mathrm{p})}{{\ell}}\big(\alpha\,\dot{\alpha}+{\ell}^{2}\nabla\alpha\cdot\nabla\dot{\alpha}\big) \quad \text{if }  \dot{\alpha}\geq 0, \\
+\infty \quad \text{otherwise}\end{dcases}$ 
\\ \bottomrule
\end{tabular}
\label{overview1}
\end{table}

\begin{table}
\renewcommand{\arraystretch}{1.1}
\small
\centering
\caption{Governing equations according to the energetic formulation.}
\begin{tabular}{lll}
\toprule
\multicolumn{3}{l}{\sffamily\textbf{Kinematic admissibility}}                                                                                                                                                       
\\ \midrule
Infinitesimal strain       &   $\Be(\Bu)=\nabla^\mathrm{s}\Bu$ & 
\\ \grayrule
Dirichlet boundary condition & $\Bu=\bar{\Bu} \ $  on $\Gamma_\mathrm{D}$
\\ \midrule
\multicolumn{2}{l}{\sffamily \textbf{Mechanical balance}}
\\ \midrule
Stress & $\Bsig(\bm{\varepsilon},\bm{\varepsilon}^\mathrm{p},\alpha)=\begin{dcases}
\bm{\mathsf{C}}^\mathrm{dam}(\alpha):\bm{\varepsilon}& \quad \text{if } \tr{\Bsp}(\bm{\varepsilon},\bm{\varepsilon}^\mathrm{p},\alpha)=0 \quad (\sf{open}),  \\
\bm{\mathsf{C}}:(\bm{\varepsilon}-\bm{\varepsilon}^\mathrm{p}) & \quad \text{if } \tr{\Bsp}(\bm{\varepsilon},\bm{\varepsilon}^\mathrm{p},\alpha)<0 \quad (\sf{closed})\end{dcases}$ 
\\ \grayrule
Equilibrium  & $\mathrm{div}\,\bm{\sigma}(\bm{\varepsilon},\bm{\varepsilon}^\mathrm{p},\alpha)+\rho\bm{b} = \boldsymbol{0} \ $  {in} $\Omega$ 
\\ \grayrule
Neumann boundary condition &  $\bm{\sigma}(\bm{\varepsilon},\bm{\varepsilon}^\mathrm{p},\alpha)\cdot\boldsymbol{n}=\bar{\bm{t}}\ $  on $\Gamma_{\mathrm{N}}$
\\ \midrule
\multicolumn{2}{l}{\sffamily \textbf{Plasticity evolution problem}}
\\ \midrule
Generalized stress & $\Bsp(\bm{\varepsilon},\bm{\varepsilon}^\mathrm{p},\alpha)=\bm{\mathsf{C}}:(\bm{\varepsilon}-\bm{\varepsilon}^\mathrm{p}) - \bm{\mathsf{H}}^\mathrm{kin}(\alpha):\bm{\varepsilon}^\mathrm{p}$ 
\\ \grayrule
Yield function & $f^\mathrm{p}(\bm{s}^\mathrm{p})=\Vert\Bsp_{{\mathrm{dev}}}\Vert + \sqrt{{2}/{3}}\,A_\varphi \mathrm{tr} \Bsp$ 
\\ \grayrule
Plastic potential & $g^\mathrm{p}(\bm{s}^\mathrm{p})=\Vert\Bsp_{{\mathrm{dev}}}\Vert + \sqrt{{2}/{3}}\,A_\theta \mathrm{tr} \Bsp$ 
\\ \grayrule
KKT system & $f^\mathrm{p}(\bm{s}^\mathrm{p})\leq0, \quad \lambda\geq 0, \quad \lambda\,f^\mathrm{p}(\bm{s}^\mathrm{p})=0 \ $  in $\Omega$
\\ \grayrule
Flow rule & $\dot{\bm{\varepsilon}}^\mathrm{p}=\lambda\,\hat{\bm{n}}, \quad \hat{\bm{n}}\in\partial g^\mathrm{p}(\bm{s}^\mathrm{p})=\partial\Vert\Bsp_{\mathrm{dev}}\Vert+\sqrt{{2}/{3}}\,A_\theta\bm{1}\ $  in $\Omega$
\\ \midrule
\multicolumn{2}{l}{\sffamily \textbf{Damage evolution problem}}
\\ \midrule
Generalized stress & $s^\mathrm{d}(\bm{\varepsilon},\bm{\varepsilon}^\mathrm{p},\alpha)=\begin{dcases}
-\frac{1}{2}\bm{\varepsilon}:{\bm{\mathsf{C}}^\mathrm{dam}}'(\alpha):\bm{\varepsilon}& \quad \text{if }  \tr{\Bsp}(\bm{\varepsilon},\bm{\varepsilon}^\mathrm{p},\alpha)=0 \quad (\sf{open}),  \\[5pt]
-\frac{1}{2}\bm{\varepsilon}^\mathrm{p}:{\bm{\mathsf{H}}^{\mathrm{kin}}}'(\alpha):\bm{\varepsilon}^\mathrm{p}  & \quad \text{if } \tr{\Bsp}(\bm{\varepsilon},\bm{\varepsilon}^\mathrm{p},\alpha)<0  \quad (\sf{closed})
\end{dcases} $ 
\\ \grayrule
Yield function & $f^\mathrm{d}(s^\mathrm{d};\Bsp)=s^\mathrm{d}-\dfrac{{G}_\mathrm{c}(\Bsp)}{{\ell}}\alpha+{\ell}\,\mathrm{div}[{G}_\mathrm{c}(\Bsp)\nabla\alpha]$ 
\\ \grayrule
KKT system & $f^\mathrm{d}(s^\mathrm{d};\Bsp)\leq0, \quad \dot{\alpha}\geq 0, \quad \dot{\alpha}\,f^\mathrm{d}(s^\mathrm{d};\Bsp)=0 \ $  in $\Omega$
\\ \grayrule
Boundary condition & $\nabla\alpha\cdot\bm{n}=0 \ $  on $\Gamma$
\\ \bottomrule
\end{tabular}
\label{overview2}
\end{table}

It is worth mentioning that, in view of the energetic formulation, the numerical solution of the multi-field coupled system can be cast as an incremental energy minimization problem. In this context, if the dissipation potential is such that the dissipated energy becomes a state function, the incremental minimization problem recovers the evolution problem exactly. However, if the dissipated energy is path-dependent, it must be approximated in incremental form. Then, given a suitable approximation, the incremental minimization problem recovers the exact evolution equations for sufficiently small (pseudo-) time increments~\citep{ulloa2021b}. For the present model, the state-dependence of the dissipation potential~\eqref{phi} renders the dissipated energy path-dependent. Therefore, for simplicity, we do not resort to incremental energy minimization and, hereinafter, we directly focus on the solution of the evolution equations summarized (in strong form) in table~\ref{overview2}. 

Finally, for post-processing purposes, let us denote the equivalent plastic strain in the compressive/shear regime by $\kappa$. Then, we set
\begin{equation}
\dot{\kappa}\coloneqq\begin{dcases} 0 & \quad \text{if } \tr{\Bsp}(\bm{\varepsilon},\bm{\varepsilon}^\mathrm{p},\alpha)=0 \quad (\sf{open}),  \\
\sqrt{2/3}\,\Vert\dot{\bm{\varepsilon}}^\mathrm{p}\Vert & \quad \text{if } \tr{\Bsp}(\bm{\varepsilon},\bm{\varepsilon}^\mathrm{p},\alpha)<0 \quad (\sf{closed}).\end{dcases}
\label{kappa}
\end{equation}
This variable is intended to quantify the amount of frictional sliding, as will become clear in the sequel.

\FloatBarrier

\subsection{{Homogeneous response}}\label{homog}

We conclude this section with an illustrative description of the different material responses embedded in the proposed model. We consider the homogeneous response of a single volume element subjected to axial loading, where axial strains $\varepsilon_{zz}$ are monotonically imposed  in either tension ($\dot{\varepsilon}_{zz}>0$) or compression ($\dot{\varepsilon}_{zz}<0$). Under compression, the effect of confining pressure is also discussed. To this end, a hydrostatic confining stress is applied, with $\sigma_{xx}=\sigma_{yy}=\sigma_{zz}$ gradually varying from $0$ to $-p_0$, with $p_0\geq 0$. Then, $\varepsilon_{zz}$ is varied with $\dot{\varepsilon}_{zz}<0$, while maintaining the lateral pressure $\sigma_{xx}=\sigma_{yy}=-p_0$. Fixed material parameters are chosen as follows: Young's modulus $E=1$ MPa, Poisson's  ratio $\nu=0.3$, mode I fracture toughness $G_\mathrm{cI}=7.5$ N/mm, and degradation constant $b=2$. Further, a low initial damage $\alpha_0=1\times10^{-5}$ is considered to trigger inelastic behavior in the compressive/shear regime.

\begin{figure}[H]
  \centering
   \includegraphics[scale=1]{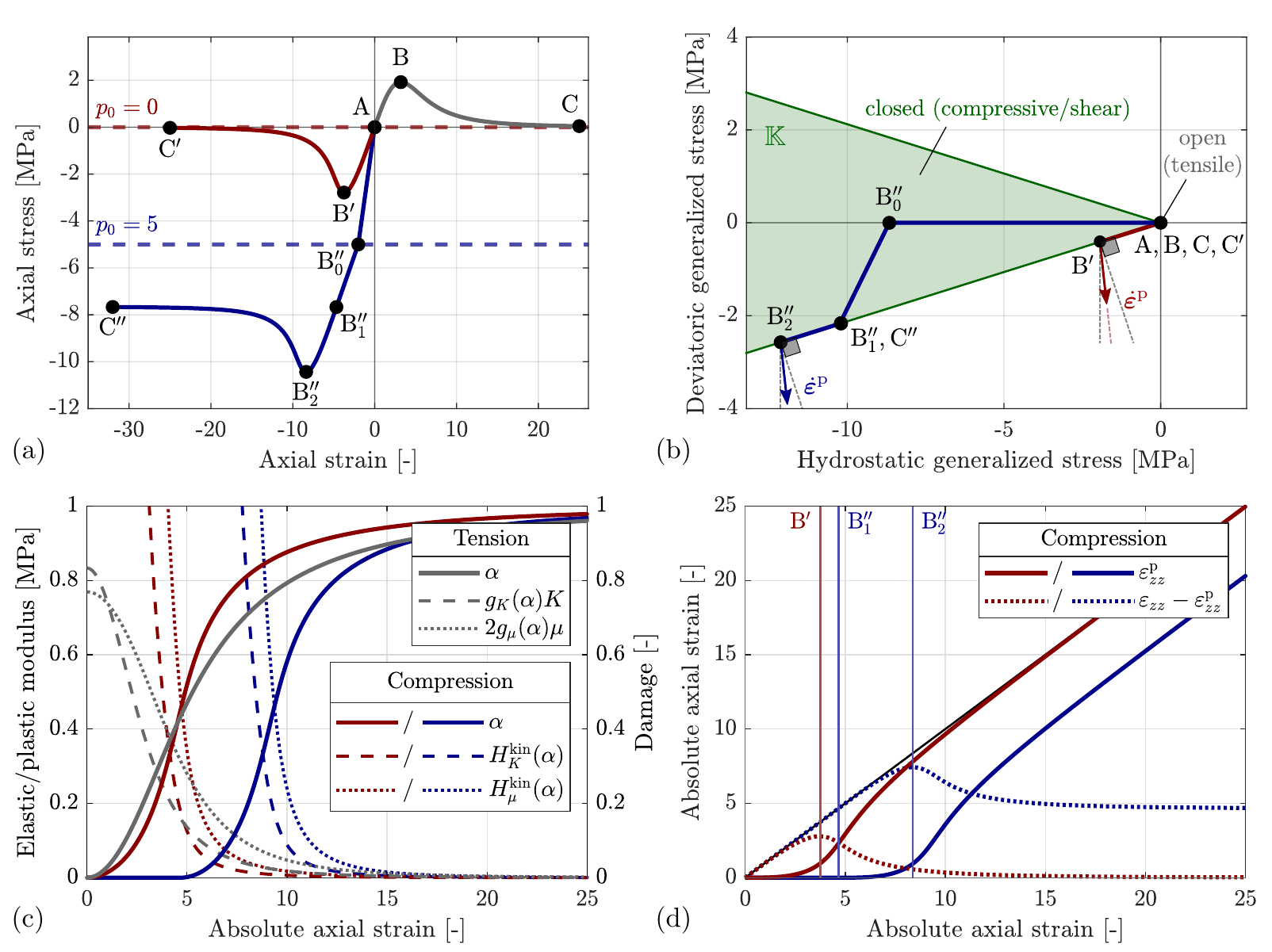}
\caption{Homogeneous response for uniaxial tension (gray), uniaxial compression (red), and triaxial compression  (blue):  (a) axial stress vs.~axial strain curves, (b) the corresponding stress path plotted in $\big(\sqrt{3}\tr\Bsp/3,\mathrm{sign}(s^\mathrm{p}_{\mathrm{dev}\,zz})\Vert\Bsp_\mathrm{dev}\Vert\big)$ space, (c) the corresponding degradation of the elastic and plastic moduli as damage evolves, and (d) the plastic strain evolution in the compressive/shear regime, causing the softening response. The compressive responses lie in the compressive/shear regime (closed microcracks), i.e., in the Drucker-Prager cone shown in green. The plastic flow directions at peak stress are plotted as arrows, showing dilation  and the non-normality condition. The unloaded states as well as the tensile response (open microcracks) correspond to a single point at the apex of the cone.}
\label{hom_mon_conf_path}
\end{figure}

In tension, the dissipative response is solely modulated by the mode I fracture toughness $G_\mathrm{cI}$. Figure~{\ref{hom_mon_conf_path}(a)} shows a softening response that resembles the behavior of standard brittle damage models. As $\alpha\to1$ (complete damage), the stress vanishes by virtue of $\bm{\mathsf{C}}^\mathrm{dam}(\alpha)\to\bm{\mathsf{0}}$ (equations~\eqref{Cdam}, \eqref{gk}, and~\eqref{gmu}), as shown in figure~{\ref{hom_mon_conf_path}(c)}. Because the microcracks remain open, the generalized stress path in $\Bsp$ space shown in figure~{\ref{hom_mon_conf_path}(b)} remains fixed at the apex (points A, B, and C). Note that here, no frictional dissipation takes~place;  energy dissipation is exclusively due to microcrack growth.

In the compressive/shear regime, the dissipative response depends on the mode II fracture toughness $G_\mathrm{cII}$, the friction constant $A_\varphi$, and the dilation constant $A_\theta$. Figure~{\ref{hom_mon_conf_path}} shows the response with $p_0=0$ and $p_0=5$ MPa, representing uniaxial compression and triaxial compression, respectively. For now, $G_\mathrm{cII}=G_\mathrm{cI}$, $A_\varphi=0.15$, and $A_\theta=0.1125$ are assumed. For uniaxial compression, an initial hardening response is observed, reaching a peak stress at point B$'$. Then, a softening stage is triggered, where the stress completely vanishes (point C$'$). In this case, no elastic stage is achieved, with the stress path in figure~{\ref{hom_mon_conf_path}}(b) always on the surface of the Drucker-Prager cone (corresponding to frictional sliding). Thus, due to the strong plastic-damage coupling, damage always increases. Note that here, as damage increases from $\alpha_0$ to $1$, the elastic moduli remain intact. Conversely, the hardening moduli are such that $H^\mathrm{kin}_K(\alpha_0)\to\infty$ and $H^\mathrm{kin}_\mu(\alpha_0)\to\infty$ for $\alpha_0\to0$ (equation~\eqref{Hkin_K_mu}). Then, as shown in figure~{\ref{hom_mon_conf_path}}(c), $H^\mathrm{kin}_K(\alpha)$ and $H^\mathrm{kin}_\mu(\alpha)$ decrease from rather high values to $0$ as $\alpha\to1$.  Consequently, a perfect plasticity stage is reached as $\bm{\mathsf{H}}^\mathrm{kin}(\alpha)\to\bm{\mathsf{0}}$, where, due to the absence of confining pressure, the stress state is such that $\Bsig=\Bsp=\bm{0}$.  Further, figure~{\ref{hom_mon_conf_path}}(d) shows that the softening stage is triggered by the increase in plastic strains as the hardening moduli decrease. Then, as $\alpha\to1$, the plastic strains and the total strains converge towards each other. 

Consider now the case of triaxial compression. While the load is applied hydrostatically, up to $p_0=5$ MPa (point B$''_0$), figures~{\ref{hom_mon_conf_path}}(a) and~{\ref{hom_mon_conf_path}}(b) show an initial elastic response. In the absence of a crushing mechanism, e.g., a compression cap, the response would indefinitely remain  elastic with further hydrostatic loading. At point B$''_0$, the (deviatoric) axial loading starts, but the response remains elastic until reaching the surface of the cone at point B$''_1$. As shown in figures~{\ref{hom_mon_conf_path}}(c) and ~{\ref{hom_mon_conf_path}}(d), $\alpha$ remains at $\alpha_0$ and $\Bep$ remains null until this point, with constant (and inactive) $H^\mathrm{kin}_K(\alpha_0)$ and $H^\mathrm{kin}_\mu(\alpha_0)$. Then, similar to the case of uniaxial compression, an initial hardening response is observed, where the plastic-damage coupling entails a competing hardening-softening mechanism. As the plastic strains increase with decreasing hardening moduli, a peak stress is reached at point B$''_2$. Then, as $\alpha\to1$ and $\bm{\mathsf{H}}^\mathrm{kin}(\alpha)\to\bm{\mathsf{0}}$, the stress drops to point C$''$, reaching a perfect plasticity stage with a finite \emph{residual strength}. The residual strength is due to the constant friction coefficient $A_\varphi$ and the fact that an initial elastic stage was achieved by imposing confining pressure.

\begin{figure}[H]
  \centering
   \includegraphics[scale=1]{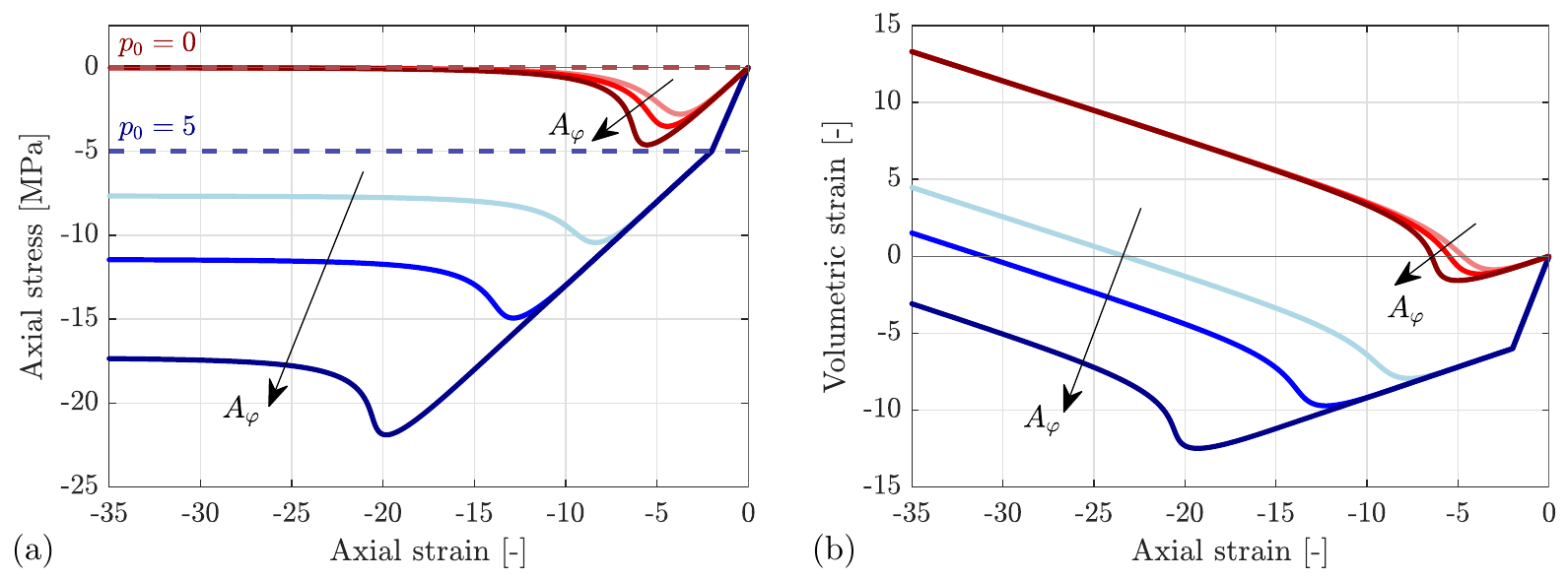}
\caption{Homogeneous response for uniaxial compression (red) and triaxial compression  (blue): (a) axial stress vs.~axial strain curves and (b) the corresponding total volumetric strains. The response is shown for $A_\varphi=0.15$ (light red/blue), $A_\varphi=0.3$ (red/blue), and $A_\varphi=4.5$ (dark red/blue).}
\label{hom_mon_conf}
\end{figure}

Let us now discuss the effect of the material parameters in the compressive/shear regime. Figure~\ref{hom_mon_conf} shows the effect of varying friction ($A_\varphi$) while fixing both $A_\theta$ and $G_\mathrm{cII}$. For uniaxial compression, the peak stress and the corresponding failure strain in figure~\ref{hom_mon_conf}(a) increase in absolute value as $A_\varphi$ increases. As $\alpha\to1$, a vanishing stress is attained, while the volumetric strains in figure~\ref{hom_mon_conf}(b) converge to the same curve. On the other hand, for triaxial compression, figure~\ref{hom_mon_conf}(a) shows that increasing $A_\varphi$ also results in higher residual strength, while only the~\emph{volumetric strain rates}, modulated by the constant dilation coefficient $A_\theta$, converge to the same values in figure~\ref{hom_mon_conf}(b). We note that an extension of the model may consider varying friction and dilation coefficients, with different peak and residual parameters. As such, we may allow the residual strength and the volumetric strain rate to decrease or even vanish. The response of such a model is presented in appendix~\ref{dam_fric_dil}.

\begin{figure}[H]
  \centering
   \includegraphics[scale=1]{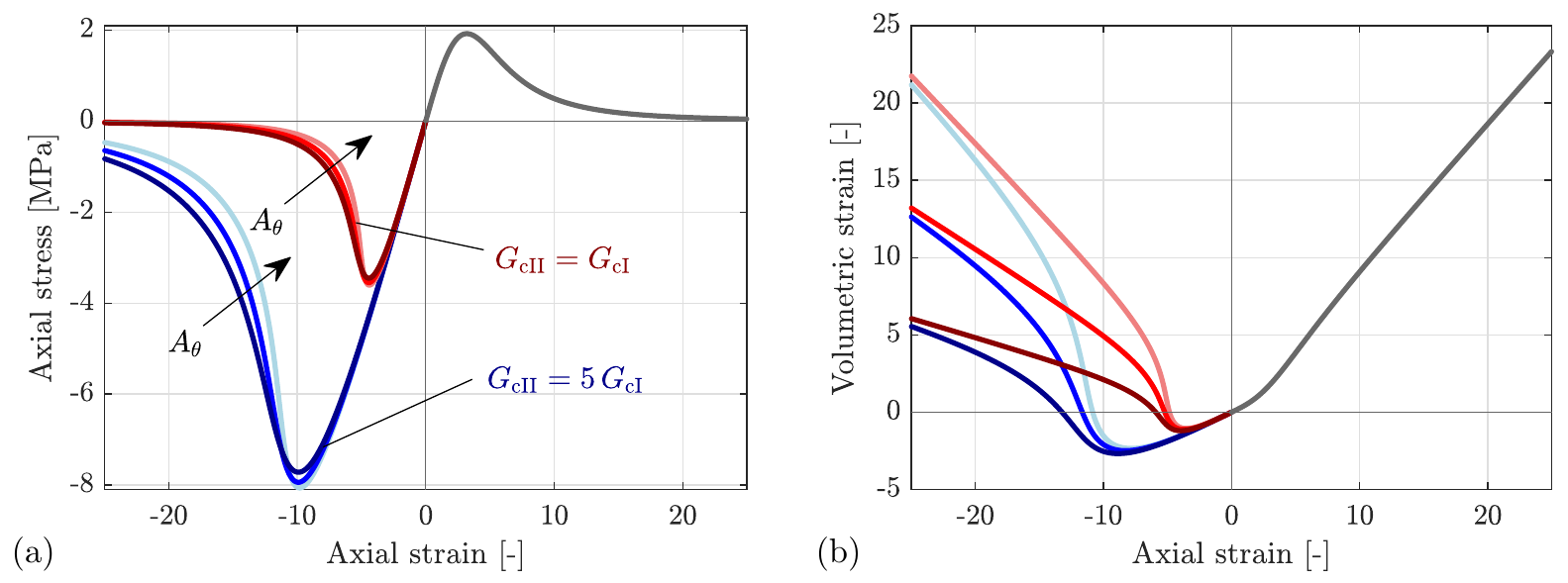}
\caption{Homogeneous response for uniaxial tension (gray) and compression at fixed $A_\varphi=0.3$ (blue and red): (a) axial stress vs.~axial strain curves and (b) the corresponding total volumetric strains. Uniaxial compression is shown for $G_\mathrm{cII}=G_\mathrm{cI}$ (red) and $G_\mathrm{cII}=5\,G_\mathrm{cI}$ (blue), with $A_\theta=0.75\,A_\varphi$ (light red/blue), $A_\theta=0.5\,A_\varphi$ (red/blue), and $A_\theta=0.25\,A_\varphi$ (dark red/blue).}
\label{hom_mon}
\end{figure}

Finally, we discuss the crucial influence of the mode II fracture toughness $G_\mathrm{cII}$ and the dilation coefficient $A_\theta$. In the responses discussed above, the tensile regime is modulated by $G_\mathrm{cI}$, while for the compressive/shear regime, $G_\mathrm{cII}=G_\mathrm{cI}$ was considered. However, a key feature of the model is that $G_\mathrm{cII}$ can be chosen independently. Figure~\ref{hom_mon} shows the response for uniaxial compression with fixed $A_\varphi=0.3$,  while both $G_\mathrm{cII}$ and $A_\theta$ vary. For a given $A_\theta$, figure~{\ref{hom_mon}(a)} shows the expected increase in strength as $G_\mathrm{cII}$ increases, while the volumetric strains in figure~{\ref{hom_mon}(b)} converge to the same curve. This increase in strength is crucial to capture the higher resistance to shear fracture  observed in rock-like materials. On the other hand, when $A_\theta$ increases, figure~{\ref{hom_mon}(a)} shows a slight increase in peak stress (in absolute value), while figure~{\ref{hom_mon}(b)} shows the expected increase in volumetric strains. Note that, as in figure~\ref{hom_mon_conf}, the volumetric strains indefinitely increase with further loading. However, if the straightforward extension presented in appendix~\ref{dam_fric_dil} is considered, the volumetric strains may be allowed to converge to a constant value.

%\clearpage

\FloatBarrier

\section{{Numerical implementation}}\label{implement}

This section is devoted to the numerical implementation of the evolution equations summarized in table~\ref{overview2}. We consider  a time discretization ${0=t_0<\dots<t_{n}<t_{n+1}<\dots<t_{n_\mathrm{t}}=t_\mathrm{max}}$, where all quantities are known up to $t_n$, and the goal is to find the state at the current time step $t_{n+1}$. For convenience, we introduce the following  notations. A quantity $\Box$ evaluated at the previous time step $t_n$ is denoted as $\Box_n$, while a quantity evaluated at $t_{n+1}$ is written without a subscript, i.e., $\Box\coloneqq \Box_{n+1}$. Moreover, the operator $\Delta\Box\coloneqq \Box-\Box_n$ is used to denote an increment of $\Box$ from $t_{n}$ to $t_{n+1}$.

\subsection{Staggered algorithm}\label{alg_stag}

As commonly done for phase-field models,  we employ a staggered solution  technique  based on an algorithmic decoupling of the governing equations. More specifically, we consider a semi-staggered scheme in which the elastoplastic problem and the damage  problem are iteratively solved for $\{\Bu,\Bep\}$ and $\alpha$, respectively. Both of these subproblems are non-linear and therefore require the use of Newton-Raphson schemes, as described in the following subsections. The overall semi-staggered procedure is summarized in algorithm~\ref{alg1}.

\begin{algorithm}[!ht]
 \hspace*{\algorithmicindent} \textbf{Input}: primary fields at the previous time step $\Bu_n$, $\Bep_n$, and $\alpha_n$. \\
 \hspace*{\algorithmicindent} \textbf{Output}: primary fields at the current time step $\Bu$, $\Bep$, and $\alpha$. 
\begin{algorithmic}[1]
\State Initialize iterations with $j\coloneqq 0$ and $\{\Bu^{(0)},\bm{\varepsilon}^{\mathrm{p}(0)},\alpha^{(0)}\}\coloneqq \{\Bu_n,\bm{\varepsilon}^{\mathrm{p}}_n,\alpha_n\}$.
\Repeat  
\State Set $j \leftarrow j+1$.
\State Solve the non-linear \underline{elastoplastic evolution problem}  for $\{\bm{u}^{(j)},\bm{\varepsilon}^{\mathrm{p}(j)}\}$ using $\alpha^{(j-1)}$ (section~\ref{alg_ep}).
 \State Solve the non-linear \underline{damage evolution problem} for $\alpha^{(j)}$ using $\{\bm{u}^{(j)},\bm{\varepsilon}^{\mathrm{p}(j)}\}$ (section~\ref{alg_dam}).
  \State Update $$\mathtt{RES}_{\mathrm{stag}}^{
(j)}\coloneqq \int_\Omega  \Big[\Bsig\big(\nabla^{\mathrm{s}}\Bu^{(j)},\bm{\varepsilon}^{\mathrm{p}(j)},\alpha^{(j)}\big):\nabla^\mathrm{s}\tilde{\Bu} - \rho\bm{b}\cdot\tilde{\bm{u}}\Big]\,\mathrm{d}\bm{x} - \int_{\Gamma_\mathrm{N}}\bar{\bm{t}}\cdot\tilde{\bm{u}}\,\mathrm{d}S \quad \forall\,\tilde{\Bu}\in\tilde{\mathscr{U}}.$$   
\Until $\big\vert \mathtt{RES}_{\mathrm{stag}}^{(j)}\big\vert\leq \mathtt{TOL}_{\mathrm{stag}}$.
\State Set $\{\Bu,\Bep,\alpha\}\coloneqq\{\Bu^{(j)},\bm{\varepsilon}^{\mathrm{p}(j)},\alpha^{(j)}\}$.
\end{algorithmic}
\caption{The (semi-) staggered solution procedure.}\label{alg1}
\end{algorithm}

\subsection{Elastoplastic evolution problem}\label{alg_ep}

According to the algorithmic decoupling shown in algorithm~\ref{alg1}, the task of finding $\{\Bu,\Bep\}$ with fixed $\alpha$ can be viewed as a classical non-linear elastoplasticity problem where the plastic strains evolve according to a non-cohesive Drucker-Prager model with kinematic hardening.  Thus, we proceed in a conventional manner by linearizing the mechanical balance equation and updating the plastic strain tensor $\Bep$ by means of a local return-mapping algorithm, as thoroughly detailed in the literature~\cite{simo1998,de2011,borja2013}. The main steps are summarized below.

The linearization of  the weak form of mechanical balance~\eqref{weak_u} yields the variational expression
\begin{equation}
\int_\Omega \nabla^\mathrm{s}[\Bu^{(k+1)}-\Bu^{(k)}]:\bm{\mathsf{C}}^{\mathrm{ep}(k)}:\nabla^\mathrm{s}\tilde{\Bu}\,\mathrm{d}\bm{x}=-\int_\Omega  \big(\Bsig^{(k)}:\nabla^\mathrm{s}\tilde{\Bu}  - \rho\bm{b}\cdot\tilde{\bm{u}}\big)\,\mathrm{d}\bm{x} + \int_{\Gamma_\mathrm{N}}\bar{\bm{t}}\cdot\tilde{\bm{u}}\,\mathrm{d}S \eqqcolon -\mathtt{RES}_{\Bu}^{(k)} \quad \forall\,\tilde{\Bu}\in\tilde{\mathscr{U}},
\label{weak_u_alg}
\end{equation}
which is solved sequentially for the displacement iterate $\Bu^{(k+1)}$ until $\big\vert\mathtt{RES}_{\Bu}^{(k)}\big\vert\leq \mathtt{TOL}_{\Bu}$, where $\mathtt{TOL}_{\Bu}$ is a small predefined tolerance.  Note that here, the index $j$ corresponding to the staggered iteration counter in algorithm~\ref{alg1} has been omitted for notational simplicity, while the superscript $k$ denotes the iteration counter for the elastoplastic problem. Moreover, $\bm{\mathsf{C}}^{\mathrm{ep}(k)}\coloneqq\partial\Bsig^{(k)}/ \partial \Be^{(k)}$ and $\Bsig^{(k)}$ denote the consistent tangent and the stress tensor at the current iteration, to be determined along with the corresponding plastic strain tensor $\bm{\varepsilon}^{\mathrm{p}(k)}$ by solving the plasticity evolution equations (table~\ref{overview2}). 

The plasticity evolution equations are solved locally by means of an implicit Euler scheme, where the flow rule~\eqref{fr_p} is approximated as
\begin{equation}
\Bep=\Bep_n + \Delta \Bep \quad \text{with} \quad \Delta\Bep=\Delta\gamma\,\hat{\bm{n}}, \quad \hat{\bm{n}}\in\partial g^\mathrm{p}(\Bsp), 
\end{equation}
where the iteration counter $k$ has been dropped for simplicity, while $\Delta\gamma$ denotes the incremental approximation  of the plastic multiplier $\dot{\gamma}\coloneqq\lambda$. Then, defining the \emph{trial state} with fixed plastic flow as
\begin{equation}
\bm{s}^\mathrm{p\,trial}\coloneqq\Bsig^\mathrm{trial} -\bm{\mathsf{H}}^\mathrm{kin}(\alpha)\Bep_n\quad \text{with} \quad \Bsig^\mathrm{trial}\coloneqq \bm{\mathsf{C}}:(\Be-\Bep_n),
\end{equation}
we may write the following incremental system to be solved for $\Delta\gamma$, $\Bep$, and $\Bsp$:
\begin{equation}
\begin{dcases}\Bep=\Bep_n + \Delta\gamma\,\hat{\bm{n}}, \quad \hat{\bm{n}}\in\partial g^\mathrm{p}(\Bsp),\\
\Bsp=\bm{s}^\mathrm{p\,trial}-\Delta\gamma\,\big[\bm{\mathsf{C}}+\bm{\mathsf{H}}^\mathrm{kin}(\alpha)\big]:\hat{\bm{n}},\\
f(\Bsp)\leq 0, \quad  \Delta\gamma\geq 0, \quad \Delta\gamma \, f(\Bsp) = 0. \end{dcases}
\label{psys}
\end{equation}
The admissibility of the trial state is verified by evaluating the yield function~\eqref{fp} as $f( \bm{s}^\mathrm{p\,trial})$, such that:
\begin{enumerate}
\item If $ f(\bm{s}^\mathrm{p\,trial}) \leq 0$, the trial state is admissible and the solution of~\eqref{psys} yields
\begin{equation}
\begin{dcases}
\Delta\gamma = 0,\\ \Bep=\Bep_n, \\
\Bsp=\bm{s}^\mathrm{p\,trial} \end{dcases} \quad \implies \quad \begin{dcases}\Bsig=\Bsig^\mathrm{trial}, \\ \bm{\mathsf{C}}^{\mathrm{ep}}=\frac{\partial\Bsig}{\partial\Be}=\bm{\mathsf{C}}.\end{dcases}
\end{equation}
\item If $f(\bm{s}^\mathrm{p\,trial})>0$, the trial state lies outside the set of admissible stresses~\eqref{sc_setK}. Then, a \emph{corrector step} is performed, where the system~\eqref{psys} takes the form
\begin{equation}
\begin{dcases}\Bep=\Bep_n + \Delta\gamma\,\hat{\bm{n}}, \quad \hat{\bm{n}}\in\partial g^\mathrm{p}(\Bsp), \\
\Bsp=\bm{s}^\mathrm{p\,trial}-\Delta\gamma\,\big[\bm{\mathsf{C}}+\bm{\mathsf{H}}^\mathrm{kin}(\alpha)\big]:\hat{\bm{n}},\\
f(\Bsp)= 0. \end{dcases}
\label{psys_2}
\end{equation}
At this point, it is left to determine if the generalized stress state lies on the smooth part of the Drucker-Prager cone or at the apex, for which we adopt a scheme similar to that proposed in~\citet{sysala2016}. We begin by  expanding $\Bsp$ in~\eqref{psys} according to the spherical-deviatoric decomposition 
\begin{equation}
\Bsp_\mathrm{dev}\equiv\bm{s}^\mathrm{p\,trial}_\mathrm{dev}-\Delta\gamma\,\big[2\mu+H^\mathrm{kin}_\mu(\alpha)\big]\hat{\bm{n}}_\mathrm{dev} \quad \text{and} \quad \tr\Bsp\equiv \tr\bm{s}^\mathrm{p\,trial} -3\sqrt{6}\Delta\gamma\,A_\theta\big[K+H^\mathrm{kin}_K(\alpha)\big],
\label{sp_sptr}
\end{equation}
where we have used equations~\eqref{Celas}, \eqref{Hkin2}, and~\eqref{fr_p}. It readily follows that 
\begin{equation}
\bm{s}^\mathrm{p\,trial}_\mathrm{dev} =\begin{dcases} \bigg(1 + \dfrac{\Delta\gamma\big[2\mu+H^\mathrm{kin}_\mu (\alpha)\big]}{\Vert\Bsp_\mathrm{dev} \Vert} \bigg) \Bsp_\mathrm{dev}  & \text{if} \quad \Vert \Bsp_\mathrm{dev} \Vert > 0, \\
\Delta\gamma\,\big[2\mu+H^\mathrm{kin}_\mu(\alpha)\big]\hat{\bm{n}}_\mathrm{dev}, \quad \Vert\hat{\bm{n}}_\mathrm{dev}\Vert \leq 1 & \text{if} \quad \Vert \Bsp_\mathrm{dev} \Vert = 0.
\end{dcases}
\label{sptr}
\end{equation}
From~\eqref{sptr}$_1$, we note that $\Bsp_\mathrm{dev}$ and $\bm{s}^\mathrm{p\,trial}_\mathrm{dev}$ are collinear if $\Vert\Bsp_\mathrm{dev} \Vert>0$, such that
$$
\hat{\bm{n}}_\mathrm{dev}=\dfrac{\Bsp_\mathrm{dev}}{\Vert\Bsp_\mathrm{dev}\Vert}=\dfrac{\bm{s}^\mathrm{p\,trial}_\mathrm{dev}}{\Vert \bm{s}^\mathrm{p\,trial}_\mathrm{dev}\Vert}\eqqcolon\hat{\bm{n}}_\mathrm{dev}^\mathrm{trial} \quad \text{if} \quad \Vert \Bsp_\mathrm{dev} \Vert > 0.
$$ 
Moreover, one may write the condition
\begin{equation}
\Vert \Bsp_\mathrm{dev} \Vert = \Big\langle \Vert\bm{s}^\mathrm{p\,trial}_\mathrm{dev} \Vert - \Delta\gamma\big[2\mu+H^\mathrm{kin}_\mu (\alpha)\big] \Big\rangle_+\,,
\label{sp_ramp}
\end{equation}
where $\langle\Box\rangle_+\coloneqq(\Box+\vert\Box\vert)/2$. Then, equations~\eqref{sp_sptr} and~\eqref{sp_ramp} allow us to express the yield function in terms of the trial stress and the plastic  multiplier only, for which we define the auxiliary function
\begin{equation}
Q(\Delta\gamma)\coloneqq\Big< \Vert\bm{s}^\mathrm{p\,trial}_\mathrm{dev}\Vert - \Delta\gamma\,\big[2\mu+H^\mathrm{kin}_\mu (\alpha)\big] \Big>_++\sqrt{\frac{2}{3}}A_\varphi\Big( \tr\bm{s}^\mathrm{p\,trial} -3\sqrt{6}\,\Delta\gamma\,A_\theta\big[K+H^\mathrm{kin}_K(\alpha) \big] \Big)
\end{equation}
and  proceed as follows:
\begin{enumerate}
\item If $Q\big({\Vert\bm{s}^\mathrm{p\,trial}_\mathrm{dev} \Vert}/{\big[2\mu+H^\mathrm{kin}_\mu (\alpha)\big]}\big)<0$, $\Bsp$ lies on the smooth part of the cone, where $\Vert\Bsp_\mathrm{dev} \Vert>0$. The solution of~\eqref{psys_2} then reads 
\begin{equation}
\begin{dcases}
\Delta\gamma = \dfrac{ \Vert\bm{s}^\mathrm{p\,trial}_\mathrm{dev} \Vert+\sqrt{2/3}\,A_\varphi\tr\bm{s}^\mathrm{p\,trial}}{2\mu+H^\mathrm{kin}_\mu (\alpha)+6A_\varphi A_\theta\big[ K+H^\mathrm{kin}_K(\alpha) \big]}\,,\\ 
\Bep=\Bep_n + \Delta\gamma\,(\hat{\bm{n}}_\mathrm{dev}^\mathrm{trial}+\sqrt{2/3}\,A_\theta \bm{1} ),\\
\Bsp=\bm{s}^\mathrm{p\,trial}- \Delta\gamma\,\big[2\mu+H^\mathrm{kin}_\mu (\alpha)\big]\hat{\bm{n}}_\mathrm{dev}^\mathrm{trial}  - \sqrt{6}\,\Delta\gamma\,A_\theta\big[K+H^\mathrm{kin}_K(\alpha) \big] \bm{1}. \end{dcases}
\end{equation}
The stress tensor is updated as
\begin{equation}
\Bsig=\Bsig^\mathrm{trial}-\Delta\gamma\,\bm{\mathsf{C}}:(\hat{\bm{n}}_\mathrm{dev}^\mathrm{trial}+\sqrt{2/3}\,A_\theta \bm{1}) =\Bsig^\mathrm{trial}-\Delta\gamma\,\big(2\mu\,\hat{\bm{n}}_\mathrm{dev}^\mathrm{trial}+\sqrt{6}\,K A_\theta\bm{1}\big),
\end{equation} 
while the consistent tangent reads
\begin{equation}
\begin{aligned}
\bm{\mathsf{C}}^\mathrm{ep}=\frac{\partial\Bsig}{\partial\Be}&=\bm{\mathsf{C}}-\big(2\mu\,\hat{\bm{n}}_\mathrm{dev}^\mathrm{trial}+\sqrt{6}\,K A_\theta\bm{1}\big)\otimes\frac{\partial(\Delta\gamma)}{\partial\Be}-2\mu\,\Delta\gamma\frac{\partial\hat{\bm{n}}_\mathrm{dev}^\mathrm{trial}}{\partial\Be} \\
&=\bm{\mathsf{C}}-\frac{\big(2\mu\,\hat{\bm{n}}_\mathrm{dev}^\mathrm{trial}+\sqrt{6}\,K A_\theta\bm{1}\big) \otimes \big(2\mu\,\hat{\bm{n}}_\mathrm{dev}^\mathrm{trial}+\sqrt{6}\,K A_\varphi\bm{1}\big)}{2\mu+H^\mathrm{kin}_\mu (\alpha)+6A_\varphi A_\theta\big[ K+H^\mathrm{kin}_K(\alpha) \big]}\\ & \hspace*{5.5cm} -\frac{4\mu^2\Delta\gamma}{\Vert\bm{s}^\mathrm{p\,trial}_\mathrm{dev} \Vert}\bigg(\bm{\mathsf{I}}-\frac{1}{3}\bm{1}\otimes\bm{1}-\hat{\bm{n}}_\mathrm{dev}^\mathrm{trial}\otimes\hat{\bm{n}}_\mathrm{dev}^\mathrm{trial}\bigg).
\end{aligned}
\end{equation}
Note that this tensor is non-symmetric in view of $A_\theta\neq A_\varphi$. 
\item If $Q\big({\Vert\bm{s}^\mathrm{p\,trial}_\mathrm{dev} \Vert}/{\big[2\mu+H^\mathrm{kin}_\mu (\alpha)\big]}\big)\geq0$, the stress state lies at the apex, where $\Bsp=\bm{0}$. As such, the plastic strain tensor may be computed via equation~\eqref{ep_reqs}. It is important to recall that in this case, no frictional sliding occurs and the generalized stress state corresponds to the tensile regime. Thus, in line with remark~\ref{rem2}, $\Bep$ is readily condensed out, while the stress tensor takes the form
\begin{equation}
\Bsig=\bm{\mathsf{C}}^\mathrm{dam}(\alpha):\bm{\varepsilon},
\end{equation}
with the corresponding consistent tangent
\begin{equation}
\bm{\mathsf{C}}^\mathrm{ep}=\frac{\partial\Bsig}{\partial\Be}=\bm{\mathsf{C}}^\mathrm{dam}(\alpha).
\end{equation}

\end{enumerate}
\end{enumerate}

The procedure described above is employed at each iteration of the elastoplastic problem to provide the stress tensor $\Bsig$ and the corresponding tangent $\bm{\mathsf{C}}^{\mathrm{ep}}$ for the solution of equation~\eqref{weak_u_alg}.

\subsection{Damage evolution problem}\label{alg_dam}

According to algorithm~\ref{alg1}, having determined $\{\Bu,\Bep\}$ from the steps described above, we are now in position to solve the damage evolution problem  to find the current estimate of $\alpha$. In view of the present pseudo-time discretization, the irreversibility condition in incremental form reads
\begin{equation}
\Delta\alpha\geq 0.
\label{irr_inc}
\end{equation}
Then, the task is to solve the following set-valued PDE in weak form:
\begin{equation}
\int_\Omega \bigg( -s^\mathrm{d}(\bm{\varepsilon},\bm{\varepsilon}^\mathrm{p},\alpha)\,\tilde{\alpha}  + \frac{{G}_\mathrm{c}(\bm{s}^\mathrm{p}_n)}{{\ell}}\big(\alpha\,\tilde{\alpha}+{\ell}^{2}\nabla\alpha\cdot\nabla\tilde{\alpha}\big) + \partial I_{\mathbb{R}_+}(\Delta \alpha)\,\tilde{\alpha} \bigg) \,\mathrm{d}\bm{x}   \ni0 \quad \forall\,\tilde{\alpha}\in\mathrm{H}^1(\Omega;\mathbb{R}),
\label{weak_d_inc}
\end{equation}
where the indicator function is employed to enforce the irreversibility condition~\eqref{irr_inc}. Recalling that $\partial I_{\mathbb{R}_+}(\Box)$ gives $0$ for $\Box>0$, $\mathbb{R}_-$ for $\Box=0$, and $\varnothing$ for $\Box<0$, it is easy to see that this expression entails a KKT system consistent with the damage evolution problem. In particular, equation~\eqref{weak_d_inc} corresponds to the weak form of the incremental approximation of the differential inclusion~\eqref{biot_model_d}, which in turn is equivalent to the KKT conditions of the damage evolution problem, derived from energy balance and local stability in section~\ref{formulation} (table~\ref{overview2}). Specifically, in the incremental approximation of~\eqref{biot_model_d}, the damage dissipation potential~\eqref{phi_d} is expressed in terms of $\Delta\alpha$, while its state-dependence through $\Bsp$ is evaluated at the previous time step, i.e., in terms of $\Bsp_n$. This treatment is consistent with the incremental approximation considered for state-dependent dissipation potentials in previous works~\cite{miehe2011,luege2018,ulloa2021b}.

The difficulty in solving~\eqref{weak_d_inc} is related to (i) the strong non-linearity imposed by the dependence of the driving force~\eqref{sd} on $\Bsp(\bm{\varepsilon},\bm{\varepsilon}^\mathrm{p},\alpha)$ through the opening/closure transition~\eqref{optrans}, (ii) the non-smoothness imposed by the irreversibility condition though $\partial I_{\mathbb{R}_+}(\Delta \alpha)$, and (iii) the non-linear terms in the driving force~\eqref{sd}, which result from the degradation functions~\eqref{gmu}, \eqref{gk}, and~\eqref{Hkin_K_mu}.

Concerning the first point, in agreement with the algorithmic decoupling of algorithm~\ref{alg1}, we assume that the opening/closure transition~\eqref{optrans} is known from the solution of the elastoplastic problem (section~\ref{alg_ep}). Thus, we evaluate the damage driving force based on $\Bsp\big(\bm{\varepsilon}^{(j)},\bm{\varepsilon}^{\mathrm{p}(j)},\alpha^{(j-1)}\big)$, with $j$ referring to the staggered iteration counter. Then, dropping $j$ at the current iteration for notational simplicity, and using equations~\eqref{Cdam} and~\eqref{Hkin2} along with degradation functions~\eqref{gmu}, \eqref{gk}, and~\eqref{Hkin_K_mu}, the driving force~\eqref{sd} takes the form
\begin{equation}
s^\mathrm{d}(\bm{\varepsilon},\bm{\varepsilon}^\mathrm{p},\alpha)=\begin{dcases}
-\frac{1}{2}g_K'(\alpha)K\,(\tr\Be)^2-g_\mu'(\alpha)\mu\,\Be_\mathrm{dev}:\Be_\mathrm{dev}& \quad \text{if } \tr{\Bsp}\big(\bm{\varepsilon},\bm{\varepsilon}^\mathrm{p},\alpha^{(j-1)}\big)=0,  \\
-\frac{g_K'(\alpha)K}{2[1-g_K(\alpha)]^2}\,(\tr\Bep)^2 - \frac{g_\mu'(\alpha)\mu}{[1-g_\mu(\alpha)]^2}\Bep_\mathrm{dev}:\Bep_\mathrm{dev} & \quad \text{if } \tr{\Bsp}\big(\bm{\varepsilon},\bm{\varepsilon}^\mathrm{p},\alpha^{(j-1)}\big)<0. 
\end{dcases}
\label{sd_alg1}
\end{equation}
In order to enforce the irreversibility condition and render equation~\eqref{weak_d_inc} an equality, we consider a generalization of the \emph{history field} method initially proposed by~\citet{MieHofWel2010}. In particular, the method is adapted to the present model by employing the maximum time history values of the individual terms in the crack driving force. To this end, we define the history fields
\begin{equation}
\begin{aligned}
&\mathcal{H}_{K\mathrm{I}}(\bm{x},t)\coloneqq \max_{s\in[0,t]}\frac{1}{2}K\big(\tr\Be(\bm{x},s)\big)^2, \quad &&\mathcal{H}_{\mu\mathrm{I}}(\bm{x},t)\coloneqq\max_{s\in[0,t]}\mu\,\Be_\mathrm{dev}(\bm{x},s):\Be_\mathrm{dev}(\bm{x},s), \\
&\mathcal{H}_{K\mathrm{II}}(\bm{x},t)\coloneqq\max_{s\in[0,t]}\frac{1}{2}K\big(\tr\Bep(\bm{x},t)\big)^2,  \quad &&\mathcal{H}_{\mu\mathrm{II}}(\bm{x},s)\coloneqq \max_{s\in[0,t]}\mu\,\Bep_\mathrm{dev}(\bm{x},s):\Bep_\mathrm{dev}(\bm{x},s),
\end{aligned}
\end{equation}
and set
\begin{equation}
s^\mathrm{d\,hist}(\bm{\varepsilon},\bm{\varepsilon}^\mathrm{p},\alpha)\coloneqq\begin{dcases}
-g_K'(\alpha)\mathcal{H}_{K\mathrm{I}}-g_\mu'(\alpha)\mathcal{H}_{\mu\mathrm{I}}& \quad \text{if } \tr{\Bsp}\big(\bm{\varepsilon},\bm{\varepsilon}^\mathrm{p},\alpha^{(j-1)}\big)=0,  \\
-\frac{g_K'(\alpha)}{[1-g_K(\alpha)]^2}\,\mathcal{H}_{K\mathrm{II}} - \frac{g_\mu'(\alpha)}{[1-g_\mu(\alpha)]^2}\mathcal{H}_{\mu\mathrm{II}} & \quad \text{if } \tr{\Bsp}\big(\bm{\varepsilon},\bm{\varepsilon}^\mathrm{p},\alpha^{(j-1)}\big)<0. 
\end{dcases}
\label{sd_alg2}
\end{equation}
Then, equation~\eqref{weak_d_inc} is replaced by the single-valued expression
\begin{equation}
\int_\Omega \bigg( -s^\mathrm{d\,hist}(\bm{\varepsilon},\bm{\varepsilon}^\mathrm{p},\alpha)\,\tilde{\alpha}  + \frac{{G}_\mathrm{c}(\bm{s}^\mathrm{p}_n)}{{\ell}}\big(\alpha\,\tilde{\alpha}+{\ell}^{2}\nabla\alpha\cdot\nabla\tilde{\alpha}\big) \bigg) \,\mathrm{d}\bm{x}  = 0 \quad \forall\,\tilde{\alpha}\in\mathrm{H}^1(\Omega;\mathbb{R}).
\label{weak_d_inc2}
\end{equation}
Finally, to handle the non-linear degradation functions in the crack driving force~\eqref{sd_alg2}, a standard Newton-Raphson scheme is employed, where the linearization of equation~\eqref{weak_d_inc2} reads
\begin{equation}
\begin{aligned}
\int_\Omega \bigg[\bigg(-\frac{\partial s^\mathrm{d\,hist}(\bm{\varepsilon},\bm{\varepsilon}^\mathrm{p},\alpha^{(k)})}{\partial\alpha^{(k)}} + \frac{{G}_\mathrm{c}(\bm{s}^\mathrm{p}_n)}{{\ell}}\bigg)\big({\alpha}^{(k+1)}-{\alpha}^{(k)}\big)\tilde{\alpha}+ {\ell}\,{G}_\mathrm{c}(\bm{s}^\mathrm{p}_n)\nabla\big[{\alpha}^{(k+1)}-{\alpha}^{(k)}\big]\cdot\nabla\tilde{\alpha} \bigg] \,\mathrm{d}\bm{x}\\ =\int_\Omega \bigg( s^\mathrm{d\,hist}(\bm{\varepsilon},\bm{\varepsilon}^\mathrm{p},\alpha^{(k)})\,\tilde{\alpha}  - \frac{{G}_\mathrm{c}(\bm{s}^\mathrm{p}_n)}{{\ell}}\big(\alpha^{(k)}\,\tilde{\alpha}+{\ell}^{2}\nabla\alpha^{(k)}\cdot\nabla\tilde{\alpha}\big) \bigg) \,\mathrm{d}\bm{x}  \eqqcolon \mathtt{RES}_{\alpha}^{(k)} \quad \forall\,\tilde{\alpha}\in\mathrm{H}^1(\Omega;\mathbb{R}).
\end{aligned}
\label{weak_d_inc2_lin}
\end{equation}
This equation is solved sequentially for the iterate $\alpha^{(k+1)}$ until $\big\vert\mathtt{RES}_{\alpha}^{(k)}\big\vert\leq \mathtt{TOL}_{\alpha}$.

At this point, it is important to mention that the implementation of the model can be greatly simplified if $\Bep$ is also updated at the  apex of the Drucker-Prager cone, i.e., in the tensile regime (case 2, point (b) of the return-mapping scheme in section~\ref{alg_ep}; see also remark~\ref{rem2}). In that case, the bottom expressions in equations~\eqref{sd_alg1} and~\eqref{sd_alg2} are valid for $\tr\Bsp\big(\bm{\varepsilon},\bm{\varepsilon}{^\mathrm{p}},\alpha^{(j-1)}\big)\leq 0$ and, consequently, the driving force~\eqref{sd_alg2} can be written in terms of $\mathcal{H}_{K\mathrm{II}}$ and $\mathcal{H}_{\mu\mathrm{II}}$ only. Indeed, this alternative treatment is equivalent to the use of $\Bsp\big(\bm{\varepsilon},\bm{\varepsilon}^{\mathrm{p}},\alpha^{(j-1)}\big)$ in equations~\eqref{sd_alg1} and~\eqref{sd_alg2} to distinguish the tensile from the compressive/shear regime.

\subsection{Viscous regularization}

According to preliminary results, certain instances of brutal crack propagation may result in convergence issues in the solution of the elastoplastic problem described in section~\ref{alg_ep}. To remedy this, we consider a numerical viscous regularization in the damage evolution problem, where equation~\eqref{weak_d_inc2} is augmented as
\begin{equation}
\int_\Omega \bigg( -s^\mathrm{d\,hist}(\bm{\varepsilon},\bm{\varepsilon}^\mathrm{p},\alpha)\,\tilde{\alpha}  + \frac{{G}_\mathrm{c}(\bm{s}^\mathrm{p}_n)}{{\ell}}\big(\alpha\,\tilde{\alpha}+{\ell}^{2}\nabla\alpha\cdot\nabla\tilde{\alpha}\big) + \frac{\eta_\mathrm{vd}}{\Delta t}(\alpha-\alpha_n)\,\tilde{\alpha} \bigg) \,\mathrm{d}\bm{x}  = 0 \quad \forall\,\tilde{\alpha}\in\mathrm{H}^1(\Omega;\mathbb{R}).
\label{weak_d_inc3}
\end{equation}
Here, $\eta_\mathrm{vd}$ is a small viscosity parameter. Following~\citet{MieHofWel2010}, we do not view the viscous term embedded in~\eqref{weak_d_inc3} as a physical mechanism, but rather as a numerical technique intended to stabilize the solution procedure. In view of this modification, the linearized form~\eqref{weak_d_inc2_lin} is replaced by
\begin{equation}
\begin{aligned}
\int_\Omega \bigg[\bigg(-\frac{\partial s^\mathrm{d\,hist}(\bm{\varepsilon},\bm{\varepsilon}^\mathrm{p},\alpha^{(k)})}{\partial\alpha^{(k)}} + \frac{{G}_\mathrm{c}(\bm{s}^\mathrm{p}_n)}{{\ell}}+\frac{\eta_\mathrm{vd}}{\Delta t}\bigg)\big({\alpha}^{(k+1)}-{\alpha}^{(k)}\big)\tilde{\alpha}+ {\ell}\,{G}_\mathrm{c}(\bm{s}^\mathrm{p}_n)\nabla\big[{\alpha}^{(k+1)}-{\alpha}^{(k)}\big]\cdot\nabla\tilde{\alpha} \bigg] \,\mathrm{d}\bm{x}\\ =\int_\Omega \bigg( s^\mathrm{d\,hist}(\bm{\varepsilon},\bm{\varepsilon}^\mathrm{p},\alpha^{(k)})\,\tilde{\alpha}  - \frac{{G}_\mathrm{c}(\bm{s}^\mathrm{p}_n)}{{\ell}}\big(\alpha^{(k)}\,\tilde{\alpha}+{\ell}^{2}\nabla\alpha^{(k)}\cdot\nabla\tilde{\alpha}\big)-\frac{\eta_\mathrm{vd}}{\Delta t}(\alpha^{(k)}-\alpha_n)\,\tilde{\alpha} \bigg) \,\mathrm{d}\bm{x}  \eqqcolon \mathtt{RES}_{\alpha}^{(k)} \\ \forall\,\tilde{\alpha}\in\mathrm{H}^1(\Omega;\mathbb{R}).
\end{aligned}
\label{weak_d_inc3_lin}
\end{equation}
In order to avoid significant deviations from the original problem, $\eta_\mathrm{vd}$ must be chosen as small as possible. In the simulations presented in section~\ref{numsim}, $\Delta t$ is defined as the loading increment, while $\eta_\mathrm{vd}=1\times10^{-7}$ MPa$\,\cdot\,$s. If required for convergence in certain time steps, $\eta_\mathrm{vd}$ is increased up to a maximum $\eta_\mathrm{vd}=1\times10^{-5}$ MPa$\,\cdot\,$s. We note that another possibility to deal with numerical difficulties associated with brutal crack propagation is the use of a dissipation-based path-following constraint~\cite{wambacq2021b}, a topic worth considering in future research.

\subsection{Spatial discretization}

The linearized forms~\eqref{weak_u_alg} and~\eqref{weak_d_inc2_lin} (or the viscous version~\eqref{weak_d_inc3_lin}) are suitable for spatial discretization. To this end, we consider  standard finite elements, where the non-local primary fields $\Bu$ and $\alpha$ are interpolated using bilinear shape functions. On the other hand, the local primary field $\Bep$ is evaluated at Gauss integration points. This procedure is straightforward and is therefore not presented for the sake of~brevity.

It is worth mentioning that the chosen finite elements cannot describe the discontinuities embedded in the function spaces~\eqref{funscspace_u} and~\eqref{funscspace_p}. In this context, the use of discontinuous finite element techniques appears as a more suitable choice to be considered for future works. Nevertheless, it has been shown that for ductile phase-field models with perfect plasticity, a strong concentration of plastic strains is attained upon damage localization, representing a regularized version of the discontinuous response, with the element size playing the role of a convergence parameter~\citep{alessi2015}. As will be shown in the following section, the same behavior is observed for the present model. Note that this remark is only relevant for the compressive/shear regime, which approaches a purely frictional sliding stage with perfect plasticity as $\alpha\to1$.

\FloatBarrier

%\clearpage

\section{{Numerical simulations}}\label{numsim}

This section presents numerical simulations that highlight the main features of the model described in section~\ref{formulation}. Specifically, 2D finite element simulations are performed, aiming to capture different failure modes including tensile, shear, and mixed-mode fracture. Depending on the example, the results are compared with numerical and experimental observations, as well as analytical results from fracture mechanics. 

In all examples, plane-strain conditions are assumed. Moreover, a low-level initial damage $\alpha_0=1\times10^{-5}$ is uniformly distributed in the domain to allow for plastic-damage evolution in the compressive/shear regime.

\subsection{Biaxial compression tests}

The first example highlights the ability of the  model to describe shear banding and shear fracture. To this end, we consider plane-strain specimens subjected to biaxial compression. These tests are of particular interest for the analysis of failure in geomaterials~\cite{ord1991, labuz1996, fakhimi2002}; plane-strain conditions are encountered often in geotechnical engineering problems, such as the analysis of underground excavations in rock~\cite{labuz1996}. 

\begin{figure}[!b]
\centering
    \includeinkscape[scale=0.95]{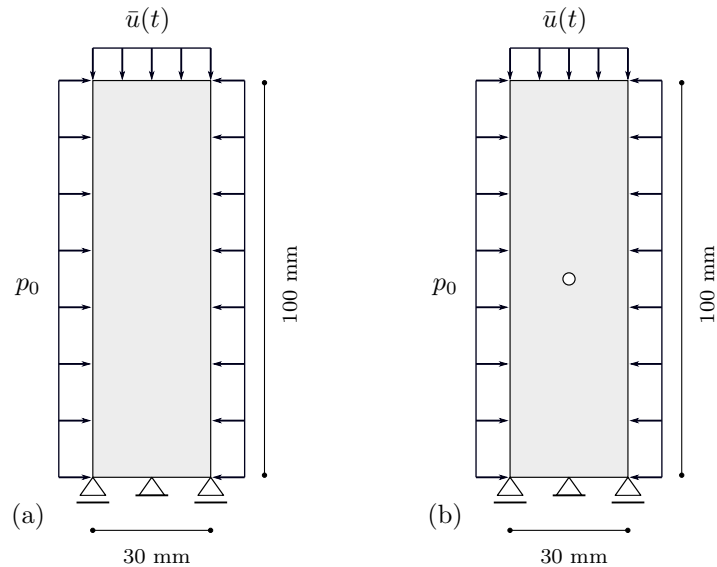}\vspace*{-0.5cm}
\caption{Schematic representation of the biaxial compression test: (a) initially homogeneous  specimen and (b)  specimen with a central hole of $3.4$ mm diameter. The confining pressure is fixed at $p_0=5$ MPa during the displacement loading stage.}
\label{bicomp_scheme}
\end{figure}

For the present study, we consider the two scenarios depicted in figure~\ref{bicomp_scheme}: an initially homogeneous specimen and a specimen with a central hole, representative of experiments on sandstone reported in~\citet{labuz1996}. In the simulations, uniform meshes of $\sim$43000 quadrilateral elements with a characteristic element size $h_\mathrm{c}=0.3$~mm are employed. The test is divided into two loading stages. In the first stage, confining pressure is gradually applied until $p_0=5$ MPa, keeping the vertical displacements at the bottom edge fixed, where only the center node is also fixed horizontally. For the second loading stage, the lateral pressure $p_0$ is  fixed while vertical displacements $\bar{u}$ are imposed downwards in increments of 0.01~mm. 

\citet{labuz1996,labuz2006} report the following experimental material parameters for the sandstone specimens: Young's modulus $E=14$ GPa; Poisson's ratio $\nu=0.31$; friction angle $\varphi=44^\circ$; dilation angle $\theta=30^\circ$, showing non-associative behavior; mode I critical stress intensity factor $K_\mathrm{cI}=0.4$ MPa$\,\cdot\,$m$^{1/2}$; and mode II critical stress intensity factor $K_\mathrm{cII}=3$ MPa$\,\cdot\,$m$^{1/2}$. The same parameters are chosen for the simulations, with the friction and dilation coefficients computed from $\varphi$ and $\theta$ according to the shear approximation of the Mohr-Coulomb failure criterion:
\begin{equation}
A_\varphi=\frac{\sqrt{3}\sin\varphi}{3}, \quad A_\theta=\frac{\sqrt{3}\sin\theta}{3}.
\end{equation}
Moreover, for in-plane self-similar crack growth, $G_\mathrm{Ic}$ and $G_\mathrm{cII}$ are computed from $K_\mathrm{cI}$ and~$K_\mathrm{cII}$:
\begin{equation}
G_\mathrm{cI}=\frac{K_\mathrm{cI}^2}{E'}, \quad G_\mathrm{cII}=\frac{K_\mathrm{cII}^2}{E'},
\label{GcIKcI}
\end{equation}
with $E'=E/(1-\nu^2)$ for plane strain. Finally, a degradation parameter $b=25$ and a length scale $\ell=0.97$ mm were chosen. As such, $\ell/h_\mathrm{c}\approx 3.23$,  which showed mesh-converging results in preliminary simulations not reproduced here for the sake of brevity. 

\begin{figure}
  \centering
   \includegraphics[scale=1]{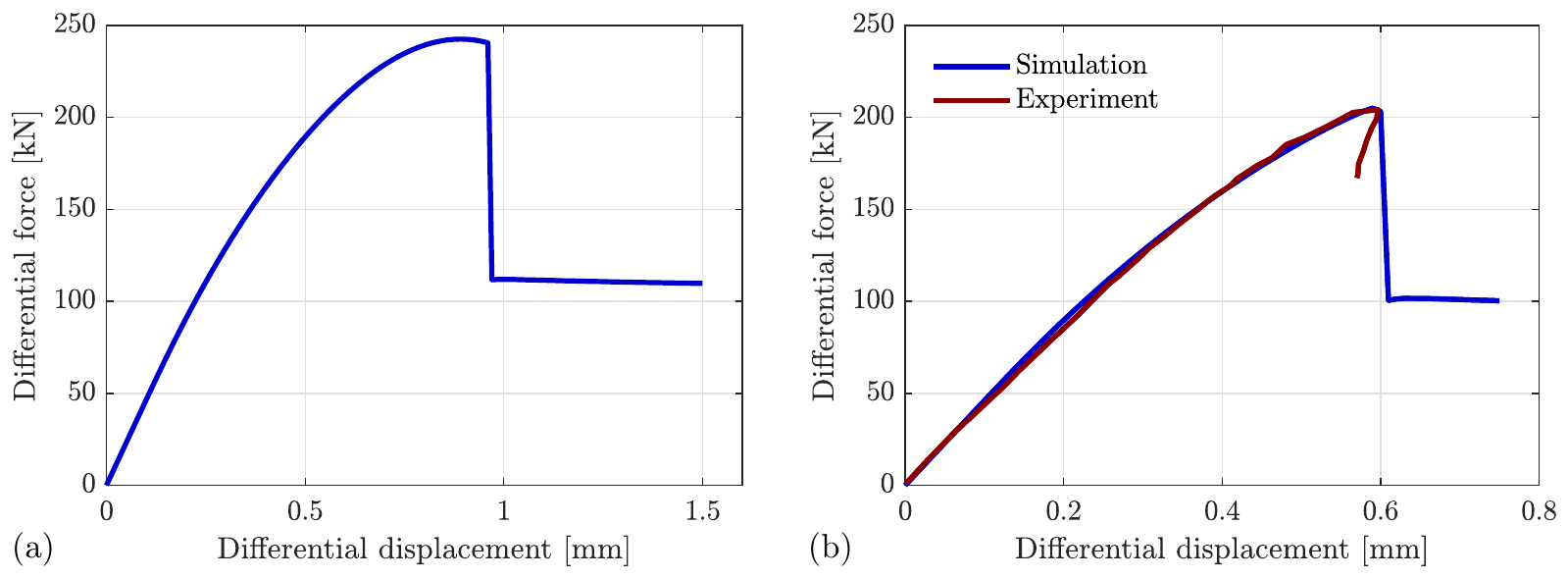}
\caption{Force-displacement curves for the biaxial compression tests: (a) initially homogeneous specimen  and  (b) perforated specimen including a comparison with the experimental curve reported in~\citet{labuz1996}. Here, the specimens are assumed to have an out-of-plane thickness of 100 mm.}
\label{bicomp_fd}
\end{figure}

\begin{figure}
  \centering
    \includegraphics[scale=1]{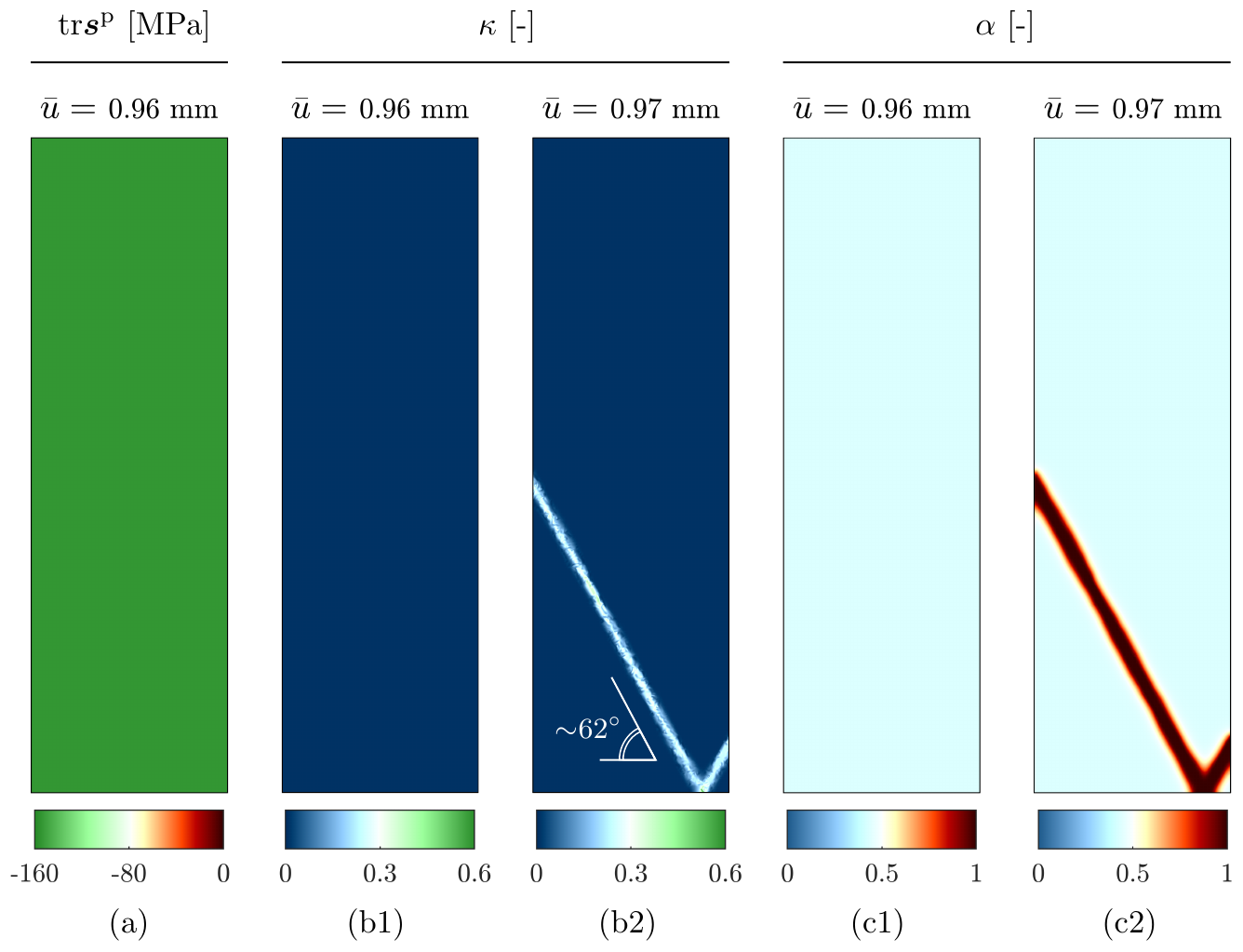}
\caption{Shear fracture process in the biaxial compression test for the initially homogeneous specimen: (a) pre-failure hydrostatic generalized stress $\tr\Bsp$ [MPa], (b) pre- and post-failure equivalent plastic strain, and (c) pre- and post-failure damage profile.}
\label{bicomp_nohole_state_contour}
\end{figure}

\begin{figure}
  \centering
    \includegraphics[scale=1]{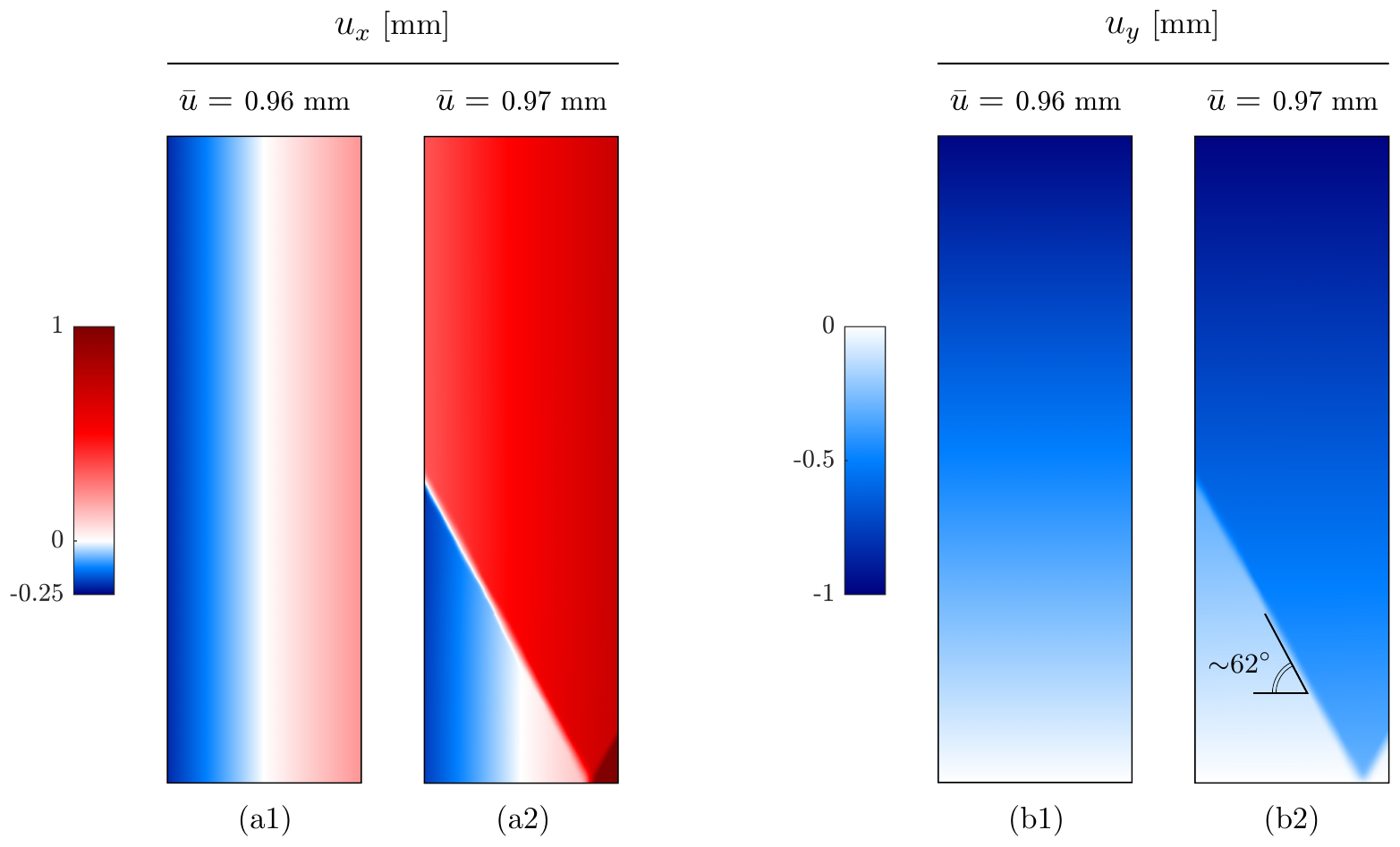}
\caption{Kinematics of the shear fracture process in the biaxial compression test for the initially homogeneous specimen, showing the pre- and post-failure values for (a) the horizontal displacement component and (b) the vertical displacement component.}
\label{bicomp_nohole_kin_contour}
\end{figure}

Figure~{\ref{bicomp_fd}(a)} shows the force-displacement curve for the initially homogeneous specimen, exhibiting failure for $\bar{u}>0.96$ mm. At $\bar{u}=0.96$ mm, figure~\ref{bicomp_nohole_state_contour} shows a homogeneous hydrostatic generalized stress $\tr\Bsp<0$, as well as homogeneous profiles for the equivalent plastic strain field $\kappa$ and the damage field $\alpha$. Further, figure~\ref{bicomp_nohole_kin_contour} shows smooth and symmetric displacements. At the next time step, the load abruptly drops, while shear fracture brutally takes place in a V-shaped pattern. Note from figure~\ref{bicomp_nohole_state_contour}(b2) that, as expected, the plastic strains concentrate in a very narrow band. This is also evident from the (mode II)  kinematics shown in figure~\ref{bicomp_nohole_kin_contour}, where the displacement field closely approximates a jump across the sliding failure planes. At this point, the material points with $\alpha\approx1$ approach a perfectly plastic stage that corresponds to a frictional sliding mechanism between the (approximately) separate specimen blocks. As such, a residual strength is observed in figure~{\ref{bicomp_fd}(a)}, modulated by the friction coefficient $A_\varphi$ and the confining~pressure.

\begin{figure}
  \centering
    \includegraphics[scale=1]{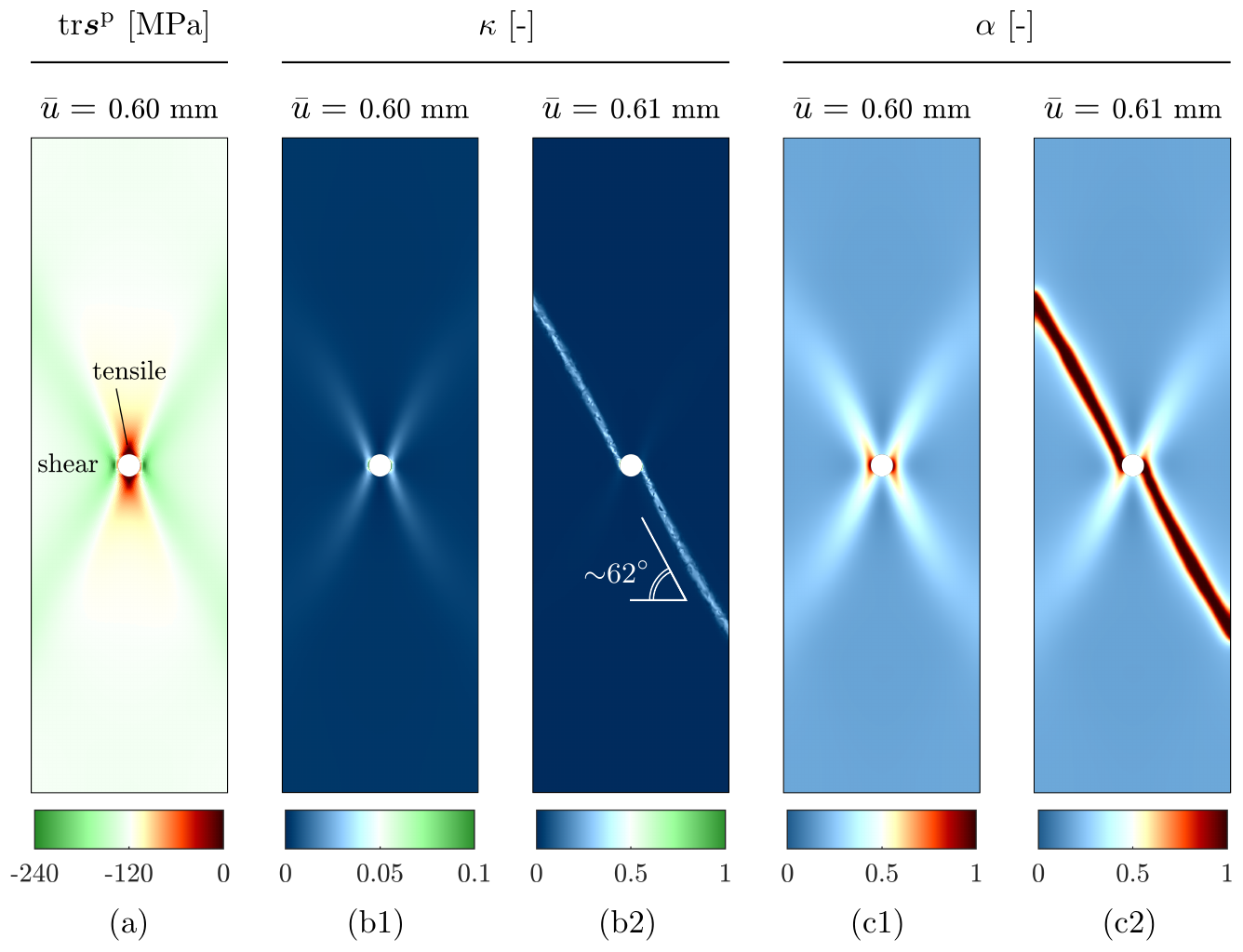}
\caption{Shear fracture process in the biaxial compression test for the perforated specimen: (a) pre-failure hydrostatic generalized stress $\tr\Bsp$ [MPa], (b) pre- and post-failure equivalent plastic strain, and (c) pre- and post-failure damage profile.}
\label{bicomp_hole_state_contour}
\end{figure}

\begin{figure}
  \centering
    \includegraphics[scale=1]{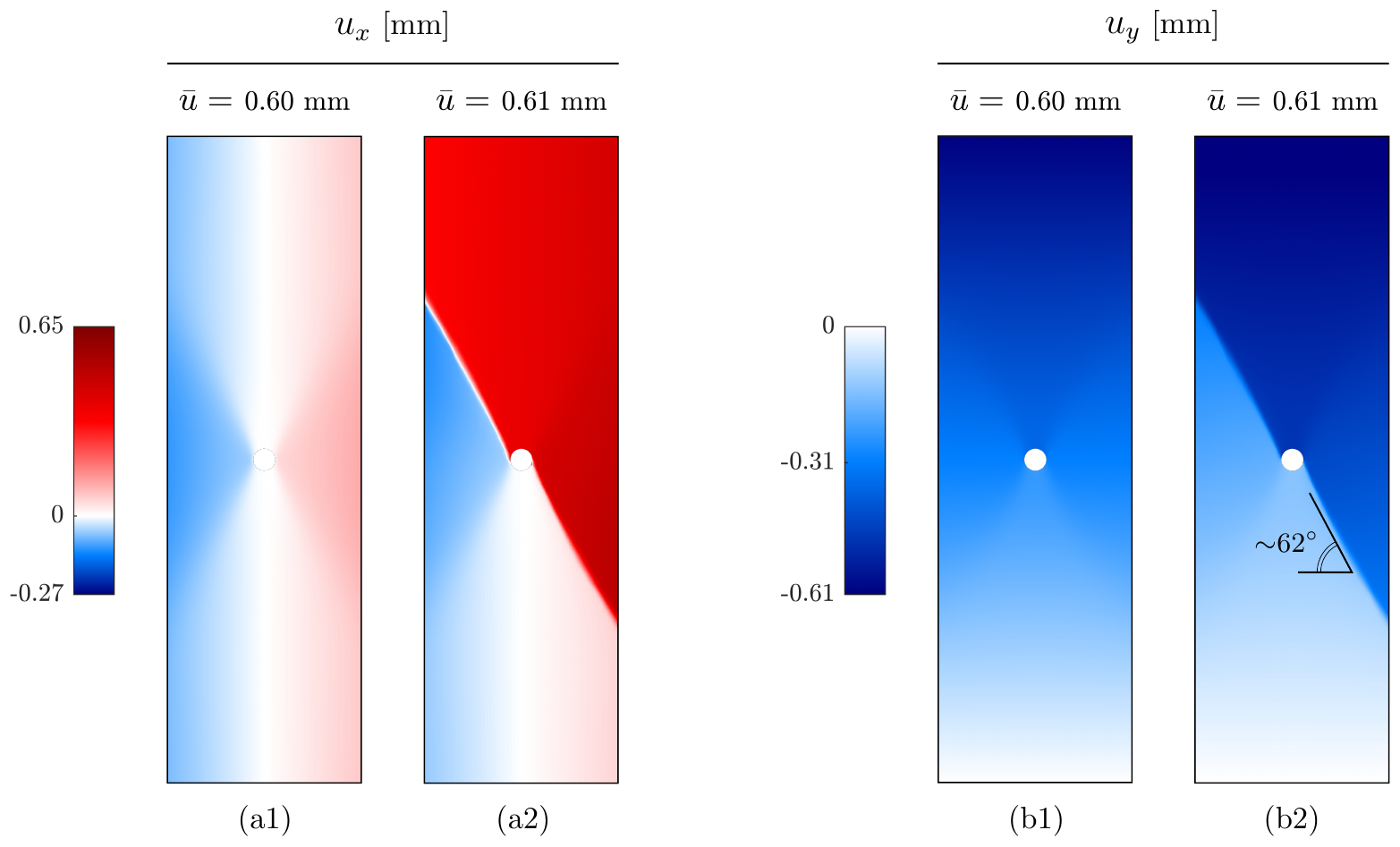}
\caption{Kinematics of the shear fracture process in the biaxial compression test for the perforated specimen, showing the pre- and post-failure values for (a) the horizontal displacement component and (b) the vertical displacement component.}
\label{bicomp_hole_kin_contour}
\end{figure}

For the perforated specimen, a lower peak load than that of the initially homogeneous specimen is observed in figure~{\ref{bicomp_fd}(b)}, with failure occurring at $\bar{u}>0.60$ mm. Therein, the force-displacement curve is compared with the experimental results reported in~\cite{labuz1996}, showing a very close agreement. \tC{We note that, having fixed all material parameters and the fracture length scale, the curve was tuned by the single degradation parameter $b=25$.} At $\bar{u}=0.60$ mm, figure~\ref{bicomp_hole_state_contour} shows a hydrostatic generalized stress $\tr\Bsp$ where the tensile region ($\tr\Bsp=0$, i.e., open microcracks) can be clearly distinguished from the compressive/shear region ($\tr\Bsp<0$, i.e., closed microcracks with frictional sliding). The corresponding equivalent plastic strain and damage profiles already hint shear banding, even prior to the peak load, with crack nucleation at the hole edges at $\bar{u}=0.60$ mm. Nevertheless, the response remains symmetric and weakly localized. The same observation can be made from the displacement profiles in figure~\ref{bicomp_hole_state_contour}. At the next time step, the load abruptly drops, while two shear fractures brutally propagate from the hole along the same failure plane. As in the previous case, figure~\ref{bicomp_hole_state_contour}(b2) shows a strong concentration of plastic strains in a very narrow band along the failure plane. Accordingly, the displacement profiles in figure~\ref{bicomp_hole_kin_contour} clearly show the kinematics of mode II failure.  

In both the initially homogeneous and the perforated specimens, the failure plane is found to have an inclination of $62^\circ$ with respect to the horizontal axis, in very close agreement with ${45^\circ +(\varphi+\theta)/4=63.5^\circ}$, as obtained using bifurcation analyses of Mohr-Coulomb--based frictional plasticity~\cite{vermeer1984} as well as in experiments on granular materials~\cite{arthur1977,vardoulakis1980}. This result further motivates the need for a non-associative law, highlighting the influence of the dilation angle in the inclination of the failure plane. Moreover, it is remarkable that the experimental results of~\citet{labuz1996,labuz2006} for these specimens also report an inclination of about $62^\circ$. Nevertheless, it is worth noting that these experiments showed a kinked fracture with a second, steeper plane of about $77^\circ$. As discussed in~\citet{labuz1996}, Mohr-Coulomb-based theories are unable to account for this steep portion, requiring an unrealistically high friction angle.

\FloatBarrier
%\clearpage

\subsection{Brazilian splitting tests}

Having shown the ability of the proposed model to describe shear fracture, we now turn our attention to the occurrence of tensile fracture under mixed-mode loading conditions. For this purpose, we consider centrally cracked Brazilian disk specimens subjected to diametral compression, representing benchmark experiments in rock-like materials~\cite{chang2002,ayatollahi2007,ayatollahi2008,aliha2010,haeri2014,xiankai2018,wang2020phase}. Particular emphasis is put on the crack propagation angle with respect to the initial flaw, as typically studied in problems of this type.

\begin{figure}[!b]
\centering
    \includeinkscape[scale=1]{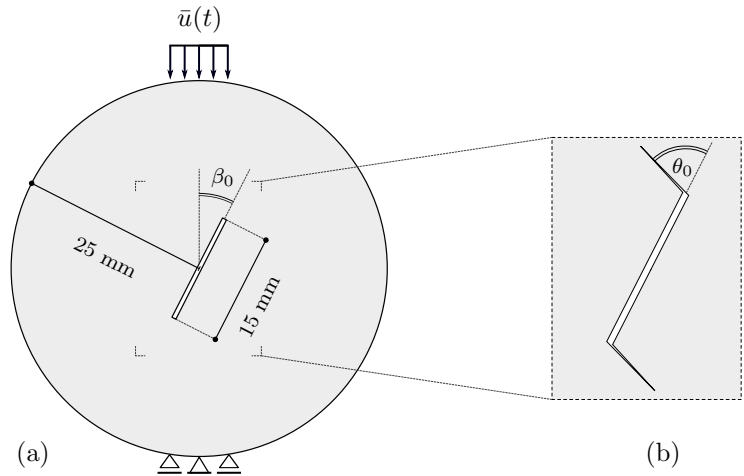}
\caption{Schematic representation of the Brazilian tests: (a) geometry and boundary conditions and (b) crack initiation angle. The initial flaw has a thickness of 0.75 mm, and displacements are imposed on a small region spanning a circular arc of~$8.5^\circ$.}
\label{brazil_scheme}
\end{figure}

Figure~\ref{brazil_scheme} shows the geometry and boundary conditions of the pre-cracked disk specimens. We consider ten different inclinations of the initial flaw, with $\beta_0\in\{0^\circ, 6.75^\circ, 13.5^\circ, 20.25^\circ, 27^\circ, 40^\circ, 52^\circ, 64^\circ, 76^\circ, 90^\circ\}$. In the simulations, a mesh of quadrilateral elements with refinement in a central region was employed, with a minimum characteristic element size $h_\mathrm{min}=0.1$ mm. This element size led to, e.g., $\sim$43000 elements for $\beta_0=0^\circ$ and $\sim$80000 elements for $\beta_0=27^\circ$. In order to avoid numerical difficulties, the displacements are imposed at the top and bottom caps of the boundary over a small region, chosen for simplicity as a circular arc of $8.5^\circ$. Vertical displacements $\bar{u}$ are imposed at the top with increments of $2\times10^{-4}$ mm, while the bottom cap is fixed vertically, with only the center node also fixed horizontally.

The material parameters for sandstone specimens are chosen as follows~\cite{wang2020phase}:  Young's modulus {$E=20$~GPa}, Poisson's ratio $\nu=0.25$, mode I fracture toughness $G_\mathrm{cI}=15.68$ N/m, and length scale $\ell=0.5$ mm. We further assume the degradation parameter $b=1$, the mode II fracture toughness $G_\mathrm{cII}=10\,G_\mathrm{cI}$, and the same friction and dilation parameters from the previous example, i.e., $\varphi=44^\circ$ and $\theta=30^\circ$.

The loading mode is characterized by the stress intensity factors ${K_\mathrm{I}}$ and ${K_\mathrm{II}}$ through the mixity parameter
\begin{equation}
M_\mathrm{e}=\frac{2}{\pi}\tan^{-1}\bigg(\frac{K_\mathrm{I}}{K_\mathrm{II}}\bigg)=\frac{2}{\pi}\tan^{-1}\bigg(\frac{Y_\mathrm{I}}{Y_\mathrm{II}}\bigg),
\label{mixity}
\end{equation}
where ${Y_\mathrm{I}}$ and ${Y_\mathrm{II}}$ are dimensionless geometry parameters depending on $\beta_0$ and on the ratio between the diameter of the specimen and the length of the initial flaw. For the present case, ${Y_\mathrm{I}}$ and ${Y_\mathrm{II}}$ were extracted from~\citet{ayatollahi2007} for all considered $\beta_0$ values, with ${Y_\mathrm{II}}=0$ for $\beta_0=0^\circ$ and ${Y_\mathrm{I}}=0$ for $\beta_0\approx 27^\circ$. Thus, pure mode I loading ($M_\mathrm{e}=1$) and pure mode II loading ($M_\mathrm{e}=0$) correspond to $\beta_0=0^\circ$ and $\beta_0=27^\circ$, respectively, while intermediate angles $0^\circ<\beta_0<27^\circ$ imply a transition from mode I to mode~II. Beyond $\beta_0=27^\circ$, mixed-mode loading is again achieved, recovering pure mode I loading at $\beta_0=90^\circ$. 

\begin{figure}[t]
  \centering
    \includegraphics[scale=0.675]{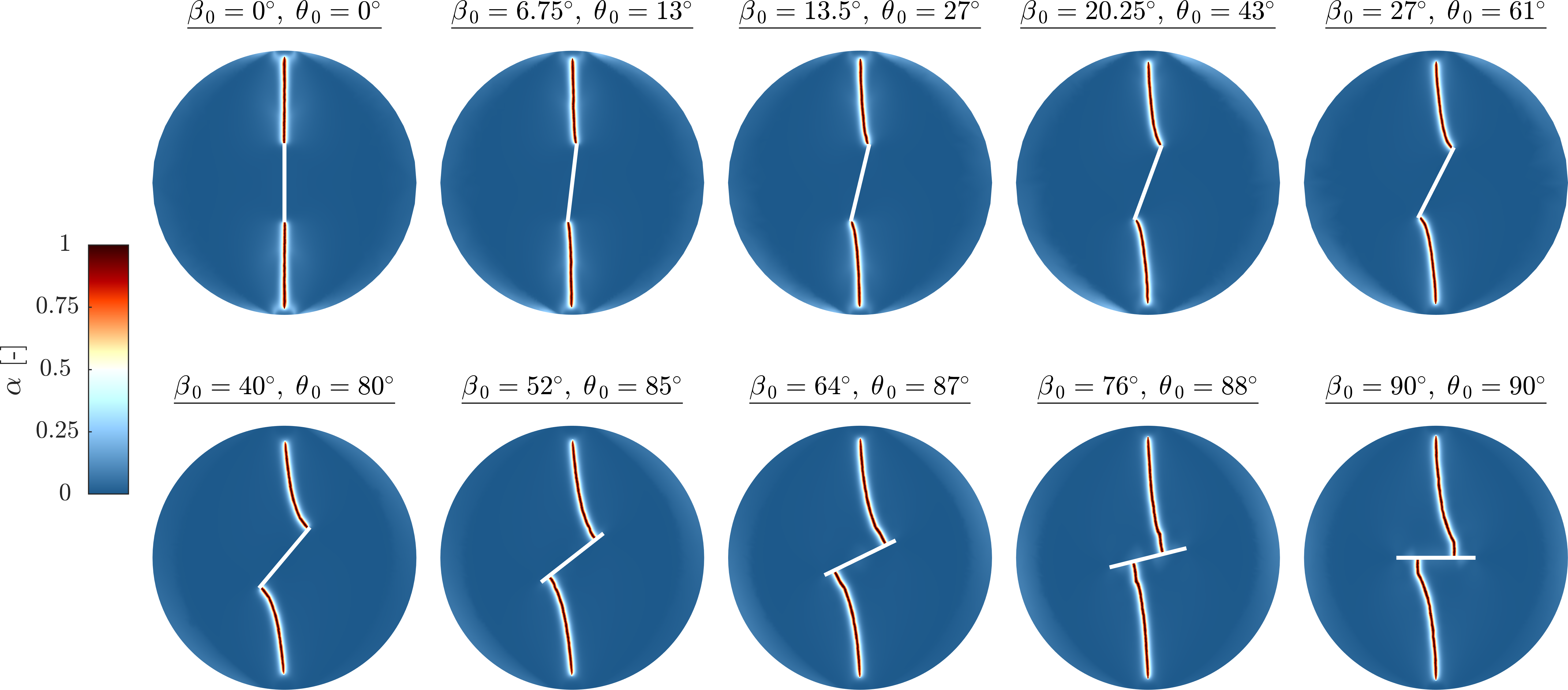}
\caption{Fractured specimens showing the final damage profiles for different load mixities obtained by varying the inclination of the initial flaw. In all cases, tensile fracture is achieved.}
\label{brazil_mix_contour}
\end{figure}

Figure~\ref{brazil_mix_contour} shows the resulting fracture patterns for varying $\beta_0$ and the corresponding initial crack propagation angle $\theta_0$. Focusing first on the transition from mode I to mode II loading (top row), we observe that cracks nucleate at the tips of the initial flaw, with $\theta_0$ increasing from $0^\circ$ at $\beta_0=0^\circ$ to  $61^\circ$ at $\beta_0=27^\circ$. Thus, pure mode I loading results in self-similar crack propagation and tensile splitting, while mixed-mode and mode II loading result in wing-shaped cracks. For higher $\beta_0$ ($40^\circ\leq\beta_0\leq90^\circ$; bottom row),  crack nucleation shifts from the tips to the edges of the flaw, with an increasing distance from the tip as $\beta_0$ increases. Moreover, the type of failure shifts from wing-shaped cracks to tensile splitting, with $\theta_0=90^\circ$ at $\beta_0=90^\circ$. 

\begin{figure}
  \centering
    \includegraphics[scale=0.95]{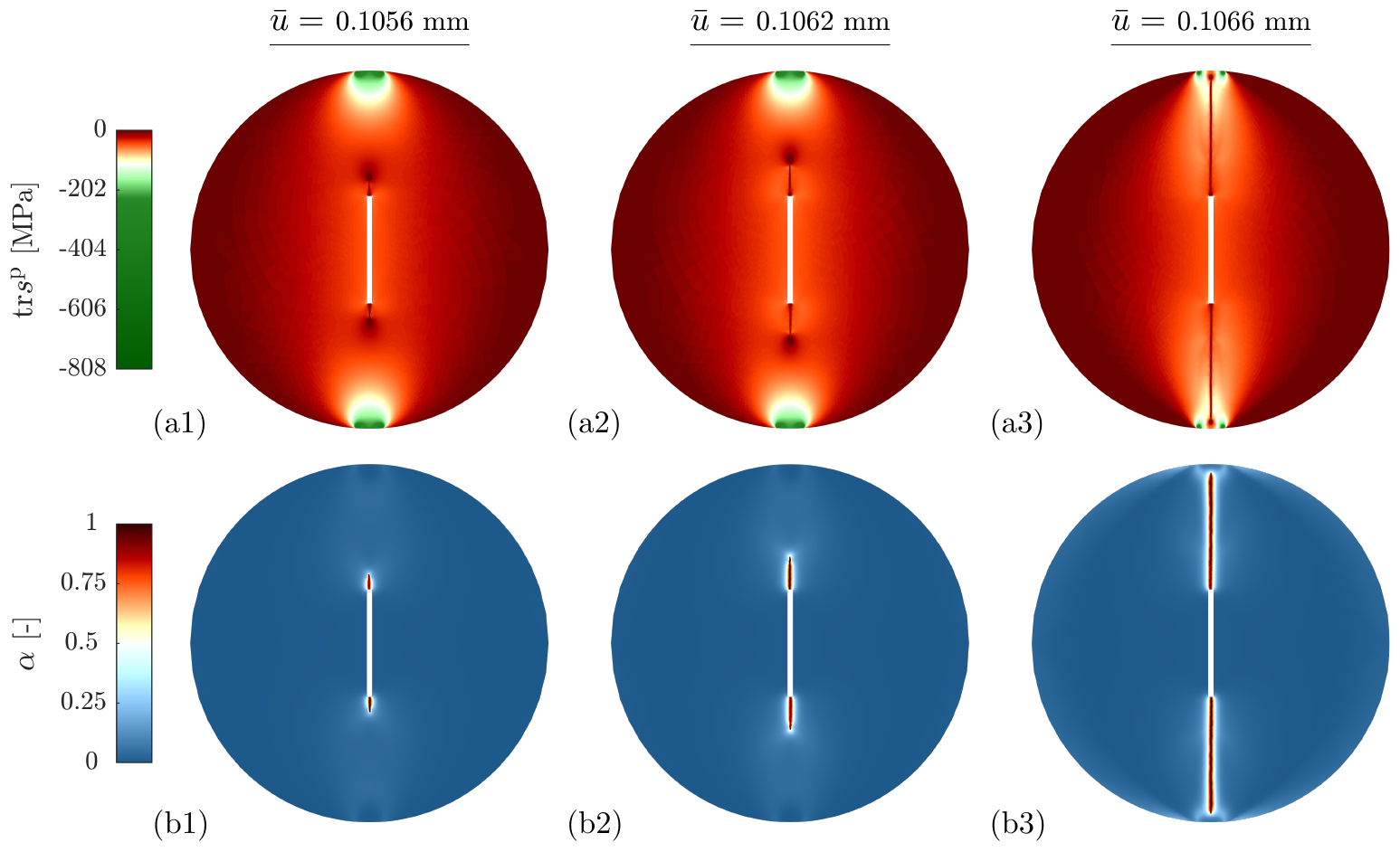}
\caption{Tensile fracture process  at different time steps for the specimen with $\beta_0=0^\circ$, showing (a) $\tr\Bsp$ in MPa and  (b) the corresponding damage profile.}
\label{brazil_0}
\end{figure}

\begin{figure}
  \centering
    \includegraphics[scale=0.95]{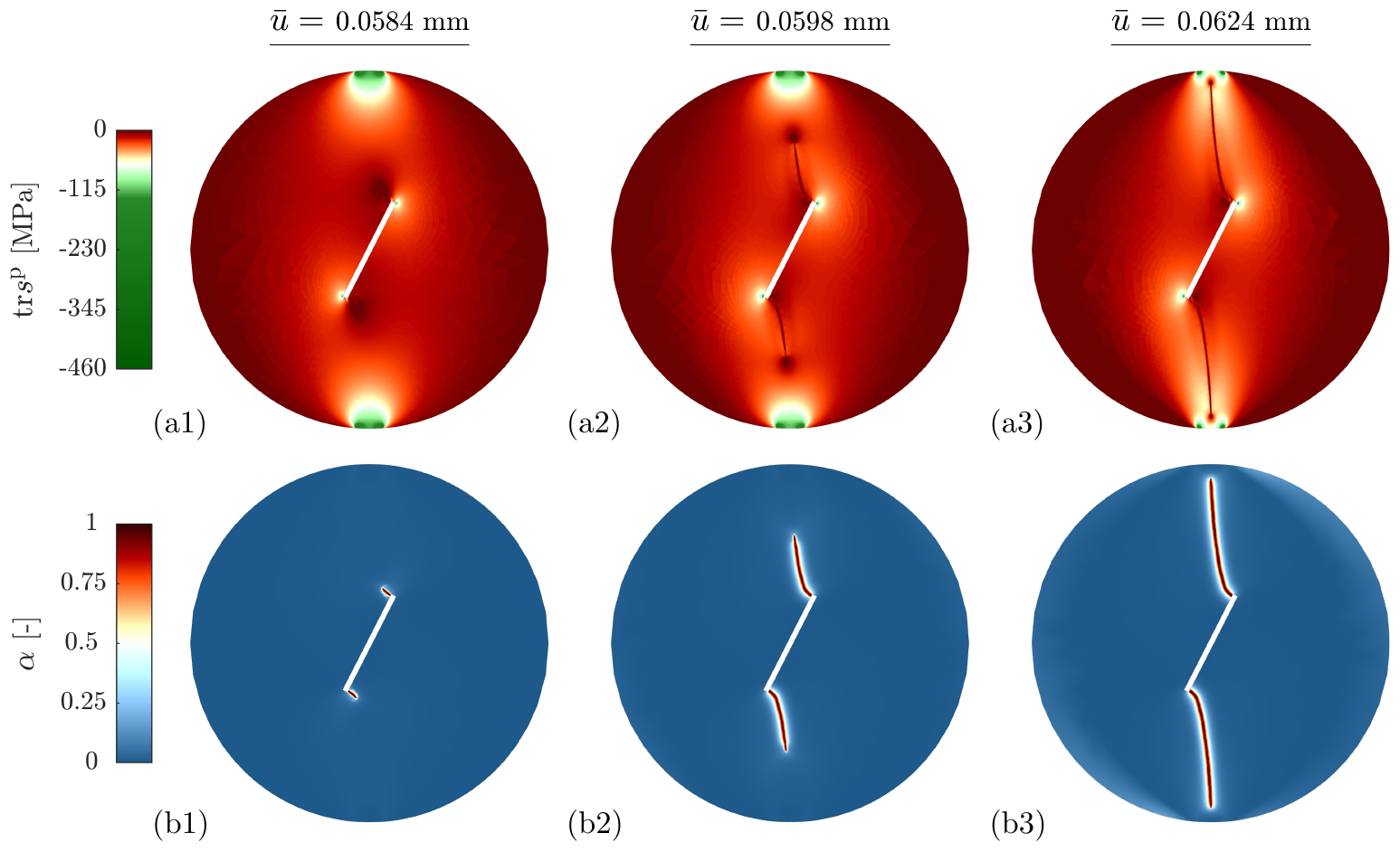}
\caption{Tensile fracture process  at different time steps for the specimen with $\beta_0=27^\circ$, showing (a) $\tr\Bsp$ in MPa and (b) the corresponding damage profile.}
\label{brazil_27}
\end{figure}

\begin{figure}
  \centering
    \includegraphics[scale=0.95]{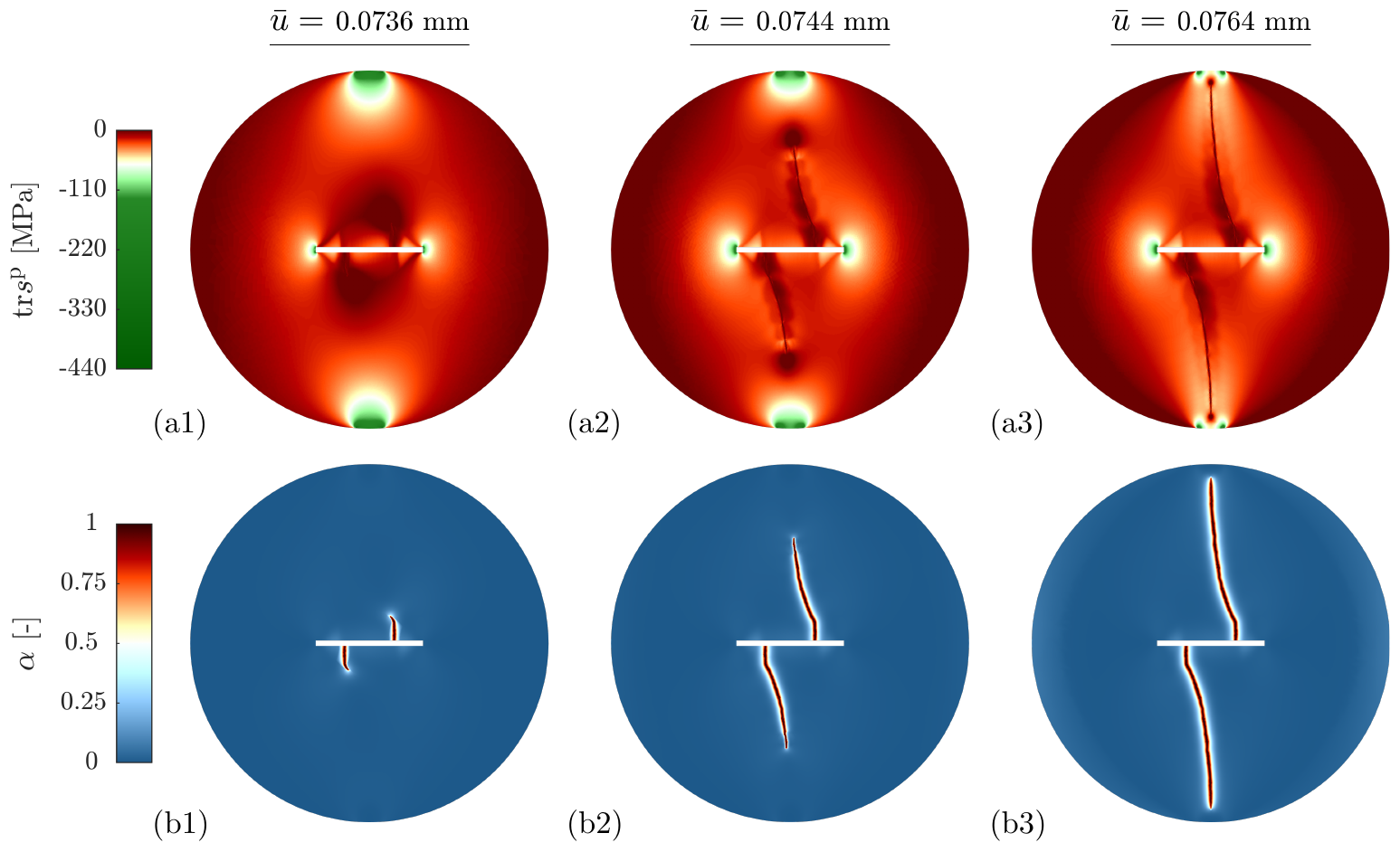}
\caption{Tensile fracture process at different time steps for the specimen with $\beta_0=90^\circ$, showing (a) $\tr\Bsp$ in MPa and (b) the corresponding damage profile.}
\label{brazil_90}
\end{figure}

A key observation is that, in all cases shown in figure~\ref{brazil_mix_contour}, the resulting fractures are tensile. Thus, mode I fracture is predicted by the model for all loading modes, with fracture patterns in agreement with  both numerical and experimental observations reported in the literature~\cite{aliha2010,haeri2014,xiankai2018,wang2020phase}. Figures~\ref{brazil_0}, \ref{brazil_27}, and~\ref{brazil_90} show the fracture process for $\beta_0=0^\circ$, $\beta_0=27^\circ$, and $\beta_0=90^\circ$, respectively. In all cases, at a given time step, a marked tensile region physically linked to opening microcracks ($\tr\Bsp=0$) can be observed ahead of the crack tip, paving the way for tensile crack propagation at the obtained orientations $\theta_0$. This region is clearly distinguished from compressive/shear regions ($\tr\Bsp<0$) in other parts of the domain, particularly at the caps where displacements are imposed, which in turn show a Boussinesq-like generalized stress distribution. Moreover, for $\beta_0=27^\circ$ and $\beta_0=90^\circ$, a marked compressive/shear region is also present at the crack tips, but in directions different from $\theta_0$. These results highlight an important feature of the model: the ability to preclude unrealistic fracture in compressive regions without resorting to the usual heuristic energy splits.

Finally, we proceed to assess the initial crack propagation angles in relation to the orientations predicted by classical fracture criteria, for which we consider the maximum tangential stress (MTS) criterion~\cite{erdogan1963} and the generalized maximum tangential stress (GMTS) criterion~\cite{smith2001}. In both cases, mode I crack propagation is predicted in the direction of maximum tangential stress $\sigma_{\theta\theta}$ near the crack tip. However, the GMTS criterion employs an approximation of $\sigma_{\theta\theta}$ that retains the non-singular term, and thus considers the maximum $\sigma_{\theta\theta}$ at a critical radial distance from the crack tip. As a consequence, the GMTS criterion takes into account the effect of both the $T$-stress and the fracture process zone. For pre-cracked Brazilian splitting tests, it has been shown experimentally~\cite{ayatollahi2008,aliha2010,xiankai2018} and numerically~\cite{wang2020phase} that the MTS criterion strongly overestimates the initial crack propagation angle, while the GMTS criterion provides a much better agreement. Therefore, it is interesting to verify if the present model agrees with these observations.

According to the MTS criterion, the initial crack propagation angle $\theta_0$ is computed from 
\begin{equation}
{Y_\mathrm{I}}\sin\theta_0+ {Y_\mathrm{II}}(3\cos\theta_0-1)=0.
\label{MTS}
\end{equation}
On the other hand, the GMTS can be expressed in terms of dimensionless parameters as~\cite{wang2020phase}
\begin{equation}
{Y_\mathrm{I}}\sin\theta_0+ {Y_\mathrm{II}}(3\cos\theta_0-1) - \frac{64}{3}T^*\sqrt{\frac{r_\mathrm{c}}{R}}\cos\theta_0\sin\frac{\theta_0}{2}=0.
\label{GMTS}
\end{equation}
In this expression, $R=25$ mm denotes the radius of the specimen (figure~\ref{brazil_scheme}), $T^*$ denotes the dimensionless form of the $T$-stress, and $r_\mathrm{c}$ is a critical fracture zone distance, which can be estimated as
\begin{equation}
r_\mathrm{c}=\frac{1}{2\pi}\bigg(\frac{{K}_\mathrm{cI}}{\sigma_\mathrm{t}}\bigg)^{\!2}.
\end{equation}
Here, ${K}_\mathrm{cI}$ is the mode I fracture toughness, computed from equation~\eqref{GcIKcI}. Moreover, $\sigma_\mathrm{t}$ is the uniaxial tensile strength, here  computed as in standard phase-field models, i.e., $\sigma_\mathrm{t}=\sqrt{27\,G_\mathrm{cI}E'/(256\,\ell)}$~\cite{marigo2016,tanne2018}. 

To compute $\theta_0$ from the MTS criterion~\eqref{MTS} and the GMTS criterion~\eqref{GMTS}, the values of $Y_\mathrm{I}$, $Y_\mathrm{II}$, and $T^*$ for varying $\beta_0$ were extracted from previous works~\cite{ayatollahi2007,wang2020phase}. Therein, $Y_\mathrm{I}$, $Y_\mathrm{II}$, and $T^*$ were computed for the same geometry (figure~\ref{brazil_scheme}) using stress intensity factors obtained from finite element analyses. 

\begin{figure}[H]
  \centering
    \includegraphics[scale=0.6]{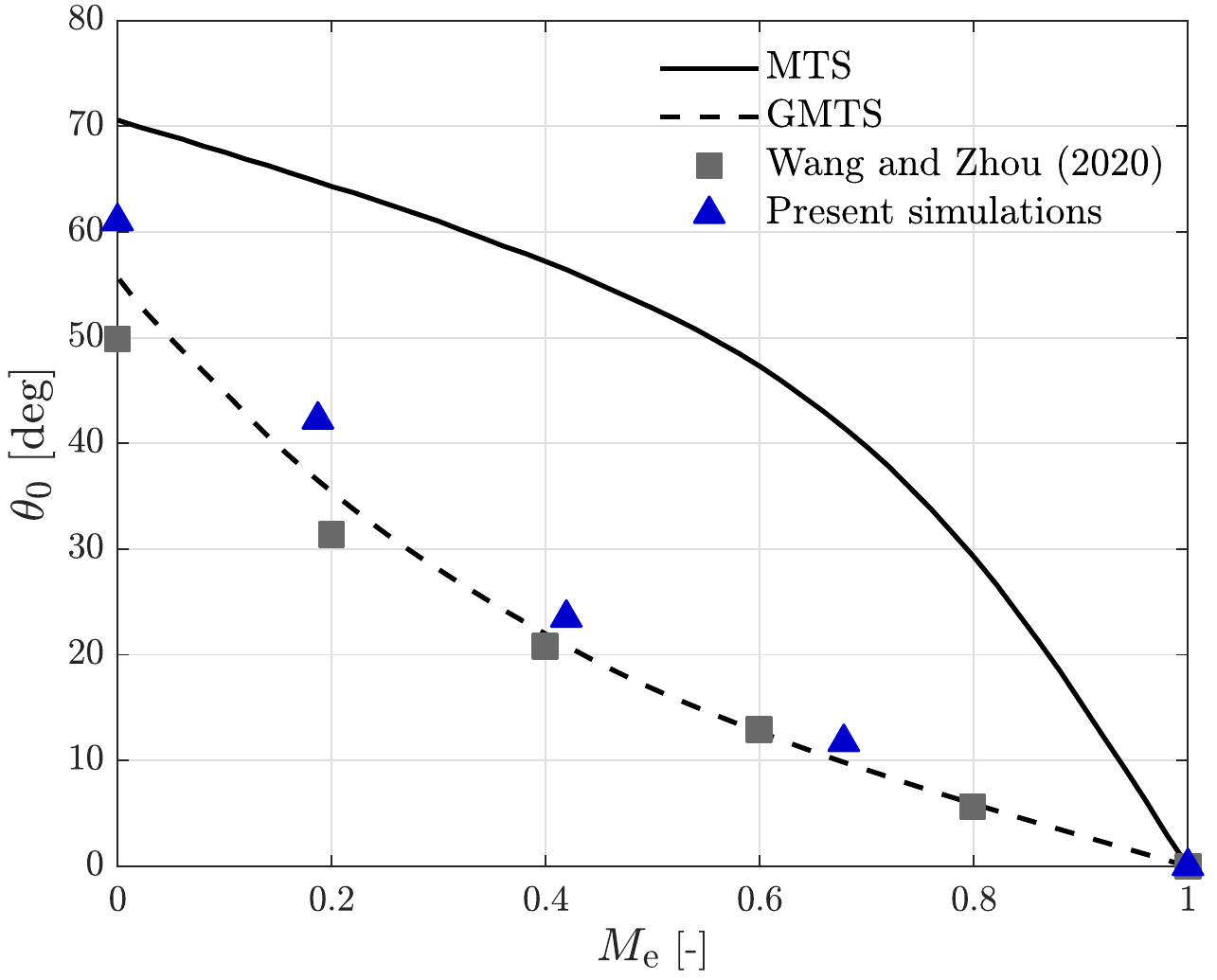}
\caption{Crack initiation angles for the different load mixities and comparison with the MTS and GMTS criteria, as well numerical results reported in~\citet{wang2020phase} for a standard phase-field model with a spectral energy split.}
\label{angles}
\end{figure}

The mode mixity for $0^\circ\leq\beta_0\leq27^\circ$ was computed from equation~\eqref{mixity} and plotted versus the initial crack propagation angle obtained from the simulations (measured from figure~\ref{brazil_mix_contour}), where $M_\mathrm{e}=1$ for $\beta_0=0^\circ$ (pure mode I), $M_\mathrm{e}=0.19$ for $\beta_0=6.75^\circ$, $M_\mathrm{e}=0.42$ for $\beta_0=13.5^\circ$, $M_\mathrm{e}=0.68$ for $\beta_0=20.25^\circ$, and  $M_\mathrm{e}=0$ for $\beta_0=27^\circ$ (pure mode II). The results are compared with the values predicted by both the MTS and the GMTS criteria, showing close agreement with the latter. As shown in figure~\ref{angles}, this result is consistent with the simulations performed in~\citet{wang2020phase}. Therein, a standard phase-field model with a spectral energy split~\cite{MieHofWel2010} was employed, \tC{as typically done in the phase-field literature for Brazilian tests~\cite{navidtehrani2021simple}}. Consequently, we conclude that, by virtue of the micromechanics-based free energy~\eqref{free}, the proposed model can recover the behavior expected from classical mixed-mode fracture mechanics, as well as the predictions of standard phase-field models that are enhanced with a typical energy split. Moreover, compared to~\citet{wang2020phase}, the profiles shown in figure~\ref{brazil_mix_contour} show much thinner cracks, although these differences may be attributed to differences in the numerical treatment of the governing equations.

\FloatBarrier

\subsection{Brittle-to-ductile transition in a dog-bone--shaped specimen}

The last example addresses the response of dog-bone--shaped specimens under tensile loading at different confining pressure levels. \citet{ramsey2004} studied this problem experimentally, revealing that tensile fracture occurs under little confining pressure, and that a continuous transition to shear fracture is observed as the confining pressure increases.  The problem was studied numerically under plane-strain conditions in~\citet{choo2018} using a non-variational phase-field model at finite strains. Therein, a spectral energy split~\cite{MieHofWel2010} was employed for the elastic energy density, while a split based on the Jacobian of the plastic deformation gradient was proposed for the plastic energy density.

\begin{figure}[t]
  \centering
    \includeinkscape[scale=1.25]{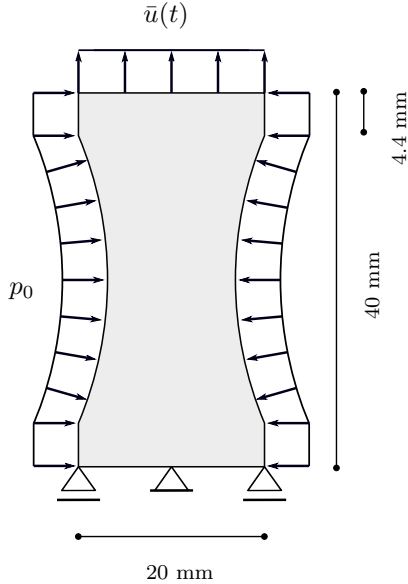}
\caption{Schematic representation of the tensile test in the dog-bone--shaped specimen under confining pressure. The cross-section decreases radially from $20$ mm to $13.6$ mm at the center of the specimen.}
\label{b2d_scheme}
\end{figure}

Figure~\ref{b2d_scheme} shows the geometry and boundary conditions considered in the present study. A mesh of $\sim$11500 quadrilateral elements with refinement in the central region was employed, with a minimum characteristic element size $h_\mathrm{min}=0.15$ mm. The test is divided into two loading stages. In the first stage, confining pressure is gradually applied up to $p_0$, keeping the vertical displacements at the bottom edge fixed, where only the center node is also fixed horizontally. For the second loading stage, the lateral pressure $p_0$ is  fixed while vertical displacements $\bar{u}$ are imposed upwards on the top edge. The loading increments are selected depending on the level of confining pressure, with $5\times10^{-4}$ mm for $p_0\leq 5$ MPa and $5\times10^{-3}$ mm otherwise.

The material parameters are chosen as follows:  Young's modulus $E=5.96$ GPa, Poisson's ratio $\nu=0.15$, mode I fracture toughness $G_\mathrm{cI}=16$ N/m, mode II fracture toughness $G_\mathrm{cII}=250$ N/m, fracture length scale $\ell=1$ mm, degradation parameter $b=1$, friction angle $\varphi=30^\circ$, and dilation angle $\theta=25^\circ$.

\begin{figure}
  \centering
    \hspace*{-2.5cm}\includegraphics[scale=1]{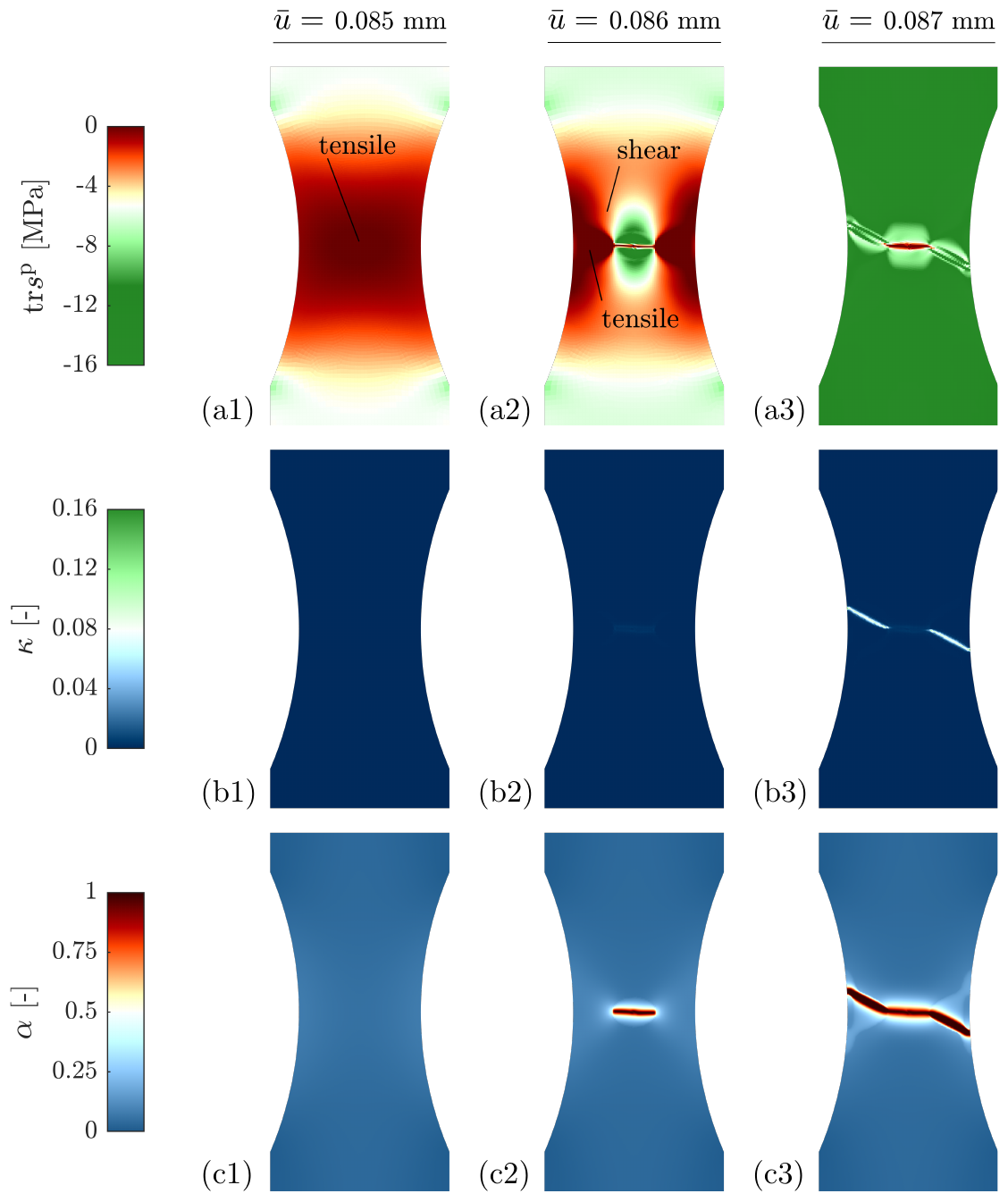}
\caption{Fracture process at different time steps for the dog-bone--shaped specimen with $p_0=5$ MPa, showing (a) $\tr\Bsp$ in MPa, (b) the corresponding equivalent plastic strains, and (c) the corresponding damage profiles.}
\label{b2d_5mpa}
\end{figure}

Figure~\ref{b2d_5mpa} shows the fracture process for $p_0=5$ MPa. The hydrostatic generalized stress $\tr\Bsp$ in figure~\ref{b2d_5mpa}(a) shows tensile regions ($\tr\Bsp=0$, i.e., open microcracks) that can be clearly distinguished from compressive/shear regions ($\tr\Bsp<0$, i.e., closed microcracks  with frictional sliding). Accordingly, the equivalent plastic strain in figure~\ref{b2d_5mpa}(b) vanishes in the tensile regions and strongly localizes in the critical compressive/shear regions, respectively corresponding to tensile fracture and shear fracture in figure~\ref{b2d_5mpa}(c). At $\bar{u}=0.085$ mm, $\tr\Bsp$ shows a clear tensile region at the center of the specimen, while $\kappa$ remains null and $\alpha$ remains low and homogeneous. In the next time step, a central tensile crack propagates horizontally in figure~\ref{b2d_5mpa}(c2). At this point, the $\tr\Bsp$ profile in figure~\ref{b2d_5mpa}(a2) shows distinctive tensile and compressive/shear regions ahead of the crack tip. Due to the influence of confining pressure, secondary shear fractures propagate from the crack tips in the compressive/shear regions. We interpret this response as a mixed or hybrid fracture pattern, exhibiting tensile (mode I) and shear (mode II) cracks in different parts of the domain. Thus, we expect a transition to purely tensile fracture at lower confining pressure, and a transition to purely shear fracture at higher~confining~pressure.

\begin{figure}
  \centering
    \includegraphics[scale=0.6]{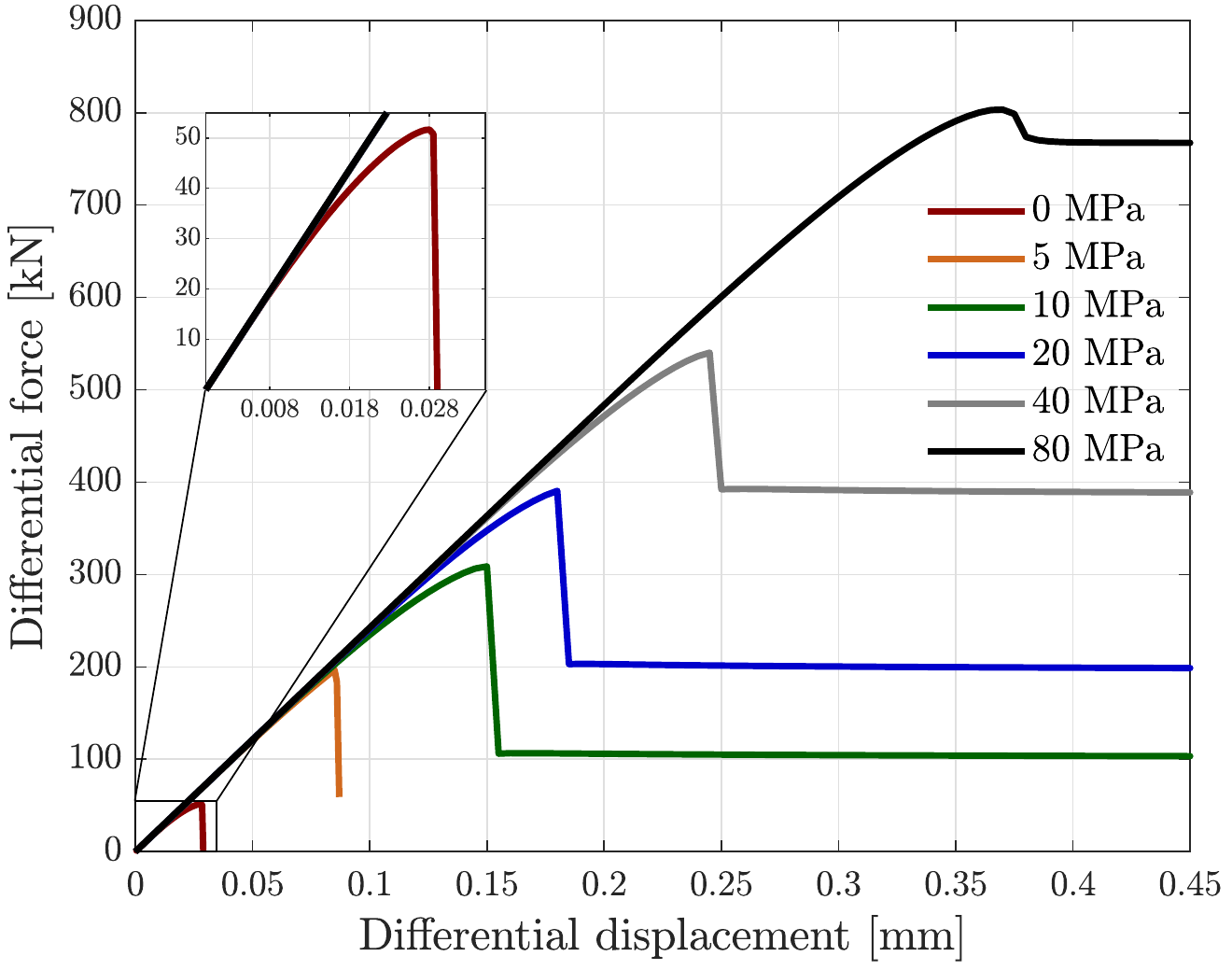}
\caption{Force-displacement diagram for varying confining pressures, exhibiting brittle-to-ductile transition.}
\label{b2d1}
\end{figure}

\begin{figure}[H]
  \centering
    \includegraphics[scale=0.84]{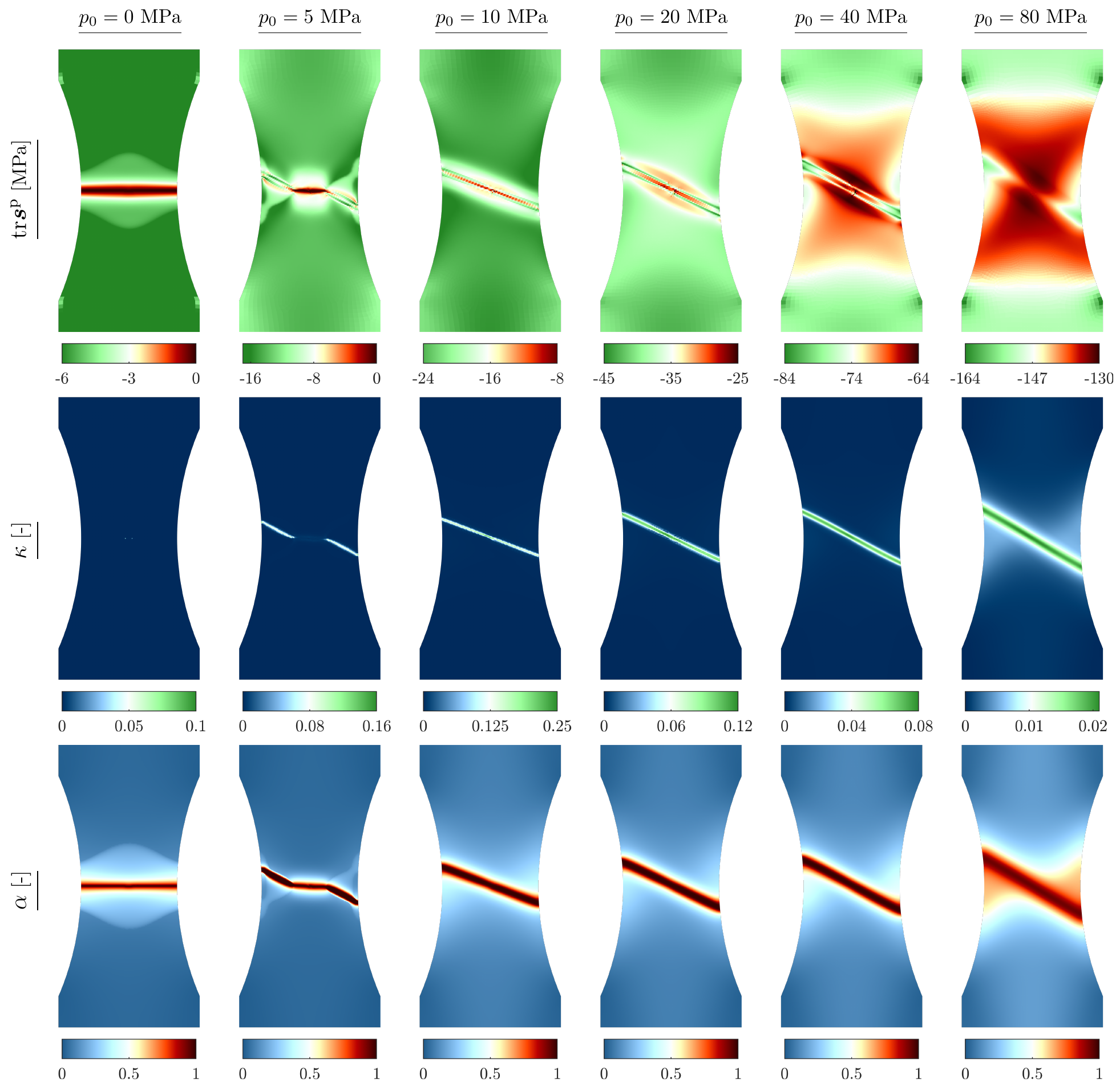}
\caption{Fractured specimens showing the post-failure  $\tr\Bsp$ [MPa] profiles (top), the corresponding equivalent plastic strain field (middle), and the corresponding damage field (bottom) for different confining pressures (see also figure~\ref{b2d1}).}
\label{b2d2}
\end{figure}

To address this hypothesis, we subject the specimen to varying confining pressures $0\leq p_0\leq80$~MPa.  Figure~\ref{b2d1} shows the resulting force-displacement curves, while figure~\ref{b2d2} shows the corresponding profiles for the hydrostatic generalized stress $\tr\Bsp$, the equivalent plastic strain field $\kappa$, and the damage field $\alpha$. Note that the main topological changes in the fracture pattern are observed between $p_0=0$ MPa, $p_0=5$ MPa, and $p_0=10$ MPa, as expected. At $p_0=0$ MPa, figure~\ref{b2d1} shows a typical brittle response, while figure~\ref{b2d2} shows a mode I fracture, which propagates brutally in the horizontal direction in the absence of plastic strains. In contrast, for $p_0>5$ MPa, the response becomes increasingly ductile, capturing the brittle-to-ductile transition. Thus, figure~\ref{b2d1} shows force-displacement curves with residual strength, while figure~\ref{b2d2} shows inclined shear fractures accompanied by localized plastic strains. As $p_0$ increases, the peak load, the corresponding failure displacement, and the residual strength increase, ending with a very limited softening stage for $p_0=80$ MPa. Accordingly, the plastic strains become increasingly diffuse (but still localized), while the fracture orientation angle with respect to the horizontal axis increases. This progressive increase in the fracture orientation angle can be viewed as a continuous transition from tensile fracture at low confining pressure to shear fracture at high confining pressure, in agreement with experimental observations~\cite{ramsey2004}.

Moreover, these results are in agreement with numerical results reported in the literature for a similar problem~\cite{choo2018}. However, in contrast with~\citet{choo2018}, the proposed model is able to predict strongly localized plastic strains and much more delineated crack profiles, allowing us to clearly distinguish tensile fracture from shear fracture. Moreover, the brittle-to-ductile transition includes a clear transitional mode at $p_0=5$ MPa. We further note that most previous works dealing with similar problems adopt a heuristic split for the strain energy and/or the crack driving force, while in the proposed model, the different failure modes are the consequence of the micromechanics-based formulation. 

\FloatBarrier

\section{{Conclusions}}

We have presented a micromechanics-based gradient-damage/phase-field model for fracture in quasi-brittle geomaterials. Two distinctive behaviors were embedded in the formulation: a brittle tensile regime, corresponding to the growth of opening microcracks and mode I fracture, and a ductile compressive/shear regime, corresponding to the growth of closed microcracks under frictional sliding and mode II fracture. By virtue of the micromechanical arguments, the model was constructed with a limited number of parameters and field variables, all of which can be linked to physical lower-scale mechanisms. A direct relation was thus established between the gradient-damage/phase-field variable and a microcrack density parameter, as well as between plastic strains and the frictional sliding of closed microcracks. Moreover, the constitutive hardening/softening functions and parameters in the free energy density were defined as functions of the elastic material properties and a single degradation function. As a key feature, a non-associative plasticity law was considered, including the effect of the dilation angle. This feature was crucial to ensure a non-vanishing energy dissipation due to frictional sliding. Moreover, the model was constructed in variational form using the energetic formulation for rate-independent systems, where a careful treatment of the non-associative law was employed by adopting a generalized principle of maximum dissipation. Finally, the numerical implementation procedure was described in detail.

Several numerical simulations were conducted, highlighting the ability of the model to capture the expected failure modes under different loading conditions. Biaxial compression tests were performed in specimens with and without imperfections. The results from these tests showed (mode II) shear fracture orientations in agreement with experimental observations and bifurcation analyses of Mohr-Coulomb--based models.  Moreover, mixed-mode loading tests were performed on pre-cracked Brazilian specimens subjected to diametral compression for a wide range of initial flaw inclinations. The results showed a first transition from self-similar crack growth at pure mode I loading to wing-shaped cracks at pure mode II loading, and a second transition to tensile splitting for  higher initial flaw inclinations with respect to the loading axis. The (mode I) tensile cracks obtained in the pre-cracked Brazilian tests were in agreement with classical mixed-mode fracture mechanics, as expected from experimental observations. Finally, dog-bone--shaped specimens were subjected to tension under different confining pressure levels. In qualitative agreement with experimental evidence, the results showed a brittle-to-ductile transition with increasing confining pressure, from tensile fracture to shear fracture, including a hybrid transitional mode. We highlight that the results of the different tests were obtained as a consequence of the micromechanics-based formulation, providing a physically meaningful alternative to the wide variety of heuristic modifications that have been proposed for the strain energy density and/or the crack driving force in the phase-field literature. 

The present study paves the way for several topics of future research. As presented in appendix~\ref{dam_fric_dil}, a straightforward extension to damage-dependent friction and dilation allows for decreasing the frictional resistance and reaching a constant volume state in the post-critical stage. Further, an enhancement of the formulation to model the response under cyclic loading is discussed in appendix~\ref{cyclic}. \tC{Another topic of interest consists of relaxing the assumption of isotropic behavior, which may be achieved by taking the more general micromechanical framework with multiple crack families~\cite{zhu2015} as a point of departure, and extending the phase-field model to multiple damage variables (cf.~\cite{bleyer2018phase}) and multiple plastic strains}. Other topics include \tC{performing 3D simulations, e.g., to capture the brittle-to-ductile transition in triaxial cylinders}, the inclusion of a compression cap mechanism to model plastic compaction, and the extension to fluid-driven fracture in porous media. Finally, a thorough experimental verification study is worth considering to quantitatively assess the performance of the model in relation to standard test procedures in both plane-strain and triaxial~conditions.

%\clearpage

\appendix

\section*{Appendix}

\section{On possible extensions}

This appendix presents extensions of the proposed model that may prove useful for future~work. 

\subsection{Variable friction and dilation}\label{dam_fric_dil}

\begin{figure}[H]
  \centering
   \includegraphics[scale=1]{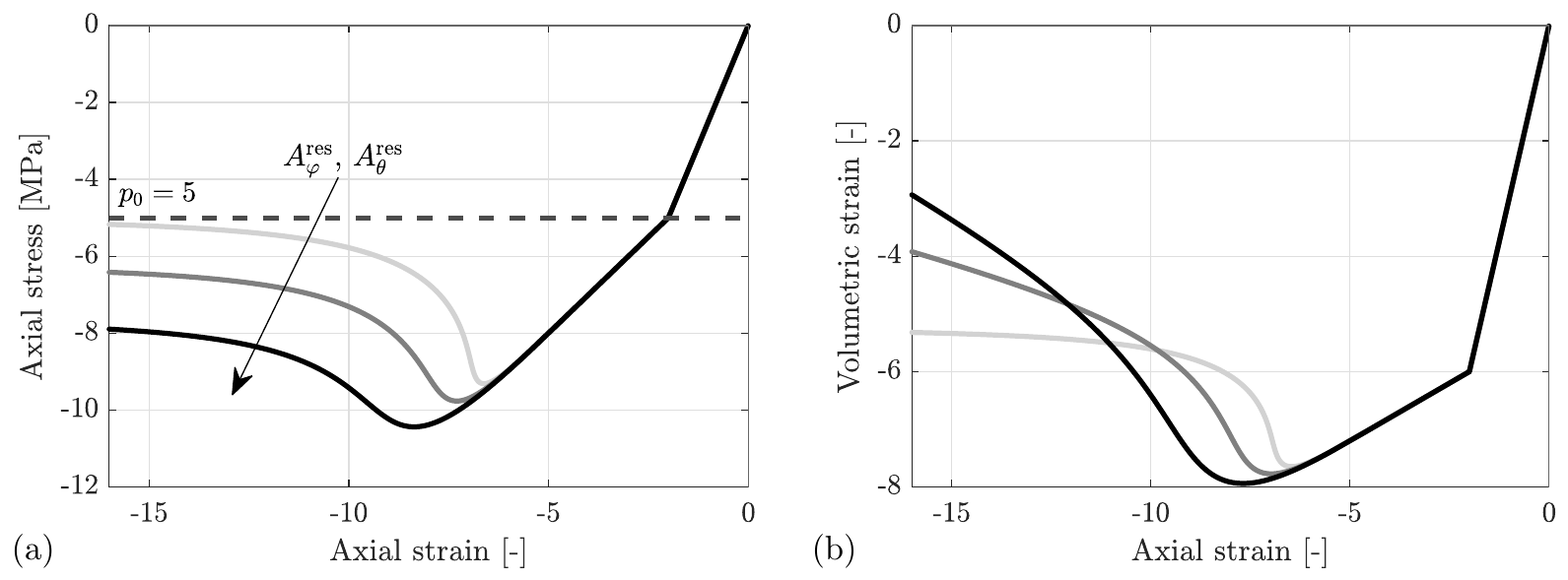}
\caption{Homogeneous response for uniaxial compression with different residual parameters $A_\varphi^\mathrm{res}$ and $A_\theta^\mathrm{res}$: (a) axial strain vs.~axial stress curves and (b) the corresponding total volumetric strains. The responses are shown for $A_{\Box}^\mathrm{res}\in\{A_{\Box}^\mathrm{peak}, 0.5\,A_{\Box}^\mathrm{peak}, 0\}$.}
\label{hom_mon_conf_soft}
\end{figure}

So far, the friction and dilation coefficients have been considered constant. A straightforward extension is to consider functions $\alpha\mapsto A_\varphi(\alpha)$ and $\alpha\mapsto A_\theta(\alpha)$, with $A_\Box(0)=A_\Box^\mathrm{peak}$ and $A_\Box(1)=A_\Box^\mathrm{res}$. A simple example is
\begin{equation}
A_\varphi(\alpha)\coloneqq A_\varphi^\mathrm{res}-(1-\alpha)^2(A_\varphi^\mathrm{res}-A_\varphi^\mathrm{peak}), \quad A_\theta(\alpha)\coloneqq A_\theta^\mathrm{res}-(1-\alpha)^2(A_\theta^\mathrm{res}-A_\theta^\mathrm{peak}).
\end{equation}
The homogeneous response for such a model is shown in figure~\ref{hom_mon_conf_soft} with fixed $A_\varphi^\mathrm{peak}=0.15$ and $A_\theta^\mathrm{peak}=0.0375$, while $A_\varphi^\mathrm{res}$ and $A_\theta^\mathrm{res}$ vary. The main observation is that as $A_\varphi^\mathrm{res}\to0$, the residual strength vanishes and the stress drops to the imposed confining pressure. We interpret this response as the degradation of the surface asperities in frictional contact. On the other hand,  as $A_\theta^\mathrm{res}\to0$, the volumetric strain approaches a constant value resembling a \emph{critical state}, a crucial concept in soil mechanics~\cite{schofield1968}.

\subsection{Opening/closure transition for cyclic loading}\label{cyclic}

As presented in this paper, the proposed model relies on the opening/closure condition~\eqref{optrans} to distinguish between the tensile and compressive/shear regimes. Thereby, $\tr\Bsp$ is known from the solution of the plasticity evolution equations (table~\ref{overview2}). Under monotonic loading, the solution of these equations properly predicts $\tr\Bsp=0$ and $\tr\Bsp<0$ in the tensile and compressive/shear regimes, respectively. The same observation can be made about unloading from a compressive/shear state and further loading in the tensile regime. 

\begin{figure}[H]
  \centering
   \includegraphics[scale=1]{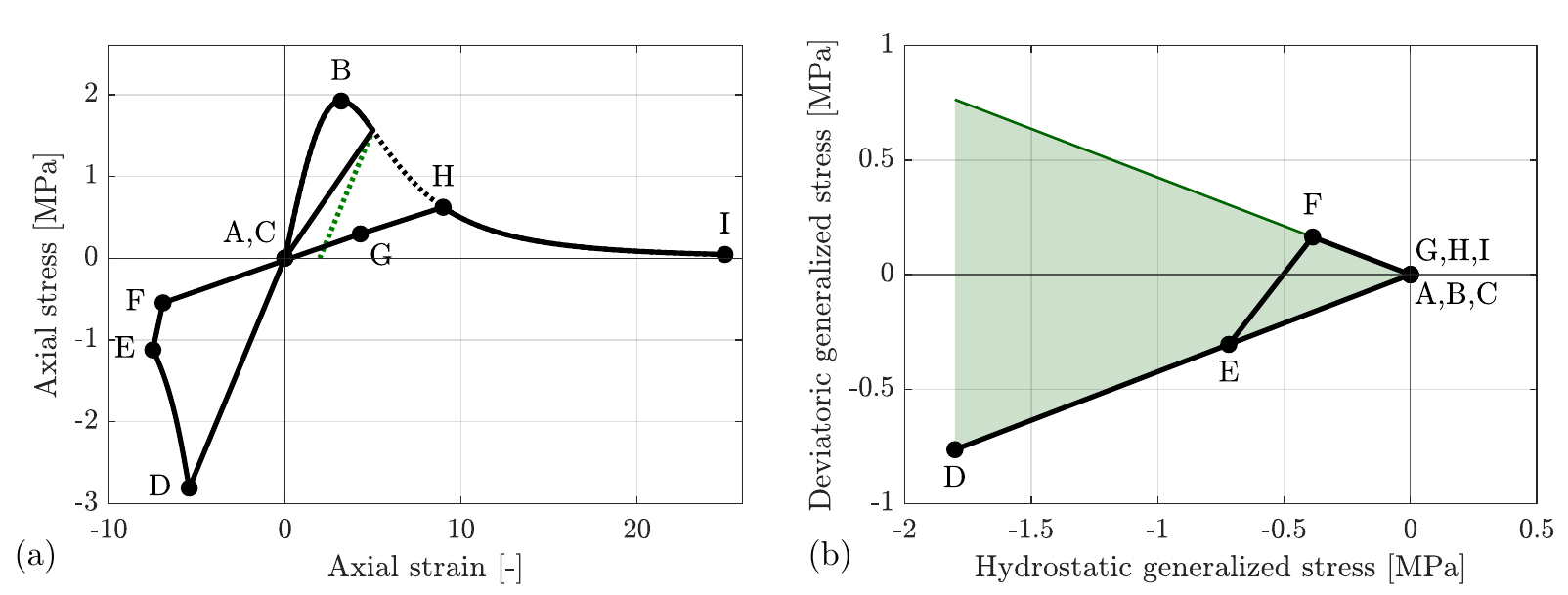}
\caption{Homogeneous response for uniaxial cyclic loading: (a) axial stress vs.~axial strain curves and (b) the corresponding stress path plotted in $\big(\sqrt{3}\tr\Bsp/3,\mathrm{sign}(s^\mathrm{p}_{\mathrm{dev}\,zz})\Vert\Bsp_\mathrm{dev}\Vert\big)$ space. The dotted green line corresponds to unloading from the tensile stage if condition~\eqref{rhobar} is not considered.}
\label{hom_tens2comp2tens}
\end{figure}

Consider now the case in which the material is first loaded in tension and then unloaded, as shown in figure~\ref{hom_tens2comp2tens} (points A, B, and C). Because the tensile regime is brittle, we expect the unloading branch to return to the origin in figure~\ref{hom_tens2comp2tens}(a). However, the plasticity evolution equations predict $\tr\Bsp<0$ as soon as unloading takes place. As a consequence, the unloading branch follows the dotted green line and yields residual strains. The reason for this unexpected behavior is that, according to the Drucker-Prager model, $\tr\Bep$ is irreversible. We thus conclude that, for cyclic loading, the KKT plasticity conditions should not be verified in the tensile regime. Of course, this renders the opening/closure condition~\eqref{optrans} insufficient to characterize the response. We propose to remedy this issue by complementing condition~\eqref{optrans} with a new history parameter $\bar{\rho}$ intended to track the microcrack opening state. To this end, during the tensile regime, we prescribe the microcrack opening measure $\tr\Bep$ from equation~\eqref{ep_reqs} and define
\begin{equation}
\bar{\rho}(\bm{x},t)\coloneqq\begin{dcases} \int_{t_*}^t\tr\dot{\bm{\varepsilon}}^\mathrm{p}(\bm{x},s)\,\mathrm{d}s, \quad \tr\Bep=\big(1-g_K(\alpha)\big)\tr\Be & \quad \text{if \sf{open}},  \\
0 & \quad \text{if \sf{closed}}.
\end{dcases}
\label{rhobar}
\end{equation}
Here, $t_*$ corresponds to the last time step in which a transition from the compressive/shear to the tensile regime took place. Therefore, in a full loading-unloading process in tension, the integral on top vanishes. Thus, we prevent the verification of the KKT plasticity conditions when $\bar{\rho}>0$. By incorporating this condition in the solution of the governing equations, unloading from the tensile regime returns to the origin (point C) in figure~\ref{hom_tens2comp2tens}(a), as expected. For illustrative purposes, the material is further loaded in compression up to point E, and then unloaded elastically to point F. Beyond this point, the stress path approaches the apex at point G, signaling the transition to the tensile regime. Then, the material is fully damaged in~tension. 

For simplicity of presentation, and in view of the monotonic loading conditions considered in the numerical examples, condition~\eqref{rhobar} was not considered in the present work, but it seems to be crucial for computing the response under cyclic loading. This topic will be considered in future studies. Moreover, the effect of residual stress~\cite{salvati2021} in the response emerges as an interesting topic for further research.

\end{document}